\documentclass[11pt]{article}
\usepackage[utf8]{inputenc}
\usepackage[T1]{fontenc}
\usepackage{graphicx}
\usepackage{amssymb,amsmath,amsbsy}
\usepackage{caption}
\usepackage{comment}
\usepackage{hyperref}
\hypersetup{breaklinks=true}
\hypersetup{colorlinks=true}
\hypersetup{urlcolor=blue}
\hypersetup{citecolor=blue}
\hypersetup{linkcolor=blue}

\DeclareMathOperator\re{Re}
\DeclareMathOperator\im{Im}

\newcommand{\thb}{\bar{\theta}}
\newcommand{\psib}{\bar{\psi}}
\newcommand{\chib}{\bar{\chi}}
\newcommand{\kk}{\mathbf{k}}

\newcommand{\qq}{\mathbf{q}}

\newcommand{\qb}{\bar{q}}
\newcommand{\kb}{\bar{k}}
\newcommand{\g}[1]{«~#1~»}
\newcommand{\be}{\begin{equation}}
\newcommand{\ee}{\end{equation}}
\newcommand{\bea}{\begin{eqnarray}}
\newcommand{\eea}{\end{eqnarray}}
\newcommand{\ii}{\mathrm{i}}
\newcommand{\ff}{\mathrm{f}}
\newcommand{\eee}{\mathrm{e}}
\newcommand{\dd}{\mathrm{d}}
\newcommand{\veps}{\varepsilon}

\begin{document}

\title{Phonon number relaxation in a 3D superfluid with a concave acoustic branch}

\author{Yvan Castin and Mariia Tsimokha \\
Laboratoire Kastler Brossel, ENS-Université PSL, CNRS, \\
Université Sorbonne et Collège de France, 24 rue Lhomond, 75231 Paris, France \\
yvan.castin@lkb.ens.fr}

\maketitle 

\begin{abstract}
We consider the collisional evolution towards equilibrium of a spatially homogeneous and isotropic phonon gas of a three-dimensional superfluid with a concave acoustic excitation branch, at a non-zero but arbitrarily low temperature $T$. Three-phonon collisions $1\phi\leftrightarrow 2\phi$ are forbidden by conservation of energy-momentum. Four-phonon collisions $2\phi\to 2\phi$ of Landau and Khalatnikov lead, after a time $\propto T^{-7}$, only to a partial thermal equilibrium, a Bose law of non-zero chemical potential for the phonons, because they conserve the total number of phonons. Relaxation towards complete thermochemical equilibrium is therefore ensured by the much slower five-phonon collisions $2\phi\leftrightarrow 3\phi$ of Khalatnikov, in a time $\propto T^{-9}$. Using kinetic equations on the occupation numbers of the phonon modes and explicitly calculating the $2\phi\to 3\phi$ collisional amplitude with quantum hydrodynamics at low temperature, we determine the corresponding evolution of the fugacity $z_\phi$ of the phonon gas from the non-degenerate regime $z_\phi=0^+$ to complete equilibrium $z_\phi=1^-$. Using the conservation of total energy, we find that the fugacity varies with a non-integer power law $\propto t^{4/5}$ at short times and an exponential law at long times; the speed of change of entropy, always positive, is asymptotically proportional to the square of the speed of change of fugacity, $(\mathrm{d}/\mathrm{d}t)S_\phi\propto[(\mathrm{d}/\mathrm{d}t)z_\phi]^2$, as Landau predicted for an arbitrarily slow adiabatic transformation. Our results bring to a close the study initiated by Khalatnikov in 1950 and could be experimentally verified in a gas of cold spin-1/2 fermionic atoms in the sector of the BEC-BCS crossover where the acoustic branch has a concave start, or in superfluid liquid helium-4 at sufficiently high pressure.

\medskip
{{\bf Keywords.} Superfluid, acoustic branch, phonons, kinetic equations, phonon-phonon collisions, chemical potential, low-temperature limit}

\end{abstract}

\tableofcontents

\section{Problem Statement}\label{sec0}

Consider an isolated three-dimensional system of bosonic or fermionic particles with short-range interactions, spatially homogeneous with a number density $\rho$ inside an arbitrarily large quantization box $[0,L]^3$, fully superfluid in its ground state, and having as its only gapless excitation branch the linearly starting acoustic branch \cite{SonWingate,Escobedo,insuffisance} 
\begin{equation}
\label{eq:brac}
\hbar\omega_\mathbf{q} \underset{q\to 0^+}{=} \hbar c q \left[1+\frac{\gamma}{8}\left(\frac{\hbar q}{mc}\right)^2+O(q^4\ln q)\right]
\end{equation}
 Here $\omega_\qq$ is the natural angular frequency of the excitation quanta--the phonons $\phi$--with wave vector $\qq$; $c$ is the speed of sound at zero temperature, $m$ is the mass of a superfluid particle, and $\gamma$ is a dimensionless curvature parameter. At a non-zero but sufficiently low temperature, no other excitation branch can be significantly populated, and the system reduces to a bosonic gas of phonons, whose dispersion relation (\ref{eq:brac}) is reminiscent of an ideal-gas quadratic Hamiltonian $\hat{H}_2=\sum_\qq \hbar\omega_\qq \hat{b}_\qq^\dagger \hat{b}_\qq$, but which in reality interact via the Hamiltonian $\hat{H}_3$ of Landau and Khalatnikov’s quantum hydrodynamics, which is cubic in the creation and annihilation operators $\hat{b}_\qq^\dagger$ and $\hat{b}_\qq$ of the phonons; see Section 2.1 in \cite{insuffisance}, Section \S 24 in \cite{LandauStat2} and references \cite{LK,QO}:\footnote{Like any low-energy effective theory, quantum hydrodynamics must be regularized by an ultraviolet cutoff and treated according to the renormalization concepts standard in quantum electrodynamics; thus, in principle, the eigenfrequency $\omega_q$ in $\hat{H}_2$ should be taken not as the experimentally measured value $cq[1+(\gamma/8)(\hbar q/mc)^2+\ldots]$--referred to as \g{effective}-- as we do in this article but as the \g{bare} value $c^{(0)}q[1+(\gamma^{(0)}/8)(\hbar q/mc^{(0)})^2]$, where $c^{(0)}$ and $\gamma^{(0)}$ are the bare (unrenormalized) speed of sound and curvature parameter. The transition from the bare value to the effective value of $\omega_q$ then arises from phonon-phonon interactions. This renormalization procedure is an unnecessary luxury here: we are calculating a damping rate using the generalized Fermi's golden rule rather than an energy shift; the entire process takes place on the collisional energy shell, and no ultraviolet cutoff is required.}${}^{,}$\footnote{In quantum hydrodynamics (see the aforementioned references), the phonon-phonon interaction Hamiltonian also includes a contribution $\hat{H}_4$ that is quartic in the operators $\hat{b}$ and $\hat{b}^\dagger$; however, in the low-temperature limit, this term provides a sub-dominant contribution to the phonon-number relaxation rate calculated here--see the discussion following equation (\ref{eq:h3m})--so we omit it in this introduction.} 
\be
\label{eq:H3}
\hat{H}_3 = \frac{1}{L^{3/2}} \sum_{\qq,\qq'} \frac{1}{2}\mathcal{C}(\qq,\qq',\qq+\qq') \left(\hat{b}^\dagger_{\qq+\qq'}\hat{b}_\qq \hat{b}_{\qq'}+\mbox{h.c.}\right) +\frac{1}{3!} \mathcal{C}(\qq,\qq',-(\qq+\qq')) \left(\hat{b}_\qq\hat{b}_{\qq'}\hat{b}_{-(\qq+\qq')}+\mbox{h.c.}\right)
\ee
with the coupling amplitude depending on the dimensionless parameter $\Lambda$ from equation (\ref{eq:defLam}):
\be
\mathcal{C}(\qq,\qq',\qq'')=mc^2\left(\frac{\hbar^3 q q'q''}{8\rho m^3c^3}\right)^{1/2} \left(3\Lambda + \frac{\qq\cdot \qq'}{q q'} + \frac{\qq\cdot \qq''}{q q''} + \frac{\qq'\cdot \qq''}{q' q''} \right)
\ee
The superfluid thus constitutes a nonlinear medium for sound \cite{Maris} and, recall, the Hamiltonian $\hat{H}=\hat{H}_2+\hat{H}_3$ applies regardless of the strength of the interactions in the underlying superfluid or its liquid or gaseous nature, provided that the wave numbers $q$ are sufficiently small).

Suppose our phonon gas is initially prepared in a spatially homogeneous and isotropic excited state of low energy but out of equilibrium, and let us ask how it will relax toward full thermodynamic equilibrium of temperature $T_{\rm eq}$ and chemical potential $\mu_\phi^{\rm eq}=0$ under the effect of collisions between phonons (we necessarily have $\mu_\phi^{\rm eq}=0$ since $\hat{H}_3$ does not conserve the total number of phonons $N_\phi$).

When the acoustic branch (\ref{eq:brac}) is convex at low $q$ ($\gamma>0$), the dominant low-energy phonon collision processes are three-phonon processes $1\phi\leftrightarrow2\phi$, Belyaev’s decay processes $1\phi\to 2\phi$ (terms $\hat{b}^\dagger\hat{b}^\dagger\hat{b}$ in $\hat{H}_3$), and Landau’s inverse merging processes $2\phi\to1\phi$ (terms $\hat{b}^\dagger\hat{b}\hat{b}$ in $\hat{H}_3$),
 see section \S 34 of \cite{LandauStat2} and reference \cite{Maris}. They alter both the wave vector distribution $n_\qq$ of the phonons and their total number $N_\phi$, and lead to an a priori simultaneous attainment of thermal $T\to T_{\rm eq}$ and chemical $\mu_\phi\to \mu_\phi^{\rm eq}=0$ equilibrium, with a characteristic relaxation rate $\Gamma_\phi^{\rm eq}\propto T_{\rm eq}^{5}$ in the vicinity of equilibrium \cite{Annalen}.\footnote{We consider here only the typical thermal phonons of energy $\hbar\omega_\mathbf{q}\propto k_{\rm B}T$, which are at sufficiently low temperature in the weakly collisional regime $\omega_\mathbf{q}\gg\Gamma_\phi^{\rm eq}$ involving well-identified elementary processes $n\phi\leftrightarrow n'\phi$. We do not discuss phonons of angular frequency $\omega_\mathbf{q}\ll\Gamma_\phi^{\rm eq}$, which are in the hydrodynamic regime described by macroscopic transport coefficients such as viscosity \cite{KhalatLivre} and corresponding to a diffusive damping of sound in the fluid \cite{diffus}. The same restriction applies to $\gamma < 0$ regarding $2\phi \to 2\phi$ collisions.} This single-scale case is not studied here.

When the acoustic branch (\ref{eq:brac}) is concave at low $q$ ($\gamma<0$),\footnote{We do not address here the delicate case $\gamma=0$, which remains largely open \cite{QO}.} three-phonon processes no longer conserve energy-momentum, and the dominant collision processes become the four-phonon processes $2\phi\to 2\phi$ of Landau and Khalatnikov \cite{LK,EPL,Annalen} (more generally, note that processes $1\phi\to n\phi$ and $n\phi\to 1\phi$ violate the conservation of energy-momentum for all $n>1$). The four-phonon processes certainly ensure the thermal relaxation of the phonon wave vector distribution toward a Bose distribution with chemical potential $\mu_\phi$, 
\begin{equation}
\label{eq:loiB}
n_\mathbf{q} \to \bar{n}_\qq(T,\mu_\phi) = \frac{1}{\exp[\beta(\hbar\omega_\qq-\mu_\phi)]-1}
\end{equation}
 where $\beta=1/k_{\rm B}T$, with an instantaneous time constant $\propto T^{-7}$ in the sense that the damping rate of phonons with energy $\approx k_{\rm B}T$ is $\propto T^7 f(\beta\mu_\phi)$ with $f(\beta\mu_\phi)\approx 1$ for $\beta|\mu_\phi|\lesssim 1$ \cite{etalement},\footnote{As mentioned, it is assumed in this work that the distribution $n_\qq$ is isotropic. Otherwise, its thermal relaxation via the processes $2\phi\to 2\phi$ may take a time $\propto T^{-9}$, which undermines the scale separation to be discussed later; see reference \cite{Khalat1950} and note 4 of reference \cite{etalement}. Furthermore, recall that, in the low-temperature limit considered here, interactions between phonons contribute a negligible correction to the equilibrium distribution of an ideal gas (\ref{eq:loiB}).} but not the chemical relaxation of $\mu_\phi$ to zero, since they conserve the total number of phonons $N_\phi$. The thermalization of $N_\phi$ thus results, in a second step, from higher-order collision processes, which are much slower at low temperatures: the subdominant five-phonon processes $2\phi\leftrightarrow 3\phi$, with an instantaneous characteristic time $\propto T^{-9}$; see reference \cite{Khalat1950} and note 57 of reference \cite{QO}.\footnote{When the phonon gas is non-degenerate, i.e., with fugacity $z_\phi$--defined later--small compared to one, the instantaneous rate of thermalization via the processes $2\phi\to 2\phi$ is in fact proportional to $z_\phi T^7 \ll T^7$ where $T$ is the instantaneous temperature, see equations (38) and (41) in reference \cite{etalement}, so that the two-step division by separation of time scales seems more difficult to achieve experimentally (the temperature should be reduced even further). Fortunately, this is not the case. In the non-degenerate limit, the instantaneous relaxation rate of $z_\phi$ via $2\phi\leftrightarrow 3\phi$ processes is $\Gamma_\phi^{\rm non\, deg}\equiv (\dd z_\phi/\dd t)/z_\phi\propto T_{\rm eq}^9/z_\phi^{5/4}$ according to (\ref{eq:mouv}); since $T \propto T_{\rm eq}/z_\phi^{1/4}$ according to (\ref{eq:elim}), it follows that $\Gamma_\phi^{\rm non\, deg}\propto z_\phi T^9$ is also reduced by a factor of $z_\phi$. Furthermore, we find that $\Gamma_{2\phi\to 2\phi}^{\rm non\, deg}/\Gamma_{2\phi\to 2\phi}^{\rm eq}\propto z_\phi T^7/T_{\rm eq}^7\propto z_\phi^{-3/4}\gg 1$: the typical thermal $2\phi\to 2\phi$ collision rate is actually higher in the non-degenerate case ($z_\phi\ll 1$) than at full equilibrium ($z_\phi=1^-$).} In the limit $T_{\rm eq}\to 0^+$, by omitting a transient in $O(T_{\rm eq}^{-7})$, we can then assume that the Bose law (\ref{eq:loiB}) applies, with a temperature $T(t)$ and a chemical potential $\mu_\phi(t)$ that depends on time $t$ on a scale $T_{\rm eq}^{-9}$.

The objective of this work is thus to determine, in the case of a concave-start acoustic branch ($\gamma<0$), the relaxation law of the phonon chemical potential toward zero under the effect of collisions $2\phi\leftrightarrow 3\phi$, by adopting the theoretical method based on kinetic equations developed qualitatively by Khalatnikov in reference \cite{Khalat1950} but making it fully quantitative, in particular (i) by explicitly calculating the collision amplitude $\mathcal{A}_{2\phi\to 3\phi}$, (ii)
 by completing the evaluation of a global numerical factor in the rate of change $\mu_\phi$, and (iii) by not limiting ourselves to the vicinity of $\mu_\phi=0$ but by describing the entire interval $\mu_\phi\in]-\infty,0]$.

In Section \ref{sec1}, we write the contribution of collisions $2\phi\leftrightarrow 3\phi$ to the kinetic equations on $n_\qq$, and we deduce the rate of change of the number of phonons $(\dd/\mathrm{d} t) N_\phi$ as a functional $\mathcal{F}$ of the $n_\qq$ and then, after replacing them with the instantaneous Bose law $\bar{n}_\qq(T(t),\mu_\phi(t))$, as a function $F$ of $T(t)$ and $\mu_\phi(t)$. In Section \ref{sec2}, we perform the calculation of the collision amplitude $\mathcal{A}_{2\phi\to 3\phi}$ at low temperature and incorporate it into the function $F$; the rate of change $(\dd/\mathrm{d} t) N_\phi$ becomes proportional to $T^{12}z_\phi^2(1-z_\phi)C_\phi(z_\phi)$ where $z_\phi=\exp(\beta\mu_\phi)$ is the phonon fugacity and $C_\phi(z_\phi)$ is the numerical factor--weakly dependent on $z_\phi$--mentioned above. Finally, in Section \ref{sec3}, using the conservation of total phonon energy $E_\phi$, we relate $T(t)$ to $\mu_\phi(t)$ and obtain a closed-form evolution equation for $\mu_\phi(t)$, more conveniently for $z_\phi(t)$; we discuss its solution and its limiting behavior for $z_\phi=0^+,1^-$.

To conclude this introduction, we note that the problem considered here is far from being purely academic. Our predictions could indeed be tested experimentally in a gas of spin-1/2 cold fermionic atoms in the sector of the BEC-BCS crossover where we have $\gamma<0$,\footnote{This regime should correspond to an interaction parameter $(k_{\rm F}a)^{-1}\in ]-\infty,(k_{\rm F}a)^{-1}_0[$, where $k_{\rm F}$ is the Fermi wave number and $a$ is the $s$-wave scattering length between atoms of opposite spins. It certainly encompasses the BCS limit $k_{\rm F}a\to 0^-$. Its endpoint $(k_{\rm F}a)^{-1}_0$, where $\gamma=0$, is likely close to the unitary limit $(k_{\rm F}a)^{-1}=0$ but is not yet known with certainty: reference \cite{concav} predicts $(k_{\rm F}a)^{-1}_0\simeq -0.14$—thus on the BCS side—using linearized time-dependent BCS theory, whereas experimental measurements in reference \cite{Moritz} yield $(k_{\rm F}a)^{-1}_0\simeq 0.2$—thus on the BEC side; however, these measurements do not allow for a definitive conclusion, as they suffer from two limitations discussed in reference \cite{commentaire}.} or even in liquid helium-4 brought to sufficiently high pressure so that we have $\gamma<0$ \cite{He1,He2} (assuming that liquid helium can be considered an isolated system).

\paragraph{Note on the two-dimensional case} It would be tempting to generalize the present work to 2D--for instance, to a two-dimensional gas of spin-1/2 fermions with an interaction strength yielding a concave-start acoustic branch \cite{VanLoon}. However, Landau and Khalatnikov's quantum hydrodynamics is insufficient to describe phonon-phonon interactions in this case. This issue was already known regarding Landau and Khalatnikov's four-phonon ($2\phi\to 2\phi$) damping rate--see note 48 in reference \cite{QO}: in the $T\to 0^+$ limit at small angles, the transition amplitude $\mathcal{A}_{2\phi\to2\phi}$ is actually identically zero on the energy shell at the leading order naively predicted by power counting ($(T^{3/2})^2/T^3=T^0$),\,\footnote{We have $\mathcal{A}_{2\phi\to2\phi}=\langle\kk_1,\kk_2|\hat{V}\hat{G}_0(E_{\rm i})\hat{V}|\qq_1,\qq_2\rangle$ with $\hat{V}=\hat{H}_3^{(+)}+\hat{H}_3^{(-)}$ using the notation from (\ref{eq:fmix}). Now, each matrix element of $\hat{H}_3^{(\pm)}$--there are two here--contributes a factor of $T^{3/2}$ to the numerator of the transition amplitude, as seen in (\ref{eq:h3m}), and each energy denominator--there is one here--yields a factor of $T^3$, as in (\ref{eq:denomt3}).} regardless of the magnitudes and $O(T)$ angles of the wave vectors of the incoming and outgoing thermal phonons satisfying energy-momentum conservation; The amplitude is therefore likely of the dominant order $T^2T^0=T^2$ (orders in $T$ proceed in steps of two, as in the dispersion relation (\ref{eq:brac}), because the interaction between fermions is short-ranged); however, the calculation at this order cannot be performed using an effective theory as simple as the one in the present work (since various effects omitted by this theory come into play). The precise value of the Landau-Khalatnikov damping rate--which depends quadratically on $\mathcal{A}_{2\phi\to2\phi}$ in the generalized Fermi's golden rule [see equation (129) of reference \cite{Annalen}]--is thus likely of order $(T^2)^2T^3=T^7$ rather than the naively expected $T^3$ \cite{QO}, and it remains unknown in 2D. We have discovered that the same problem arises for $2\phi\leftrightarrow 3\phi$ collisions that relax the phonon number toward its equilibrium value: on the energy shell, the scattering amplitude $\mathcal{A}_{2\phi\to 3\phi}$ is actually identically zero at the order $(T^{3/2})^3/(T^3)^2=T^{-3/2}$--an order predicted by power counting on its expression (\ref{eq:fmix}) and indeed observed in 3D (\ref{eq:resann}).\footnote{As in the 3D case, the reduced angular deviations $\bar{\psi}_{\alpha i}$ and $\bar{\chi}_{ij}$ are expressed in terms of the independent variables $\bar{\theta}_{12}$, $Z_1$, and $Z_2$ according to equation (\ref{eq:tangred})--the only difference being that $Z_1$, $Z_2$, and consequently $Z_3$ are real in 2D (with $\phi_1=0[\pi]$ and $\phi_2=0[\pi]$ in (\ref{eq:chvar}) because $\kk_1$ and $\kk_2$ must lie in the same plane as $\qq_1$ and $\qq_2$). These are then substituted into the energy difference (\ref{eq:eimef}) and the $2\phi\to 3\phi$ transition amplitude (\ref{eq:resann}), written at the expected leading orders. It is thus found that the reduced energy defect $u-v$ is a second-degree polynomial in $\bar{\theta}_{12}$ and that the reduced amplitude $\bar{\mathcal{A}}_{\ii\to\ff}$ (\ref{eq:Abar}) is a rational function $P(\thb_{12})/Q(\thb_{12})$ of the variable $\thb_{12}$, with polynomials $P$ of degree 12 and $Q$ of degree 14. Using computer algebra software, one then verifies that the remainder of the Euclidean division of $P(\thb_{12})$ by $u-v$ is zero for all $Z_1, Z_2$. Thus, there exists a polynomial $R$ of degree 10 in $\thb_{12}$ such that $\bar{\mathcal{A}}_{\ii\to\ff}=R(\thb_{12})(u-v)/Q(\thb_{12})$.} The amplitude is therefore likely of leading order $T^2T^{-3/2}=T^{1/2}$ in 2D, which would lead to a phonon-number relaxation rate scaling as $T^7$--a result that, again, would require calculation using a theory more precise than our quantum hydrodynamics;\footnote{Let us be more precise about this $T^7$ law. By considering, for simplicity, the long-time asymptotic limit $z_\phi=\exp(\mu_\phi/k_{\rm B}T)\to 1^-$, we derive from the 2D version of (\ref{eq:dnphi}) in the small-angle limit the scaling law $(\dd/\dd t) N_\phi/\mathcal{V}\propto T^2(T^3)^3(T^{1/2})^2T^{-3}(1-z_\phi)C_\phi^{\rm 2D}(z_\phi)\propto T^9(1-z_\phi)C_\phi^{\rm 2D}(z_\phi)$, where the first factor arises from integration over $q_1$, the second from integration over the magnitudes and angles of $\qq_2,\kk_1,\kk_2$, the third from the square of $\mathcal{A}_{\ii\to\ff}$, the fourth from the energy conservation delta function, and the fifth from statistical factors; the sixth term, $C_\phi^{\rm 2D}(z_\phi)$, is the 2D equivalent of the dimensionless coefficient (\ref{eq:cphi}). However, in two dimensions, the energy and number of thermal phonons take the form $E_\phi\propto \mathcal{V}T^3e(z_\phi)$ and $N_\phi\propto\mathcal{V} T^2 n(z_\phi)$, where $\dd e/\dd z_\phi$ has a finite limit $>0$ at $z_\phi=1^-$, whereas $\dd n/\dd z_\phi$ diverges as $|\ln(1-z_\phi)|$. From energy conservation $\dd E_\phi/\dd t=0$, we deduce $\dd T/\dd t \propto T\dd(1-z_\phi)/\dd t$; then $(\dd/\dd t) N_\phi/\mathcal{V} \propto T^2 |\ln(1-z_\phi)| \dd (z_\phi-1)/\dd t$ and ultimately $\dd(z_\phi-1)/\dd t\propto -(z_\phi-1)[T^7 C_\phi^{\rm 2D}(z_\phi)/|\ln(1-z_\phi)|]$, yielding the $T^7$ relaxation rate for $N_\phi$ in 2D announced in the main text. However, it is clearly impossible to be so definitive without knowing the behavior of $C_\phi^{\rm 2D}(z_\phi)$ near $z_\phi=1^-$ (the main text implicitly--and baselessly--assumes that $C_\phi^{\rm 2D}(z_\phi)\propto |\ln(1-z_\phi)|$).} to top it all off, even assuming these estimates are correct, there would be no separation of time scales in 2D between four-phonon and five-phonon collisions--both scaling as $T^{-7}$--which would invalidate the hypothesis (\ref{eq:loiB}).

\section{ Kinetic equations for collisions $2\phi\leftrightarrow 3\phi$ and first expression of $(\dd/\mathrm{d} t) N_\phi$}\label{sec1}

Let us write, following the usual procedure (such as that in Section 6.3.2.2 of \cite{livre}), the kinetic equations (or quantum rate or Boltzmann equations) giving the evolution of the number of phonons $n_{\qq_1}$ of the fluid in the mode of wave vector $\qq_1$ under the effect of five-phonon collisions $2\phi\leftrightarrow 3\phi$: 
\begin{multline}
\label{eq:dnq}
\frac{\mathrm{d}}{\mathrm{d} t} n_{\qq_1}\Big|_{2\phi\leftrightarrow 3\phi} = \\
-\sum_{\qq_2} \frac{1}{3!}\sum_{\kk_1,\kk_2,\kk_3} \frac{2\pi}{\hbar} \left|\mathcal{A}(\qq_1,\qq_2\to\kk_1,\kk_2,\kk_3)\right|^2\delta_{\qq_1+\qq_2,\kk_1+\kk_2+\kk_3} \delta\left(\sum_{\alpha=1}^{2}\hbar\omega_{\qq_\alpha}-\sum_{i=1}^{3}\hbar\omega_{\kk_i}\right) \\
\times \left[\left(\prod_{\alpha=1}^{2} n_{\qq_\alpha}\right)\left(\prod_{i=1}^{3}(1+n_{\kk_i})\right)-\left(\prod_{i=1}^{3} n_{\kk_i}\right) \left(\prod_{\alpha=1}^{2} (1+n_{\qq_\alpha})\right)\right] \\
-\frac{1}{2!} \sum_{\qq_2,\qq_3} \frac{1}{2!} \sum_{\kk_1,\kk_2} \frac{2\pi}{\hbar} \left|\mathcal{A}(\qq_1,\qq_2,\qq_3\to\kk_1,\kk_2)\right|^2 \delta_{\qq_1+\qq_2+\qq_3,\kk_1+\kk_2} \delta\left(\sum_{\alpha=1}^{3}\hbar\omega_{\qq_\alpha}-\sum_{i=1}^{2}\hbar\omega_{\kk_i}\right) \\
\times \left[\left(\prod_{\alpha=1}^{3} n_{\qq_\alpha}\right)\left(\prod_{i=1}^{2}(1+n_{\kk_i})\right)-\left(\prod_{i=1}^{2} n_{\kk_i}\right) \left(\prod_{\alpha=1}^{3} (1+n_{\qq_\alpha})\right)\right]\hspace{25mm}
\end{multline}
 The elementary collision rates are expressed in terms of the transition amplitudes $2\phi\to3\phi$ or $3\phi\to2\phi$ using the generalized Fermi golden rule, briefly justified in Appendix \ref{ann:orgen}, in which one recognizes the factor $2\pi/\hbar$ and the Dirac delta for energy conservation (the Kronecker deltas ensure momentum conservation). In the first term on the right-hand side of (\ref{eq:dnq}), vector $\qq_1$ is entering a collision $2\phi\to 3\phi$, which leads to a depopulation of the mode, hence the overall minus sign; we must, of course, sum over the second possible incoming vector $\qq_2$ and over all possible outgoing vectors $\kk_1,\kk_2,\kk_3$, dividing the second sum by $3!$ to account for the invariance of the final state under permutation of the $\kk_i$. But we have cleverly combined this depopulation with the inverse microscopic process $3\phi\to2\phi$, obtained by swapping all incoming phonons and all outgoing phonons. The inverse process is a mode-feeding process since $\qq_1$ is now outgoing, hence the minus sign in square brackets in the third line; it naturally has the same energy and momentum conservation deltas, but also a transition amplitude with the same squared modulus given the principle of microreversibility: 
\begin{equation}
\label{eq:rever}
\left|\mathcal{A}(\qq_1,\qq_2\to\kk_1,\kk_2,\kk_3)\right|^2 = \left|\mathcal{A}(\kk_1,\kk_2,\kk_3\to\qq_1,\qq_2)\right|^2
\end{equation}
However, its bosonic amplification factors are different, since a factor of $n$ must be applied for each incoming phonon and a factor of $1+n$ for each outgoing phonon. 

The second contribution to the right-hand side of (\ref{eq:dnq}), in the fourth and fifth lines, is analyzed in the same way. This time, vector $\qq_1$ is incoming in a collision $3\phi\to2\phi$ or outgoing in the reverse collision $2\phi\to 3\phi$. There are two free wave vectors to sum over in the asymptotic state containing $\qq_1$, and there are also two in the other asymptotic state; so we must divide twice by $2!$ to avoid double-counting states.

The time derivative of the total number of phonons $N_\phi$ is obtained by summing (\ref{eq:dnq}) over the wave vector $\qq_1$. In the second contribution, we rename $\qq_\alpha\leftrightarrow \kk_i$ on the dummy variables and again use the principle of microreversibility (\ref{eq:rever}) so that we have only one transition amplitude $2\phi\to 3\phi$ everywhere. Once this change is made, the two contributions of (\ref{eq:dnq}) take the same form in $(\dd/\mathrm{d} t)N_\phi$:
 they differ only by a sign and by their symmetry numbers $2!\,2!$ or $3!$ in the denominator; everything else factors out. What remains is 
\begin{multline}
\label{eq:dnphi}
\frac{\mathrm{d}}{\mathrm{d} t} N_\phi = \frac{3-2}{2!\, 3!} \sum_{\qq_1,\qq_2} \sum_{\kk_1,\kk_2,\kk_3} \frac{2\pi}{\hbar} \left|\mathcal{A}(\qq_1,\qq_2\to\kk_1,\kk_2,\kk_3)\right|^2\delta_{\qq_1+\qq_2,\kk_1+\kk_2+\kk_3} \\
\times \delta\left(\sum_{\alpha=1}^{2}\hbar\omega_{\qq_\alpha}-\sum_{i=1}^{3}\hbar\omega_{\kk_i}\right) \left[\left(\prod_{\alpha=1}^{2} n_{\qq_\alpha}\right)\left(\prod_{i=1}^{3}(1+n_{\kk_i})\right)-\left(\prod_{i=1}^{3} n_{\kk_i}\right) \left(\prod_{\alpha=1}^{2} (1+n_{\qq_\alpha})\right)\right]
\end{multline}
As stated in section \ref{sec0}, we can assume here that the phonon gas is in a state of partial equilibrium, thermal but not chemical, that is, the occupation numbers $n_{\qq_\alpha}$ and $n_{\kk_i}$ follow the Bose-Einstein distribution (\ref{eq:loiB}) for temperature $T$ and chemical potential $\mu_\phi<0$ and can be replaced by $\bar{n}_{\qq_\alpha}$ and $\bar{n}_{\kk_i}$. Let’s eliminate the resulting factors $\bar{n}$ in (\ref{eq:dnphi}) using relation 
\begin{equation}
\bar{n}_{\qq_\alpha}=\eee^{-\beta(\hbar\omega_{\qq_\alpha}-\mu_\phi)}(1+\bar{n}_{\qq_\alpha}) \quad\mbox{or}\quad
\bar{n}_{\kk_i}=\eee^{-\beta(\hbar\omega_{\kk_i}-\mu_\phi)}(1+\bar{n}_{\kk_i})
\end{equation}
This brings out a factor $z_\phi^2\exp[-\beta(\hbar\omega_{\qq_1}+\hbar\omega_{\qq_2})]$ in the first term in brackets, and a factor $z_\phi^3 \exp[-\beta(\hbar\omega_{\kk_1}+\hbar\omega_{\kk_2}+\hbar\omega_{\kk_3})]$ in the second, with--let us recall--$z_\phi=\exp(\beta\mu_\phi)$. Given the Dirac distribution for energy conservation, the exponential terms of the energies have the same value in these factors, which we choose to write as $\exp[-\beta(\hbar\omega_{\qq_1}+\hbar\omega_{\qq_2})]$. To avoid multiplying equations, let’s eliminate a wave vector via momentum conservation, by agreeing that 
\begin{equation}
\label{eq:k3}
\kk_3\equiv \qq_1+\qq_2-(\kk_1+\kk_2)
\end{equation}
 and, in the process, move on to the thermodynamic limit of a quantization volume $\mathcal{V}=L^3$ tending to infinity, anticipating the fact (established in section \ref{sec2}, see its equation (\ref{eq:resann})) that the collision amplitude varies as 
\begin{equation}
\mathcal{A}(\qq_1,\qq_2\to \kk_1,\kk_2,\kk_3) \sim \frac{\check{\mathcal{A}}_{\mathrm{i}\to\textrm{f}}}{\mathcal{V}^{3/2}}
\end{equation}
 Here, for brevity, $|\mathrm{i}\rangle$ represents the initial state with two phonons $|\qq_1,\qq_2\rangle$ and $|\textrm{f}\rangle$ the final state with three phonons $|\kk_1,\kk_2,\kk_3\rangle$, and the amplitude $\check{\mathcal{A}}_{\mathrm{i}\to\textrm{f}}$ does not depend on the size of the system. We also note $E_{\mathrm{i}}=\hbar\omega_{\qq_1}+\hbar\omega_{\qq_2}$ as the energy of the initial state and $E_{\textrm{f}}=\hbar\omega_{\kk_1}+\hbar\omega_{\kk_2}+\hbar\omega_{\kk_3}$ as that of the final state. Replacing the sums over the remaining four wave vectors with integrals yields a factor $\mathcal{V}^4$, partially offset by the factor $1/\mathcal{V}^3$ arising in the square of the amplitude, as expected ($N_\phi$ is an extensive quantity $\propto \mathcal{V}$). We finally obtain 
\begin{equation}
\label{eq:res1}
\frac{\mathrm{d}}{\mathrm{d} t} \frac{N_\phi}{\mathcal{V}} = \frac{1}{12} z_\phi^2(1-z_\phi) \int_{\mathbb{R}^{12}} \frac{\dd^3q_1\dd^3q_2 \dd^3k_1\dd^3k_2}{[(2\pi)^3]^4} \frac{2\pi}{\hbar} \left|\check{\mathcal{A}}_{\mathrm{i}\to\textrm{f}}\right|^2 \delta(E_\ii-E_\ff) \eee^{-\beta E_\mathrm{i}} \prod_{\alpha=1}^{2} (1+\bar{n}_{\qq_\alpha}) \prod_{i=1}^{3} (1+\bar{n}_{\kk_i})
\end{equation}
This expression is positive and vanishes only for $z_\phi=1$, that is, for a zero phonon chemical potential $\mu_\phi=0$, as one might expect.

Finally, let us reduce the multiple integral in (\ref{eq:res1}) using rotational symmetry: 
\begin{itemize}\item[(i)] due to global isotropy, the direction of vector $\qq_1$ is irrelevant. We can therefore place $\qq_1$ along the $Oz$ axis of a Cartesian coordinate system $Oxyz$ and directed “upward” (direction of increasing $z$). We factor out a total solid angle of $4\pi$, and it remains to integrate over the modulus $q_1$ of $\qq_1$ using the usual spherical Jacobian $q_1^2$; \item[(ii)] with the direction of $\qq_1$ thus fixed, there remains rotational symmetry about axis $Oz$; in spherical coordinates with polar axis $Oz$, the integrand does not depend on the azimuthal angle of the vector $\qq_2$. We can therefore place $\qq_2$ in the half-plane $xOz$, $x>0$ (the origin plane of the azimuthal angles). We factor out a total azimuthal angle factor $2\pi$, and it remains to integrate over the magnitude $q_2$ and the polar angle $\theta_{12}$ of the vector $\qq_2$, with the usual Jacobian $q_2^2\sin\theta_{12}$;
\item[(iii)] Since the directions of $\qq_1$ and $\qq_2$ are fixed, there is no longer any rotational symmetry. We integrate over the remaining vectors $\kk_1$ and $\kk_2$ in spherical coordinates of polar axis $Oz$ and reference azimuthal half-plane $xOz$, $x>0$; we denote $k_i$, $\psi_{1i}$, and $\phi_i$ as the magnitude, polar angle, and azimuthal angle of the corresponding vector $\kk_i$. 
\end{itemize}
In other words, the independent wave vectors are written as follows in the Cartesian coordinate system $Oxyz$: 
\begin{equation}
\label{eq:vecs}
\qq_1=\begin{pmatrix}0 \\ 0 \\q_1\end{pmatrix} \quad \qq_2=\begin{pmatrix} q_2\sin\theta_{12} \\ 0 \\ q_2 \cos\theta_{12}\end{pmatrix}
\quad
\kk_1=\begin{pmatrix} k_1\sin\psi_{11}\cos\phi_1 \\ k_1\sin\psi_{11}\sin\phi_1 \\ k_1 \cos\psi_{11}\end{pmatrix}
\quad
\kk_2=\begin{pmatrix} k_2\sin\psi_{12}\cos\phi_2 \\ k_2\sin\psi_{12}\sin\phi_2 \\ k_2 \cos\psi_{12}\end{pmatrix}
\end{equation}
 the vector $\kk_3$ is always given by (\ref{eq:k3}), and we have reduced (\ref{eq:res1}) to a ninefold integral: 
\begin{multline}
\label{eq:res2}
\frac{\mathrm{d}}{\mathrm{d} t} \frac{N_\phi}{\mathcal{V}} = \frac{z_\phi^2(1-z_\phi)}{6(2\pi)^9\hbar} \int_0^{+\infty}\mathrm{d} q_1\, q_1^2 \int_0^{+\infty} \mathrm{d} q_2\, q_2^2 \int_0^\pi \mathrm{d}\theta_{12} \,\sin \theta_{12} \int_0^{+\infty}\mathrm{d} k_1\, k_1^2 \int_0^\pi \mathrm{d}\psi_{11} \,\sin \psi_{11} \int_0^{2\pi}\mathrm{d}\phi_1\\
\times \int_0^{+\infty}\mathrm{d} k_2\, k_2^2 \int_0^\pi \mathrm{d}\psi_{12} \,\sin \psi_{12} \int_0^{2\pi}\mathrm{d}\phi_2\, \left|\check{\mathcal{A}}_{\mathrm{i}\to\textrm{f}}\right|^2 \delta(E_\ii-E_\ff) \eee^{-\beta E_\mathrm{i}} \prod_{\alpha=1}^{2} (1+\bar{n}_{\qq_\alpha}) \prod_{i=1}^{3} (1+\bar{n}_{\kk_i})
\end{multline}

\section{Collision amplitude $2\phi\rightarrow 3\phi$ and expression of $(\dd/\mathrm{d} t)N_\phi$ at low temperature}\label{sec2}

The missing ingredient in expression (\ref{eq:res2}) for the time derivative of the number of phonons is the collision amplitude $2\phi\to 3\phi$ between thermal phonons, that is, the transition amplitude $\mathcal{A}_{\mathrm{i}\to\textrm{f}}$ between the initial two-phonon state $|\mathrm{i}\rangle=|\qq_1,\qq_2\rangle$ and the final three-phonon state $|\textrm{f}\rangle=|\kk_1,\kk_2,\kk_3\rangle$ under the effect of the phonon-phonon interaction Hamiltonian $\hat{V}$.

Here we carry out our calculation to the temperature-dominant order. We can then reduce $\hat{V}$ to the cubic interaction described in section \ref{sec0}, further restricting it to triplets of wave vectors making a small angle $O(T)$ between them \cite{LK,Annalen,Khalat1950}. We therefore take in practice $\hat{V}=\hat{H}_3^{(+)}+\hat{H}_3^{(-)}$, with $\hat{H}_3^{(+)}=[\hat{H}_3^{(-)}]^\dagger$ and 
\begin{equation}
\label{eq:h3m}
\hat{H}_3^{(-)} \stackrel{\rm small}{\underset{\rm angles}{=}} \frac{3mc^2(1+\Lambda)}{2\rho^{1/2}L^{3/2}} \sum_{\qq,\qq'}^{\widehat{(\qq,\qq')}=O(T)} \left(\frac{\hbar}{2mc}\right)^{3/2}\left[q q'(q+q')\right]^{1/2} \hat{b}^\dagger_{\qq+\qq'} \hat{b}_\mathbf{q} \hat{b}_{\qq'} 
\end{equation}
 as in reference \cite{QO}. In particular, as shown in references \cite{LK,Annalen,Khalat1950}, we can ignore the non-resonant processes $\hat{b}^\dagger\hat{b}^\dagger\hat{b}^\dagger$ and $\hat{b}\hat{b}\hat{b}$ naturally present in the cubic Hamiltonian $\hat{H}_3$ (\ref{eq:H3}) of quantum hydrodynamics, as well as the quartic Hamiltonian $\hat{H}_4$ predicted by this theory; see equations (24,12) and (24,13) of \cite{LandauStat2}. In expression (\ref{eq:h3m}) as in (\ref{eq:H3}), we have introduced the dimensionless parameter 
\begin{equation}
\label{eq:defLam}
\Lambda= \frac{\rho\frac{\dd^2\mu_{0}}{\mathrm{d}\rho^2}}{3\frac{\mathrm{d}\mu_{0}}{\mathrm{d}\rho}}= \frac{\rho^2\frac{\dd^2\mu_{0}}{\mathrm{d}\rho^2}}{3 mc^2}
\end{equation}
where $\mu_{0}$ is the chemical potential of the underlying superfluid in the ground state and, it should be recalled, $\rho$ is its number density, $m$ is the mass of a constituent particle and $c$ is the speed of sound at zero temperature, effectively given by $mc^2=\rho\mathrm{d}\mu_{0}/\mathrm{d}\rho$. 

\begin{figure}[t]
\centerline{\includegraphics[width=12cm,clip=]{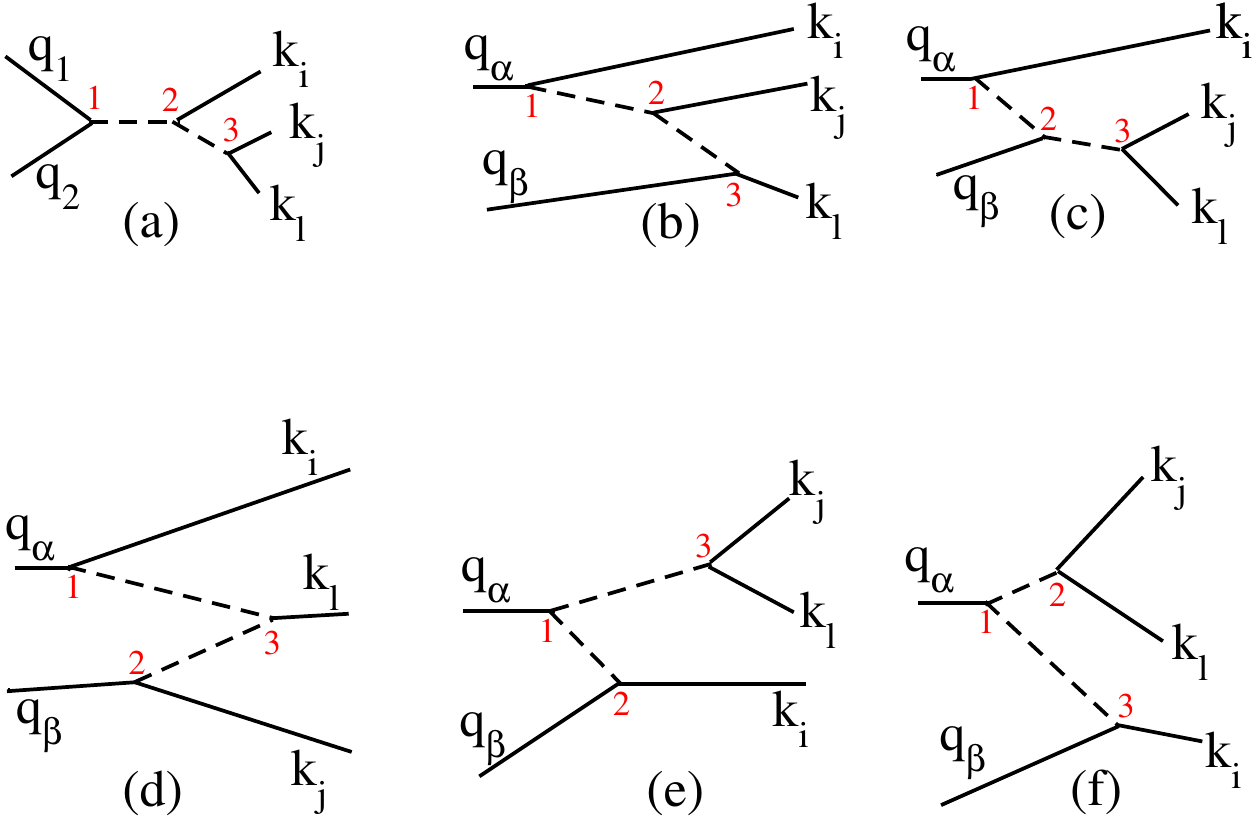}}
\caption{The six diagrams involved in the five-phonon collision amplitude $2\phi\to3\phi$ of equation (\ref{eq:fmix}). Here $\{\alpha,\beta\}=\{1,2\}$ and $\{i,j,k\}=\{1,2,3\}$ are the indices of the incoming and outgoing wave vectors to be taken in the various possible orders in the successive actions--shown from left to right--of $\hat{H}_3^{(\pm)}$, but without double counting in cases where the diagram is invariant under the exchange of two indices. The lines of the internal phonons are dashed. The numbering of the actions in $\hat{H}_3^{(\pm)}$ in red corresponds to that in Table \ref{tab:arbre}. }
\label{fig:diag}
\end{figure}

Furthermore, in 3D, again thanks to the low-temperature limit, we need only perform the calculation of the transition amplitude to the first non-zero order in $\hat{V}$, here the third order as in Appendix \ref{ann:orgen}; since we are on the energy shell $E_\ii=E_\ff$ in equation (\ref{eq:res2}), we use the following mixed form, which leads to the simplest energy denominators : 
\begin{equation}
\label{eq:fmix}
\mathcal{A}_{\mathrm{i}\to\textrm{f}}=\langle \kk_1,\kk_2,\kk_3|\hat{V}\hat{G}_0(E_\ff)\hat{V}\hat{G}_0(E_i)\hat{V}|\qq_1,\qq_2\rangle = \sum_{\lambda,\mu}\frac{\langle \ff|\hat{V}|\mu\rangle \langle\mu|\hat{V}|\lambda\rangle \langle\lambda|\hat{V}|\mathrm{i}\rangle}{\Delta E_2(\mu)\, \Delta E_1(\lambda)}
\end{equation}
 where $\hat{G}_0(z)=(z\,\mathrm{Id}-\hat{H}_2)^{-1}$ is the resolvent of the unperturbed Hamiltonian $\hat{H}_2=\sum_\mathbf{q} \hbar\omega_\mathbf{q}\hat{b}_\qq^\dagger \hat{b}_\qq$, and we have introduced a closure relation on the intermediate eigenstates $\hat{H}_2$ of $\lambda,\mu$ with the energy differences $\Delta E_1(\lambda)=E_\ii-E_\lambda$ and $\Delta E_2(\mu)=E_\ff-E_\mu$. By exploring all possible cases using Table \ref{tab:arbre}, we arrive at the six diagrams in Figure \ref{fig:diag}, numbered (a) through (f), whose contributions to the amplitude are explicitly written as follows: 
\begin{eqnarray}
\label{eq:ampa}
\mathcal{A}_{\mathrm{i}\to\textrm{f}}^{(a)} &=& \sum_i \frac{\langle\kk_j,\kk_l|\hat{V}|\kk_j+\kk_l\rangle \langle \kk_j+\kk_l,\kk_i|\hat{V}|\qq_1+\qq_2\rangle \langle \qq_1+\qq_2|\hat{V}|\qq_1,\qq_2\rangle}{(\veps_{\kk_j}+\veps_{\kk_l}-\veps_{\kk_j+\kk_l})(\veps_{\qq_1}+\veps_{\qq_2}-\veps_{\qq_1+\qq_2})} \\
\label{eq:ampb}
\mathcal{A}_{\mathrm{i}\to\textrm{f}}^{(b)} &=& \sum_{\alpha}\sum_{i\neq j}\frac{\langle\kk_l|\hat{V}|\qq_\beta,\kk_l-\qq_\beta\rangle \langle \kk_j,\kk_l-\qq_\beta|\hat{V}|\qq_\alpha-\kk_i\rangle \langle \kk_i,\qq_\alpha-\kk_i|\hat{V}|\qq_\alpha\rangle}{(\veps_{\kk_l}-\veps_{\qq_\beta}-\veps_{\kk_l-\qq_\beta})(\veps_{\qq_\alpha}-\veps_{\kk_i}-\veps_{\qq_\alpha-\kk_i})} \\
\label{eq:ampc}
\mathcal{A}_{\mathrm{i}\to\textrm{f}}^{(c)} &=& \sum_\alpha\sum_i\frac{\langle \kk_j,\kk_l|\hat{V}|\kk_j+\kk_l\rangle \langle\kk_j+\kk_l|\hat{V}|\qq_\alpha-\kk_i,\qq_\beta\rangle \langle \kk_i,\qq_\alpha-\kk_i|\hat{V}|\qq_\alpha\rangle}{(\veps_{\kk_j}+\veps_{\kk_l}-\veps_{\kk_j+\kk_l})(\veps_{\qq_\alpha}-\veps_{\kk_i}-\veps_{\qq_\alpha-\kk_i})} \\
\label{eq:ampd}
\mathcal{A}_{\mathrm{i}\to\textrm{f}}^{(d)} &=& \sum_\alpha\sum_{i\neq j}\frac{\langle\kk_l|\hat{V}|\qq_\alpha-\kk_i,\qq_\beta-\kk_j\rangle \langle\kk_j,\qq_\beta-\kk_j|\hat{V}|\qq_\beta\rangle\langle\kk_i,\qq_\alpha-\kk_i|\hat{V}|\qq_\alpha\rangle}{(\veps_{\kk_l}-\veps_{\qq_\alpha-\kk_i}-\veps_{\qq_\beta-\kk_j})(\veps_{\qq_\alpha}-\veps_{\kk_i}-\veps_{\qq_\alpha-\kk_i})} \\
\label{eq:ampe}
\mathcal{A}_{\mathrm{i}\to\textrm{f}}^{(e)} &=& \sum_\alpha\sum_i \frac{\langle\kk_j,\kk_l|\hat{V}|\kk_j+\kk_l\rangle \langle\kk_i|\hat{V}|\qq_\beta,\kk_i-\qq_\beta\rangle \langle\kk_i-\qq_\beta,\kk_j+\kk_l|\hat{V}|\qq_\alpha\rangle}{(\veps_{\kk_j}+\veps_{\kk_l}-\veps_{\kk_j+\kk_l})(\veps_{\qq_\alpha}-\veps_{\kk_j+\kk_l}-\veps_{\kk_i-\qq_\beta})} \\
\label{eq:ampf}
\mathcal{A}_{\mathrm{i}\to\textrm{f}}^{(f)} &=& \sum_\alpha\sum_i \frac{\langle\kk_i|\hat{V}|\qq_\beta,\kk_i-\qq_\beta\rangle \langle\kk_j,\kk_l|\hat{V}|\kk_j+\kk_l\rangle \langle\kk_i-\qq_\beta,\kk_j+\kk_l|\hat{V}|\qq_\alpha\rangle}{(\veps_{\kk_i}-\veps_{\qq_\beta}-\veps_{\kk_i-\qq_\beta})(\veps_{\qq_\alpha}-\veps_{\kk_j+\kk_l}-\veps_{\kk_i-\qq_\beta})}
\end{eqnarray}
 We have denoted $\veps_\qq=\hbar\omega_\qq$ and $\veps_\kk=\hbar\omega_\kk$ as the eigenenergies of the phonons and have taken into account momentum conservation at each vertex. Furthermore, as in Figure \ref{fig:diag}, we have set (i) $\{\beta\}=\{1,2\}\setminus\{\alpha\}$ when there is a sum over $\alpha\in\{1,2\}$,
 (ii) $\{j,l\}=\{1,2,3\}\setminus\{i\}$ when there is a simple sum over $i\in\{1,2,3\}$ (the order chosen between $j$ and $l$ is irrelevant since the summand is a symmetric function of these indices, for example we can impose $j<l$) and (iii) $\{l\}=\{1,2,3\}\setminus\{i,j\}$ when there is a double sum over $i\in\{1,2,3\}, j\in\{1,2,3\}, i\neq j$. In other words, $(\alpha,\beta)$ is a permutation of the indices $(1,2)$ in the initial state and $(i,j,l)$ a permutation of the indices $(1,2,3)$ in the final state. It is easy to verify that the amplitude is a function of the incoming and outgoing wave numbers, as well as of all the non-oriented angles between the incoming and outgoing wave vectors: 
\begin{equation}
\label{eq:angles}
\theta_{12}=\widehat{(\qq_1,\qq_2)} \quad ; \quad \psi_{\alpha i}=\widehat{(\qq_\alpha,\kk_i)} \quad ; \quad
\chi_{ij}=\widehat{(\kk_i,\kk_j)}
\end{equation}
 the angles $\theta_{12},\psi_{11},\psi_{12}$ thus defined reproduce those of the spherical coordinates in (\ref{eq:vecs}).

 \begin{table}[t]
\centerline{\includegraphics[width=12cm,clip=]{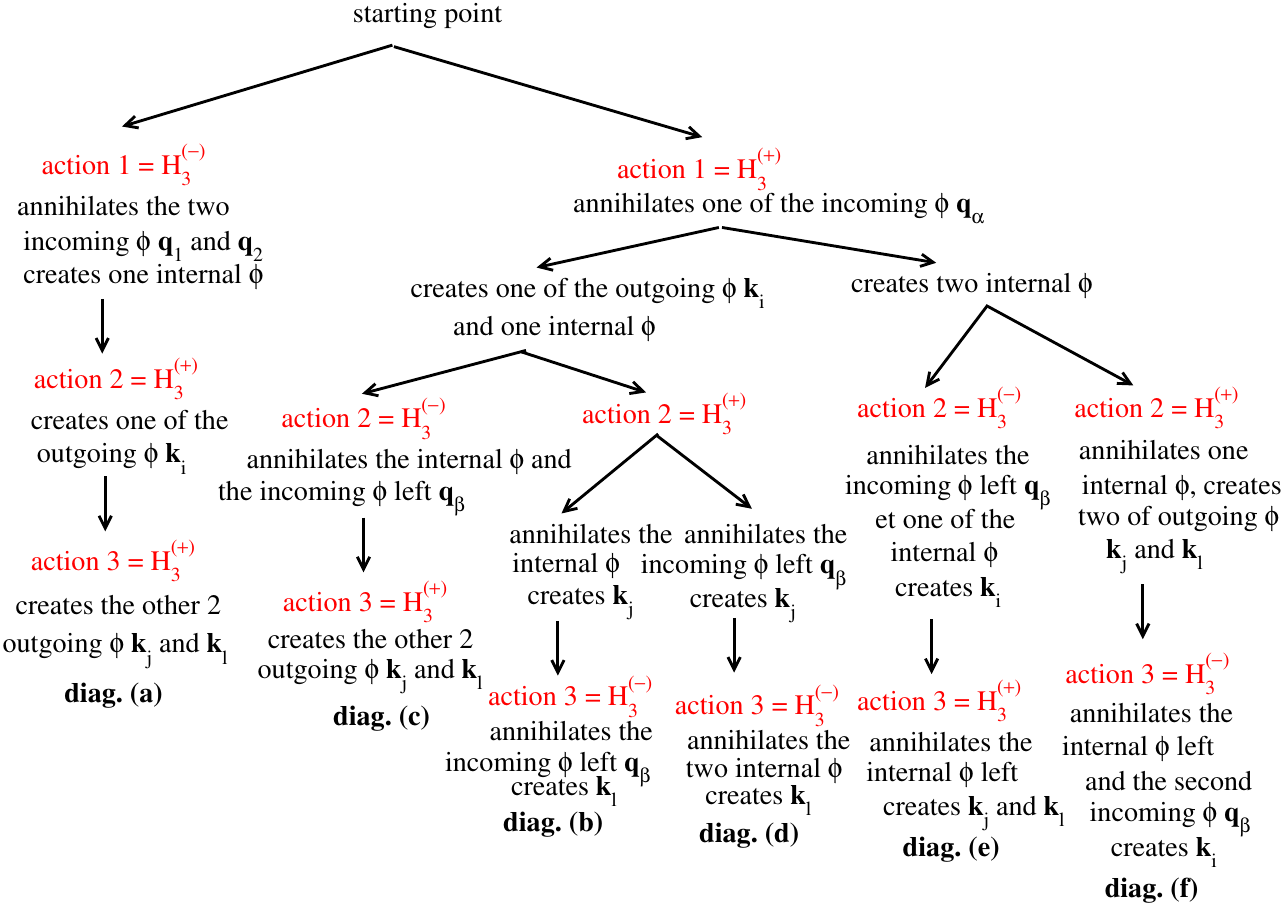}}
\caption{ Decision tree identifying all transition diagrams (\ref{eq:fmix}) of order three in $\hat{V}=\hat{H}_3^{(+)}+\hat{H}_3^{(-)}$ between the initial two-phonon state $|\mathrm{i}\rangle=|\qq_1,\qq_2\rangle$ and the final three-phonon state $|\textrm{f}\rangle=|\kk_1,\kk_2,\kk_3\rangle$. Recall that an action of $\hat{H}_3^{(+)}$ splits a phonon into two (terms $\hat{b}^\dagger\hat{b}^\dagger\hat{b}$), and that an action of $\hat{H}_3^{(-)}$ merges two phonons into one (terms $\hat{b}^\dagger\hat{b}\hat{b}$). We have excluded the cases of zero measure at the thermodynamic limit where simple relations exist between these different wave vectors, such as $\qq_\alpha=\kk_j+\kk_l$, etc., other than the conservation of total momentum. }
\label{tab:arbre}
\end{table}

 From now on, we consider the low-temperature limit, controlled by the small parameter 
\begin{equation}
\label{eq:defeps}
\epsilon\equiv \frac{k_{\rm B}T}{mc^2}\to 0^+
\end{equation}
The wave numbers of the incoming and outgoing thermal phonons and all the angles (\ref{eq:angles}) tend linearly toward zero with temperature \cite{Khalat1950}--this was already the case in four-phonon collisions \cite{LK,Annalen}; this means that the scaled wave numbers and the scaled angles, marked by a bar and defined by 
\begin{equation}
\label{eq:ech}
q_\alpha=\frac{k_{\rm B}T}{\hbar c} \qb_\alpha\ , \ \mbox{etc.} \quad ; \quad \theta_{12}= \epsilon |\gamma|^{1/2} \bar{\theta}_{12}\ ,\ \mbox{etc.}
\end{equation}
have a finite, non-zero limit, and that the dispersion relation (\ref{eq:brac}) of our concave acoustic branch (curvature parameter $\gamma<0$) is written as 
\begin{equation}
\label{eq:bacred}
\hbar\omega_\mathbf{q}\stackrel{\bar{q}\,\mbox{\scriptsize fixed}}{\underset{T\to 0^+}{=}}k_{\rm B} T \left[\bar{q} -\frac{|\gamma|}{8}\epsilon^2\qb^3 + O\left(\epsilon^4\ln\epsilon\right)\right]
\end{equation}
The limit of expressions (\ref{eq:ampa})-(\ref{eq:ampf}) on the energy shell is calculated in Appendix \ref{ann:lim}; here we give only the result: 
\begin{equation}
\label{eq:resann}
\boxed{
\mathcal{A}_{\mathrm{i}\to\textrm{f}}\stackrel{\mathrm{angles}=O(T)}{\underset{T\to 0^+}{\sim}} \frac{3^3\sqrt{2}}{2^3\mathcal{V}^{3/2}} \frac{mc^2(1+\Lambda)^3}{\epsilon^{3/2}\rho^{3/2}\gamma^2} \frac{1}{(\bar{q}_1\bar{q}_2\bar{k}_1\bar{k}_2\bar{k}_3)^{1/2}}\bar{\mathcal{A}}_{\mathrm{i}\to\textrm{f}}
}
\end{equation}
 with the reduced amplitude 
\begin{multline}
\label{eq:Abar}
\bar{\mathcal{A}}_{\mathrm{i}\to\textrm{f}}= 
\sum_{1\leq i\leq 3} \frac{\kb_i}{\Big[\frac{\bar{\theta}^2_{12}}{(\qb_1+\qb_2)^2}+\frac{3}{4}\Big]\Big[\frac{\bar{\chi}^2_{jl}}{(\kb_j+\kb_l)^2}+\frac{3}{4}\Big]}
+\sum_{1\leq\alpha\leq 2}\ \sum_{1\leq i\leq 3} \frac{(-\qb_\beta)}{\Big[\frac{\bar{\psi}^2_{\alpha i}}{(\qb_\alpha-\kb_i)^2}+\frac{3}{4}\Big]\Big[\frac{\bar{\chi}^2_{jl}}{(\kb_j+\kb_l)^2}+\frac{3}{4}\Big]} \\
+\sum_{1\leq \alpha\leq 2}\ \sum_{1\leq i< j\leq 3} \frac{\kb_l}{\Big[\frac{\bar{\psi}^2_{\alpha i}}{(\qb_\alpha-\kb_i)^2}+\frac{3}{4}\Big] \Big[\frac{\bar{\psi}^2_{\beta j}}{(\qb_\beta-\kb_j)^2}+\frac{3}{4}\Big]}
\end{multline}
Similarly, the energy difference $E_\ii-E_\ff$ must be expanded as an argument of the Dirac delta in (\ref{eq:res2}). The form (\ref{eq:bacred}) immediately yields: 
\begin{equation}
\label{eq:diff}
E_\ii-E_\textrm{f} \underset{\epsilon\to 0^+}{=} k_{\rm B}T\left[\qb_1+\qb_2-\kb_1-\kb_2-\kb_3-\frac{|\gamma|}{8}\epsilon^2 (\qb_1^3+\qb_2^3-\kb_1^3-\kb_2^3-\kb_3^3)+O(\epsilon^4\ln\epsilon)\right]
\end{equation}
Only the reduced wave number $\kb_3$ is unknown, but it is easily obtained at order $\epsilon^2$ by taking the norm of the two sides of equation (\ref{eq:k3}), using the writing trick that exploits the fact that we have $q_1+q_2>k_1+k_2$ when all incoming and outgoing vectors are nearly collinear and in the same direction: 
\begin{equation}
k_3=(q_1+q_2-k_1-k_2)\left[1+\frac{(\qq_1+\qq_2-\kk_1-\kk_2)^2-(q_1+q_2-k_1-k_2)^2}{(q_1+q_2-k_1-k_2)^2}\right]^{1/2}
\end{equation}
 After expanding the squares in the numerator of the large fraction, only differences between double products of vectors and scalars remain, such as $-2\qq_\alpha\cdot\kk_i+2q_\alpha k_i=2q_\alpha k_i(1-\cos\psi_{\alpha i})\sim q_\alpha k_i \psi_{\alpha i}^2$. The terms linear in wave number cancel out in (\ref{eq:diff}), and we retain that 
\begin{equation}
\label{eq:eimef}
E_\ii-E_\textrm{f} \underset{T\to 0^+}{\sim} \frac{1}{2} k_{\rm B} T |\gamma|\epsilon ^2 (u-v)
\end{equation}
 with 
\begin{eqnarray}
\label{eq:defu}
u &=& \frac{\qb_1\qb_2\bar{\theta}_{12}^2+\kb_1\kb_2\bar{\chi}_{12}^2-\sum_{\alpha,i=1}^{2} \qb_\alpha\kb_i \bar{\psi}_{\alpha i}^2}{\qb_1+\qb_2-\kb_1-\kb_2} \\
\label{eq:defv}
v &=& \frac{1}{4}\left(\qb_1^3+\qb_2^3-\kb_1^3-\kb_2^3-\kb_3^3\right)
\end{eqnarray}
Since the dominant behaviors in $\epsilon$ are now determined, we will assume the following relation to be satisfied in what follows, which holds for zero order: 
\begin{equation}
\label{eq:defk3b}
\qb_1+\qb_2 = \kb_1+\kb_2+\kb_3
\end{equation}

 In all these formulas, the various angular variables are of course not independent, but are expressed in terms of the polar angles $\theta_{12},\psi_{11},\psi_{12}$ and azimuthal angles $\phi_1,\phi_2$ of the spherical parameterization (\ref{eq:vecs}). The required trigonometric calculations are somewhat lengthy but straightforward within the small-angle limit.\footnote{The sine of the desired angle between two vectors is obtained through the norm of their vector product, noting that here we can linearize the sines of angles other than azimuthal angles and replace their cosines with one. For example, the exact relation $k_ik_3\sin\chi_{i3}=|\kk_i\wedge\kk_3| \ (i\in\{1,2\})$ yields, after applying (\ref{eq:k3},\ref{eq:vecs}) and trigonometric linearization, $k_i k_3\chi_{i3}\sim k_i|(\psi_{1i}\cos\phi_i,\psi_{1i}\sin\phi_i,1)\wedge (q_2\theta_{12}-k_j\psi_{1j}\cos\phi_j,-k_j\psi_{1j}\sin\phi_j,q_1+q_2-k_j\simeq k_3+k_i)|$ where $j\in\{1,2\}\setminus\{i\}$ and account has been taken of (\ref{eq:defk3b}) and the fact that $\kk_i\wedge\kk_i=\mathbf{0}$. This results in 
\begin{equation}
\chi_{i3}\sim \left|\left((k_3+k_i)\psi_{1i}\sin\phi_i+k_j\psi_{1j}\sin\phi_j, -((k_3+k_i)\psi_{1i}\cos\phi_i+k_j\psi_{1j}\cos\phi_j-q_2\theta_{12}),0\right)\right|/k_3
\end{equation}
 or, after scaling the wave numbers and angles, $\bar{\chi}_{i3}=|(\kb_3+\kb_i)z_i+\kb_jz_j-\qb_2\bar{\theta}_{12}|/\kb_3$ with the complex numbers $z_n=\bar{\psi}_{1n}\exp(\mathrm{i}\phi_n),\ n\in\{1,2\}$. This reduces to $\bar{\chi}_{i3}=|Z_i-Z_3|$ thanks to (\ref{eq:defk3b},\ref{eq:chvar}).} The results obtained are greatly simplified if we consider introducing complex variables encapsulating $\psi_{11},\psi_{12}$ and $\phi_1,\phi_2$: 
\begin{equation}
\label{eq:chvar}
Z_1\equiv \bar{\psi}_{11}\exp(\mathrm{i}\phi_1)-\frac{\qb_2}{\qb_1+\qb_2}\bar{\theta}_{12} \quad ; \quad Z_2\equiv \bar{\psi}_{12}\exp(\mathrm{i}\phi_2) -\frac{\qb_2}{\qb_1+\qb_2}\bar{\theta}_{12} \quad ; \quad Z_3=-\frac{\kb_1 Z_1+\kb_2 Z_2}{\kb_3}
\end{equation}
 The shift proportional to $\bar{\theta}_{12}$ cleverly decouples this angular variable from the others in the quadratic form $u$, see below, and the supernumerary variable $Z_3$, which is not independent, allows us to restore a certain symmetry between the outgoing vectors $\kk_i$ (broken in the parameterization (\ref{eq:k3},\ref{eq:vecs})). Thus, for all $i\in\{1,2,3\}, j\in\{1,2,3\}\setminus\{i\}$: 
\begin{equation}
\label{eq:tangred}
\bar{\psi}_{1i} = \left|Z_i+\frac{\qb_2}{\qb_1+\qb_2}\bar{\theta}_{12}\right| \quad ; \quad \bar{\psi}_{2i} = \left|Z_i-\frac{\qb_1}{\qb_1+\qb_2}\bar{\theta}_{12}\right| \quad ; \quad \bar{\chi}_{ij} = |Z_i-Z_j|
\end{equation}
 and 
\begin{equation}
u=\frac{\qb_1\qb_2}{\qb_1+\qb_2} \bar{\theta}_{12}^2 
-\frac{1}{\kb_3} (Z_1^*,Z_2^*) M \binom{Z_1}{Z_2} \quad\mbox{where}\quad M=\begin{pmatrix} \kb_1(\kb_1+\kb_3) & \kb_1\kb_2\\ \kb_1\kb_2 & \kb_2(\kb_2+\kb_3)\end{pmatrix}
\end{equation}
 Note that the matrix $M$ thus introduced is positive definite and has a determinant of 
\begin{equation}
\mathrm{det}\, M = \kb_1 \kb_2 \kb_3 (\kb_1+\kb_2+\kb_3)
\end{equation}

 By substituting the results of this section into equation (\ref{eq:res2}), we finally obtain the rate of change of the number of phonons at the dominant order in temperature, 
\begin{equation}
\label{eq:res3}
\boxed{
\frac{\mathrm{d}}{\mathrm{d} t} \frac{N_\phi}{\mathcal{V}} \underset{T\to 0^+}{\sim} k_{\rm B}T\frac{mc(1+\Lambda)^6}{\hbar^2\gamma^2\rho^3} \left(\frac{k_{\rm B}T}{\hbar c}\right)^{11} \frac{3^5}{4^7\pi^9}z_\phi^2(1-z_\phi)C_\phi(z_\phi)}
\end{equation}
 with a dimensionless coefficient depending only on the phonon fugacity, 
\begin{multline}
\label{eq:cphi}
C_\phi(z_\phi)=\int_0^{+\infty} \prod_{\alpha=1}^{2} \mathrm{d}\qb_\alpha \prod_{i=1}^{3}\mathrm{d}\kb_i\, \delta\left(\sum_{\alpha=1}^{2}\qb_\alpha-\sum_{i=1}^{3}\kb_i\right)
I_{\rm ang}(\qb_1,\qb_2,\kb_1,\kb_2,\kb_3)  \\
\times\eee^{-(\qb_1+\qb_2)} \prod_{\alpha=1}^{2} (1+\bar{n}_{q_\alpha}^{\rm lin}) \prod_{i=1}^{3}
(1+\bar{n}_{k_i}^{\rm lin})
\end{multline}
 Here, $\bar{n}^{\rm lin}$ is the Bose law for a linearized phonon dispersion relation, 
\begin{equation}
\label{eq:nqlin}
\bar{n}^{\rm lin}_{q}=\frac{1}{\eee^{\beta(\hbar c q-\mu_\phi)}-1}=\frac{1}{z_\phi^{-1}\exp \qb-1}
\end{equation}
 and the function $I_{\rm ang}$ is the angular integral \footnote{The integral in $\mathbb{C}$ of a function of the complex variable $f(z)$, denoted here as $J=\int_{\mathbb{C}}\dd^2z\, f(z)$, is defined as the integral in $\mathbb{R}^2$ over the real and imaginary parts of $z$, $J=\int_{\mathbb{R}^2}\mathrm{d} x\,\mathrm{d} y\, f(x+\mathrm{i} y)$. With a polar coordinate system in the complex plane, i.e., a representation of $z$ by its modulus $\psi>0$ and its phase $\phi\in[0,2\pi[$ as in (\ref{eq:chvar}), we obtain instead--but equivalently--$J=\int_0^{+\infty}\mathrm{d}\psi\, \psi \int_0^{2\pi} \mathrm{d} \phi\, f(\psi\exp(\mathrm{i}\phi))$, which is the form appearing in (\ref{eq:res2}) after linearization of $\sin\psi$ at small angles. Since the difference between the variables $z$ and $Z$ in (\ref{eq:chvar}) is merely a simple translation (with a Jacobian that is clearly unity), this explains how to go from the spherical integral over the direction of $\kk_1$ or $\kk_2$ in (\ref{eq:res2}) to an integral over the complex plane in (\ref{eq:Iang}), since a neighborhood near the north pole on the sphere can be approximated by the tangent plane.} 
\begin{equation}
\label{eq:Iang}
I_{\rm ang}(\qb_1,\qb_2,\kb_1,\kb_2,\kb_3) = \frac{\qb_1\qb_2\kb_1\kb_2}{\kb_3} \int_0^{+\infty} \mathrm{d}\bar{\theta}_{12}\, \bar{\theta}_{12} \int_{\mathbb{C}} \dd^2Z_1 \int_{\mathbb{C}} \dd^2Z_2 \, |\bar{\mathcal{A}}_{\mathrm{i}\to\textrm{f}}|^2 \delta(u-v)
\end{equation}
 A more elegant form of the integrals (\ref{eq:cphi}) and (\ref{eq:Iang}), and also easier to handle numerically, is given in Appendix \ref{ann:simp}. The coefficient $C_\phi$ is, like the thermal occupation numbers, an increasing function of $z_\phi$--$I_{\rm ang}$, the integral of a positive function, is positive--but, as we shall see in Section \ref{sec3}, it actually depends on it quite little.

Let us conclude this section with an important remark. We expect Landau and Khalatnikov’s quantum hydrodynamics to be an exact effective theory at low energy, and thus that (\ref{eq:res3}) is an exact equivalent of the rate of change of the number of phonons, regardless of the strength of the interactions in the superfluid (we have no reason to doubt the kinetic equations in the regime of weak coupling between phonons). 

\section{Relaxation kinetics of phonon fugacity}\label{sec3}

Bose’s law (\ref{eq:loiB}) involves two time-dependent variables: the temperature $T(t)$ and the phonon fugacity $z_\phi(t)=\exp(\beta(t)\mu_\phi(t))$, where $\mu_\phi(t)$ is the chemical potential of the phonons. We have only a single evolution equation, namely (\ref{eq:res3}) for the total number of phonons $N_\phi$. Fortunately, a second equation is provided by the conservation of total phonon energy $E_\phi$, since our fluid is isolated within the quantization volume $\mathcal{V}$. However, we still need to relate these quantities to $z_\phi$ and $T$.

To do this, the simplest approach is to perform the calculation of the grand potential $\Omega_\phi(T,\mu_\phi,\mathcal{V})$ of the phonon gas, still at the thermodynamic limit and--which allows us to linearize the acoustic branch--at the dominant order in temperature: 
\begin{equation}
\frac{\Omega_\phi}{\mathcal{V}} \underset{T\to 0^+}{\sim} k_{\rm B}T \int_{\mathbb{R}^3} \frac{\dd^3q}{(2\pi)^3} \ln\left[1-\eee^{-\beta(\hbar c q-\mu_\phi)}\right]= -mc^2\frac{\epsilon^4}{\pi^2\xi^3} g_4(z_\phi)
\end{equation}
 where $\xi=\hbar/mc$ is the so-called healing length, $g_\alpha(z)=\sum_{n\geq 1} z^n/n^\alpha$ is a Bose function or polylogarithm, and we set $\epsilon=k_{\rm B}T/mc^2$ as in equation (\ref{eq:defeps}). The various thermodynamic quantities are derived from the definition $\Omega_\phi=E_\phi-T S_\phi -\mu_\phi N_\phi$ and the associated differential form, $\mathrm{d}\Omega_\phi=-S_\phi \mathrm{d} T - N_\phi \mathrm{d}\mu_\phi -P_\phi\mathrm{d}\mathcal{V}$, where $S_\phi$ is the entropy and $P_\phi$ is the pressure of the phonon gas.\footnote{Note that, in our low-temperature limit, $S_\phi$ is also the entropy of the system as a whole, insofar as phonons are the only thermally populated excitations. We could therefore denote it as $S$.} Recalling that $(\dd/\mathrm{d}\nu) g_\alpha(\eee^\nu)=g_{\alpha-1}(\eee^\nu)$, we obtain 
\begin{equation}
\label{eq:grand}
\frac{N_\phi}{\mathcal{V}} \sim \frac{\epsilon^3}{\pi^2\xi^3} g_3(z_\phi) \quad ; \quad \frac{E_\phi}{\mathcal{V}} \sim mc^2 \frac{3\epsilon^4}{\pi^2\xi^3} g_4(z_\phi) \quad ; \quad \frac{S_\phi}{\mathcal{V}k_{\rm B}} \sim \frac{\epsilon^3}{\pi^2\xi^3}\left[4 g_4(z_\phi)-g_3(z_\phi)\,\ln z_\phi\right]
\end{equation}
The pressure is proportional to the volumetric energy, $P_\phi\sim E_\phi/3\mathcal{V}$; this is a well-known property of ultrarelativistic ideal gases.

Energy conservation allows us to eliminate $\epsilon(t)$ in favor of fugacity $z_\phi(t)$, most conveniently in terms of its value $\epsilon_{\rm eq}=k_{\rm B} T_{\rm eq}/mc^2$ at complete thermochemical equilibrium, reached for $z_\phi=1$ (asymptotically with respect to time) and denoted by the index “eq”: 
\begin{equation}
\label{eq:elim}
\epsilon^4 g_4(z_\phi) = \mbox{const} \underset{t\to +\infty}{=} \epsilon_{\rm eq}^4 g_4(1) = \epsilon_{\rm eq}^4 \zeta(4)\quad\mbox{so}\quad \epsilon(t)=\epsilon_{\rm eq} \left(\frac{\zeta(4)}{g_4(z_\phi(t))}\right)^{1/4}
\end{equation}
 where $\zeta$ is the Riemann zeta function. We then derive from expressions (\ref{eq:grand}) and (\ref{eq:res3}) of $N_\phi$ and its derivative the desired equation of motion for fugacity: 
\begin{equation}
\label{eq:mouv}
\boxed{
\frac{1}{\Gamma_\phi^{\rm eq}}\frac{\mathrm{d}}{\mathrm{d} t}z_\phi = F(z_\phi) \equiv \frac{4\zeta(2)\zeta(4)-3\zeta^2(3)}{4g_2(z_\phi)g_4(z_\phi) - 3 g_3^2(z_\phi)} \left(\frac{\zeta(4)}{g_4(z_\phi)}\right)^{5/4} \frac{C_\phi(z_\phi)}{C_\phi(1)} z_\phi^3 (1-z_\phi)\geq 0
}
\end{equation}
 We have introduced the relaxation rate $\Gamma_\phi^{\rm eq}$ for $z_\phi$ in the vicinity of complete equilibrium, 
\begin{equation}
\label{eq:gamphieq}
\Gamma^{\rm eq}_\phi\underset{T_{\rm eq}\to 0^+}{\sim}\frac{3^5}{(4\pi)^7} \frac{4\zeta(4) C_\phi(1)}{4\zeta(2)\zeta(4)-3\zeta^2(3)} \frac{(1+\Lambda)^6}{\gamma^2(\rho\xi^3)^3} \left(\frac{k_{\rm B}T_{\rm eq}}{mc^2}\right)^9 \frac{mc^2}{\hbar}
\end{equation}
 to bring us back to a dimensionless function $F(z_\phi)$ that depends only on the fugacity. This rate is proportional to the temperature raised to the ninth power, as we stated in section \ref{sec0}. It enters the linear relaxation equation to which any quantity $G$ varying--after eliminating $\epsilon$ as in (\ref{eq:elim})--linearly with the fugacity at $z_\phi=1$ (in the sense that $(\dd/\dd z_\phi)G(z_\phi)\neq 0$ at $z_\phi=1$), 
\begin{equation}
\frac{\mathrm{d}}{\mathrm{d} t} G(z_\phi(t)) \underset{\Gamma_\phi^{\rm eq}t\to+\infty}{\sim}  -\Gamma_\phi^{\rm eq} \left[G(z_\phi(t))-G_{\rm eq}\right]
\end{equation}
 such that the temperature $T$, the chemical potential $\mu_\phi$, or the density $\rho_\phi=N_\phi/\mathcal{V}$ of the phonon gas, 
\begin{equation}
\frac{\mathrm{d}}{\mathrm{d} t} T(t) \underset{\Gamma_\phi^{\rm eq}t\to+\infty}{\sim} -\Gamma_\phi^{\rm eq}[T(t)-T_{\rm eq}] \quad ; \quad 
\frac{\mathrm{d}}{\mathrm{d} t} \mu_\phi \underset{\Gamma_\phi^{\rm eq}t\to+\infty}{\sim} -\Gamma_\phi^{\rm eq} \mu_\phi \quad ; \quad
\frac{\mathrm{d}}{\mathrm{d} t} \rho_\phi \underset{\Gamma_\phi^{\rm eq}t\to+\infty}{\sim} -\Gamma_\phi^{\rm eq} [\rho_\phi-\rho_\phi^{\rm eq}]
\end{equation}
At long times $\Gamma_\phi^{\rm eq}t\gg 1$, one can indeed replace $G$ with its affine approximation $G(z_\phi)\simeq G_{\rm eq}+G'(1)(z_\phi-1)$ and use the fact that,
 by definition, $(\dd/\mathrm{d} t)z_\phi\sim -\Gamma_\phi^{\rm eq} (z_\phi-1)$. Here $G'(1)$ is the derivative of $G$ with respect to $z_\phi$ at $z_\phi=1$. If, on the other hand, the quantity $G$ varies quadratically--the first derivative $G'(1)$ is zero, the second derivative $G''(1)$ is nonzero--, we rather use the approximation $G(z_\phi)\simeq G_{\rm eq}+(1/2)G''(1)(z_\phi-1)^2$ and relaxation occurs asymptotically at twice the rate, 
\begin{equation}
\frac{\mathrm{d}}{\mathrm{d} t} G(z_\phi) \underset{\Gamma_\phi^{\rm eq}t\to+\infty}{\sim} G''(1) (z_\phi-1) \frac{\mathrm{d}}{\mathrm{d} t} z_\phi \underset{\Gamma_\phi^{\rm eq}t\to+\infty}{\sim} -2\Gamma_\phi^{\rm eq} \left[G(z_\phi)-G_{\rm eq}\right]
\end{equation}
where the time dependency of $z_\phi$ is implicit, but it makes more physical sense to write 
\begin{equation}
\label{eq:insp}
\frac{\mathrm{d}}{\mathrm{d} t} G(z_\phi) \underset{\Gamma_\phi^{\rm eq}t\to+\infty}{\sim} -\frac{G''(1)}{\Gamma_\phi^{\rm eq}} \left(\frac{\mathrm{d} z_\phi}{\mathrm{d} t}\right)^2
\end{equation}
This case of quadratic variation corresponds to entropy $S_\phi$, whose expression (\ref{eq:grand}) and elimination (\ref{eq:elim}) lead to 
\be 
\frac{S_\phi(z_\phi)}{S_\phi^{\rm eq}} = \left(\frac{\zeta(4)}{g_4(z_\phi)}\right)^{3/4} \frac{4 g_4(z_\phi)- g_3(z_\phi)\, \ln z_\phi}{4\zeta(4)}
\ee 
and thus to\footnote{We indeed have the Taylor expansions at $\nu=0$: $g_4(e^\nu)=\zeta(4)+\zeta(3)\nu+(1/2)\zeta(2)\nu^2+\ldots$ and $g_3(e^\nu)=\zeta(3)+\zeta(2)\nu+\ldots$, with $\nu=\ln z_\phi$.} $S'_\phi(1)=0$ and 
\begin{equation}
\frac{S''_\phi(1)}{S_\phi^{\rm eq}} = -\frac{4\zeta(2)\zeta(4)-3\zeta^2(3)}{16\,\zeta^{2}(4)} <0 
\end{equation}
Equation (\ref{eq:insp}) applied to $S_\phi$ nicely reproduces the first formula outside the main text of section \S 11 of reference \cite{LandauStat}, derived from an argument “à la Landau” in the slightly different context of an adiabatic transformation driven from the outside by the arbitrarily slow variation of a physical parameter of the system. More generally, in our problem, we find that entropy is always an increasing function of time, in accordance with the second law of thermodynamics, even far from the state of complete equilibrium $z_\phi=1$: 
\begin{equation}
\frac{1}{\Gamma_\phi^{\rm eq}}\frac{\mathrm{d}}{\mathrm{d} t} \frac{S_\phi(t)}{S_\phi^{\rm eq}}  = \frac{4\zeta(2)\zeta(4)-3\zeta^2(3)}{16\,\zeta^2(4)}\left(\frac{\zeta(4)}{g_4(z_\phi)}\right)^3 (-\ln z_\phi) z_\phi^2(1-z_\phi) \frac{C_\phi(z_\phi)}{C_\phi(1)} \geq 0
\end{equation}

 Formally, the equation of motion (\ref{eq:mouv}) integrates easily since it is separable. Furthermore, since it is first-order in time, its particular solution corresponding to the initial state $z_\phi(t=0)=0$ of a Boltzmann phonon gas--infinitely non-degenerate because infinitely sparse--contains all possible trajectories, up to a redefinition at the origin of time. This particular solution is given by 
\begin{equation}
\label{eq:sol0}
\int_0^{z_\phi(t)} \frac{\mathrm{d} z_\phi}{F(z_\phi)} = \Gamma_\phi^{\rm eq}t
\end{equation}
We now need to perform a numerical calculation of $C_\phi(z_\phi)$, as explained in Appendix \ref{ann:num} and shown in Figure \ref{fig2}a, to plot said particular solution in Figure \ref{fig2}b. Simple analytical expressions can be obtained in limiting cases. At short times, property $g_\alpha(z)\sim z, \ z\to 0$ leads to 
\begin{equation}
F(z_\phi) \underset{z_\phi\to 0^+}{\sim} \frac{4}{5} A z_\phi^{-1/4} \quad \mbox{with}\quad A=\frac{5}{4} [\zeta(4)]^{5/4} [4\zeta(2)\zeta(4)-3\zeta^2(3)]\frac{C_\phi(0)}{C_\phi(1)}
\end{equation}
hence the non-integer power law 
\begin{equation}
\label{eq:courts}
z_\phi(t) \underset{\Gamma_\phi^{\rm eq}t\to 0^+}{\sim} (A \Gamma_\phi^{\rm eq}t)^{4/5}
\end{equation}
 shown in red in Figure \ref{fig2}b. Numerically, we find that 
\begin{equation}
\label{eq:val01}
C_\phi(0)= 3.528(1)\times 10^4 \quad \mbox{and}\quad C_\phi(1)= 5.422(3)\times 10^4 \quad\mbox{so}\quad A= 2.502(2)
\end{equation}
At long times, property \footnote{The function $g_2(z_\phi)$ has the singular term $-(z_\phi-1)\ln(1-z_\phi)$ in its expansion at $z_\phi=1^-$, since its derivative at this point diverges as $g_1(z_\phi)=-\ln(1-z_\phi)$. Numerically, the derivative of $C_\phi(z_\phi)$ also appears to diverge logarithmically at $z_\phi=1^-$ (analytically, this results from the fact that, for $z_\phi=1$, the statistical factor $\prod_{\alpha,i}(1+\bar{n})$ in (\ref{eq:cphi}) diverges when one or more reduced wave numbers tend to zero, whereas $I_{\rm ang}$ tends to zero always fast enough to make the integral giving $C_\phi(1)$ converge but not always fast enough to make the one giving its derivative $C_\phi'(1)$ converge). Hence the logarithmic bound on the remainder in (\ref{eq:devinvF}). Let us be more precise regarding the behavior of $I_{\rm ang}(\qb_1,\qb_2,\kb_1,\kb_2,\kb_3)$ (\ref{eq:Iang}) at low wavenumbers, restricting ourselves directly to the manifold $\mathcal{W}$ (\ref{eq:defW}) for simplicity: (a) if $\qb_1\to 0^+$ with $\kb_1$ and $\kb_2$ fixed, the vanishing is quartic, $I_{\rm ang}\propto \qb_1^4$; (b) if $\kb_1\to 0^+$ with $\qb_1$ and $\kb_2$ fixed, it is linear, $I_{\rm ang}\propto \kb_1$; (c) if $\kb_1\propto \kb_2\to 0^+$ with $\qb_1$ fixed, it is quintic, $I_{\rm ang}\sim C(\qb_1)\kb_1\kb_2 (\kb_1^3+\kb_2^3)+(4\pi/3)^2\qb_1 \qb_2 \kb_1 \kb_2 (\kb_1+\kb_2)^3$, where $C(\qb_1)$ is some function of $\qb_1$; (d) if $\qb_1\propto \kb_1\propto \kb_2\to 0^+$, it is sextic, $I_{\rm ang}\sim (4\pi/3)^2\kb_1\kb_2\qb_1\left[\min(\qb_1^3,(\kb_1+\kb_2)^3)+Y(\qb_1-\kb_1) \,\min(\kb_2^3,(\qb_1-\kb_1)^3)+Y(\qb_1-\kb_2)\,\min(\kb_1^3,(\qb_1-\kb_2)^3)\right]$, with $Y$ being the Heaviside function. We obtain these results using the representation described in Appendix \ref{ssec:chvar}--specifically, the parameterization (\ref{eq:cgz12}) of the complex variables $Z_1$ and $Z_2$ in terms of $\zeta_+$ and $\zeta_-$, the formulation (\ref{eq:Iangjol}) of $I_{\rm ang}$ as an integral over $\zeta_+$ and $\zeta_-$ in the complex plane, and the expression (\ref{eq:thb120num}) for the angle $\thb_{12}$ on the energy shell. Result (a) is obtained by taking the limit $\qb_1\to 0^+$ inside the integral in (\ref{eq:Iangjol})--with $\zeta_+$ and $\zeta_-$ held fixed--and showing that $\bar{\mathcal{A}}_{\ii\to\ff} \propto \qb_1^2$ (the leading terms in (\ref{eq:Abar}) are $\propto 1/\thb_{12}^{(0)2}\propto \qb_1$, but they cancel out at this order). Result (b) is obtained by performing the change of variable $\zeta_-=\kb_1^{1/2}\check{\zeta}_-$ and then taking the limit $\kb_1\to 0^+$ with $\zeta_+$ and $\check{\zeta}_-$ held fixed (one verifies that $\bar{\mathcal{A}}_{\ii\to\ff} \propto \kb_1^0$, with the factor $\kb_1$ arising from the Jacobian $\dd^2\zeta_-/\dd^2\check{\zeta}_-$). If we set $\kb_2 > \kb_1$ to simplify the notation, result (c) is obtained by taking the limit (i) with $\zeta_+$ and $\check{\zeta}_- = \zeta_-/\kb_1^{1/2}$ fixed ($\mathcal{A}_{\ii\to\ff} \propto \kb_2^2$, with the leading terms $\propto \kb_2$ cancelling out), then (ii) with $\zeta_-$ and $\check{\zeta}_+ = \zeta_+/\kb_2^{1/2}$ fixed ($\mathcal{A}_{\ii\to\ff} \propto \kb_1^2$, with the leading terms $\propto \kb_1$ cancelling out), and finally (iii) with $\zeta_+$ and $w \in \mathbb{C}$ fixed, using the change of variable $\zeta_- = \zeta_-^{\chib_{12}=0}(\zeta_+) + w \kb_2^{3/2}$ (where $\zeta_-^{\chib_{12}=0}(\zeta_+) = [(a-b)/(a+b)](m_-/m_+)^{1/2}\zeta_+$ is the value of $\zeta_-$ that makes $\chib_{12}$ vanish for a fixed $\zeta_+$). Still assuming $\kb_2>\kb_1$, result (d) is obtained first by taking the limit with $\check{\zeta}_+=\zeta_+/\kb_2^{1/2}$ and $w$ fixed, after performing the changes of variables (i) $\zeta_-=\zeta_-^{\chib_{12}=0}(\check{\zeta}_+)+\kb_2^{3/2} w$ and (ii) $\zeta_-=\zeta_-^{\psib_{11}=0}(\check{\zeta}_+)+\kb_2^{3/2} w$, and then by taking the limit with $\check{\zeta}_-=\zeta_-/\kb_1^{1/2}$ and $w$ fixed after the change of variable (iii) $\zeta_+=\zeta_+^{\psib_{12}=0}(\check{\zeta}_-)+\kb_2^{3/2} w$; unlike $\zeta_-^{\chib_{12}=0}(\check{\zeta}_+)$, the zero of $\psib_{11}$ or $\psib_{12}$ for fixed $\check{\zeta}_+$ or $\check{\zeta}_-$ does not have a simple explicit expression, but here it suffices to know it at leading order: $\zeta_-^{\psib_{11}=0}(\check{\zeta}_+)\sim [\kb_1/(\qb_1-\kb_1)]^{1/2}[\kb_2|\check{\zeta}_+|^2+(3/4)(\kb_1+\kb_2-\qb_1)]^{1/2}$ or $\zeta_+^{\psib_{12}=0}(\check{\zeta}_-)\sim [\kb_2/(\qb_1-\kb_2)]^{1/2}[\kb_1|\check{\zeta}_-|^2+(3/4)(\kb_1+\kb_2-\qb_1)]^{1/2}$ (it exists only when $\qb_1>\kb_1$ or $\qb_1>\kb_2$). As is evident from (d), the regions where the angles $\psib_{\alpha i}$ or $\chib_{jl}$ vanish contribute to $I_{\rm ang}$ at leading order only if, in the denominators of (\ref{eq:Abar}), these angles are divided by "Lorentzian widths" $|\qb_\alpha-\kb_i|$ or $\kb_j+\kb_l$ that tend to zero (the angle $\psib_{23}$ is an exception, as it tends to zero everywhere for fixed $\check{\zeta}_+$ and $\check{\zeta}_-$).} 
\begin{equation}
\label{eq:devinvF}
\frac{1}{F(z_\phi)} \underset{z_\phi\to 1^-}{=} \frac{1}{1-z_\phi} +O(\ln(1-z_\phi))
\end{equation}
 shows that the deviation $1/F(z_\phi)-1/(1-z_\phi)$ has a finite integral over the interval $[0,1]$ and motivates a simple plus-minus trick on the integrand of (\ref{eq:sol0}), $1/F(z_\phi)=1/(1-z_\phi)+[1/F(z_\phi)-1/(1-z_\phi)$]. Integration yields an exponential law as expected, 
\begin{equation}
\label{eq:longs}
z_\phi(t) \underset{\Gamma_\phi^{\rm eq}t\to+\infty}{=} 1-\exp(B-\Gamma_\phi^{\rm eq}t) +O\left(\Gamma_\phi^{\rm eq}t\exp(-2\Gamma_\phi^{\rm eq}t)\right)
\end{equation}
 with 
\begin{equation}
B=\int_0^{1} \mathrm{d} z_\phi \left[\frac{1}{F(z_\phi)}-\frac{1}{1-z_\phi}\right] = -1.269(3)
\end{equation}
 The law (\ref{eq:longs}) is shown in blue--of course without the $O$--in Figure \ref{fig2}b.
\begin{figure}[t]
\begin{center}
\includegraphics[width=6cm,clip=]{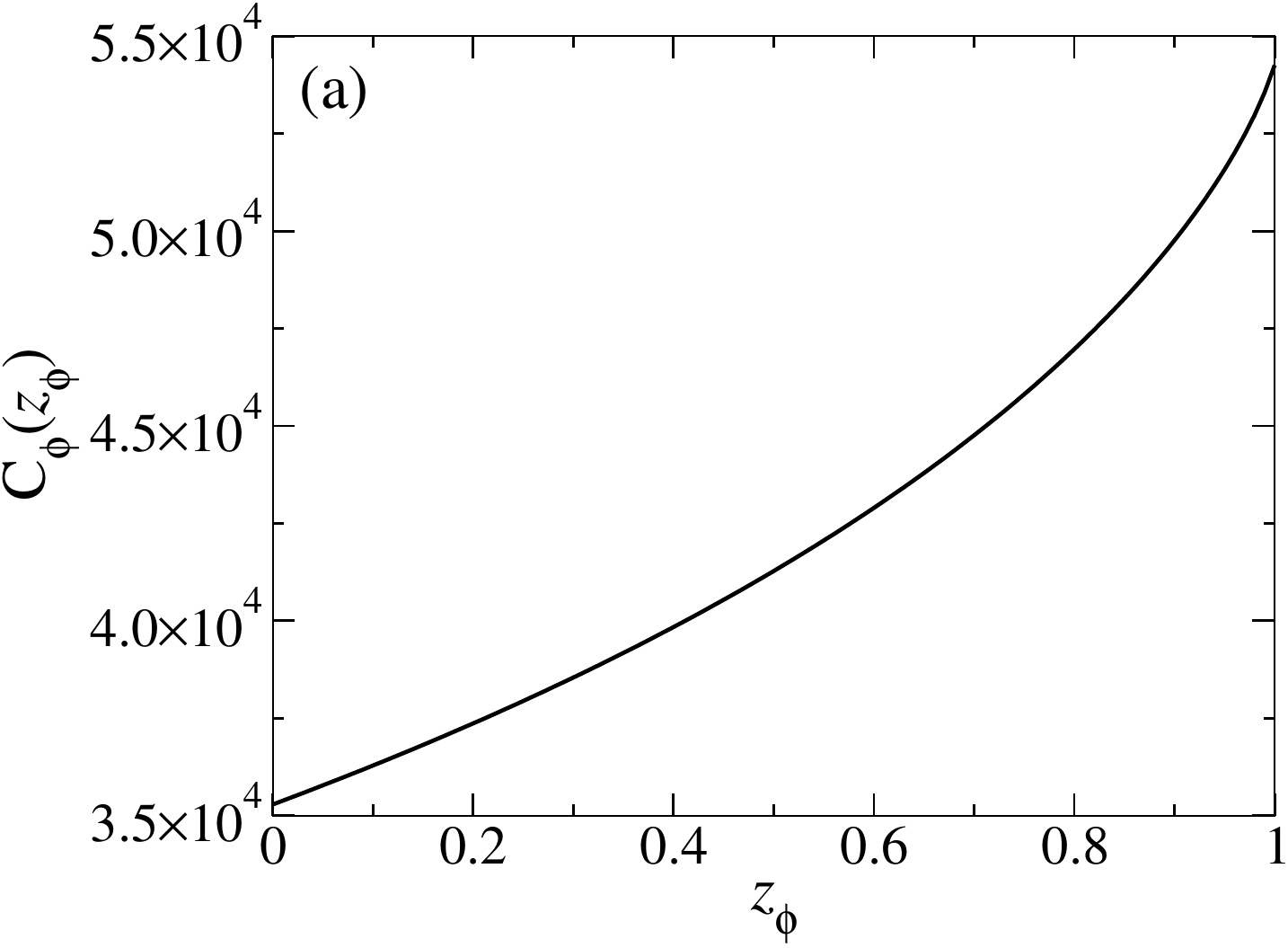} \hspace{5mm} \includegraphics[width=6cm,clip=]{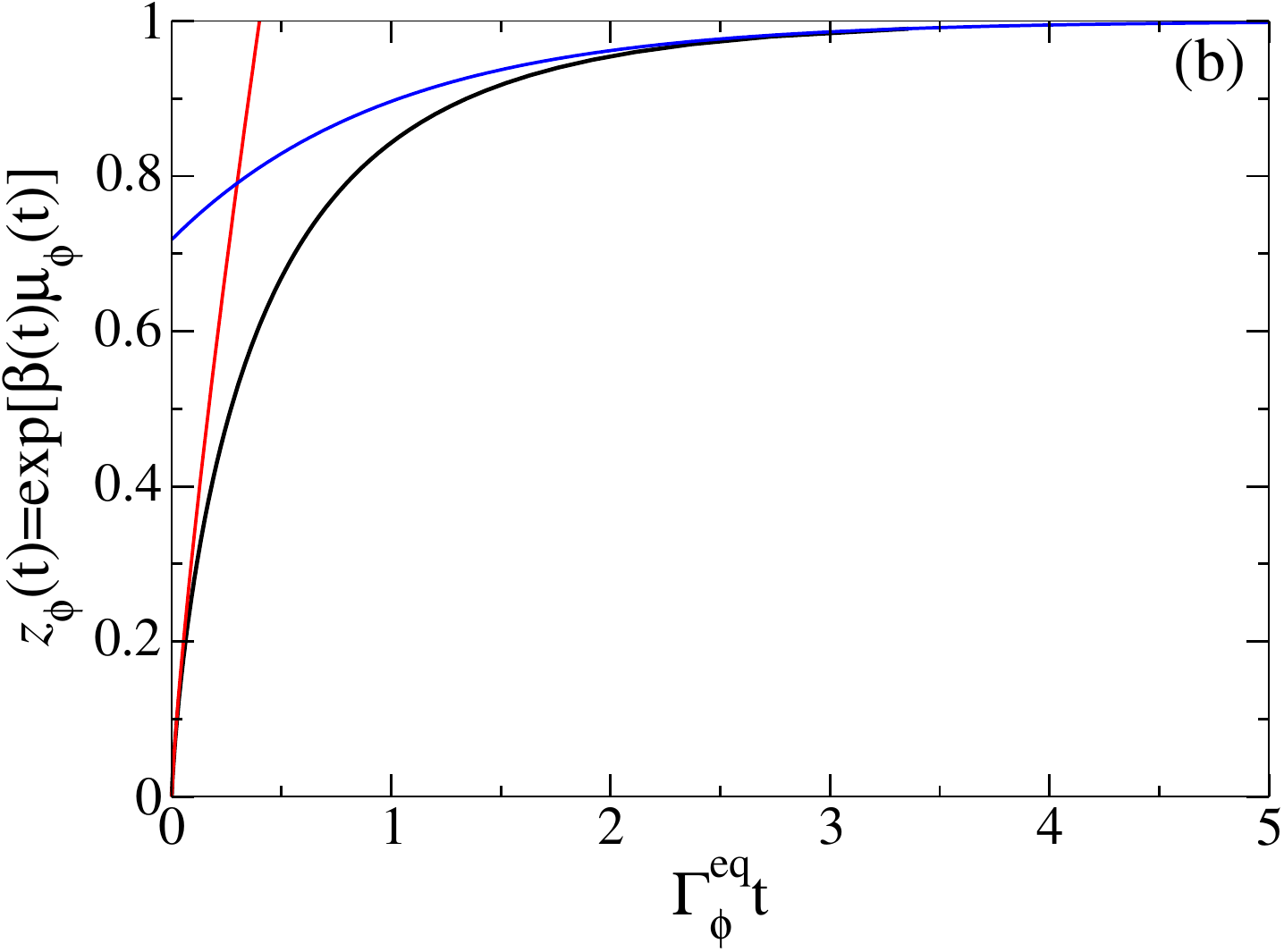} 
\end{center}
\caption{For a gas of $N_\phi$ phonons at thermal but not chemical equilibrium--that is, with fugacity $z_\phi < 1$--in a spatially homogeneous, isolated superfluid with a concave acoustic branch: (a) dependence on $z_\phi$ of the dimensionless coefficient $C_\phi$ in the rate of change (\ref{eq:res3}) of $N_\phi$ under the effect of Khalatnikov’s five-phonon collisions $2\phi\leftrightarrow 3\phi$, and (b) corresponding relaxation of $z_\phi$ toward its value at complete thermodynamic equilibrium according to equation (\ref{eq:sol0}). In (b), shown in color, the limiting behaviors at short times (\ref{eq:courts}) and at long times (\ref{eq:longs}). The asymptotic relaxation rate $\Gamma_\phi^{\rm eq}$ is given by (\ref{eq:gamphieq}).}
\label{fig2}
\end{figure}

 \appendix

\section{The generalized Fermi golden rule}\label{ann:orgen}

The generalized Fermi golden rule is usually applied at the minimal order $n_0$ in the perturbation $\hat{V}$ at which it predicts a non-zero transition rate between unperturbed eigenstates, taking into account every possible initial state $|\mathrm{i}\rangle$ of energy $E_\ii$ and every possible final state $|\textrm{f}\rangle$ of energy $E_\ff$. The intermediate states $|\mu\rangle$ involved in the perturbative expression of the corresponding order-$n_0$ transition amplitude $\mathcal{A}_{\mathrm{i}\to\textrm{f}}$ then all have energies $E_\mu$ different from $E_\ii$ (and $E_\ff$), otherwise the transition $|\mathrm{i}\rangle\to |\mu\rangle$ would be allowed at an order in $\hat{V}$ lower than $n_0$. Here, on a concave acoustic branch, the perturbation $\hat{V}$ is the sum of the cubic interaction (\ref{eq:h3m}) between phonons and its Hermitian conjugate; we have $n_0=2$ (the dominant processes $2\phi\to 2\phi$ have an order-two amplitude in $\hat{V}$), and our use of the golden rule in section \ref{sec1} falls outside this framework, since the kinetic equation (\ref{eq:dnq}) describes the subdominant processes $2\phi\leftrightarrow3\phi$, whose amplitude is of order three in $\hat{V}$. We therefore expect to encounter some subtleties, as explained in section \ref{ssec:crise}.

Another unsatisfactory aspect of our use of the golden rule is that the transition amplitude $2\phi\rightarrow3\phi$ (\ref{eq:fmix}) is calculated in vacuum, whereas collisions actually occur in the thermal bath of the phonon gas; thus, in the contributions (\ref{eq:ampa})--
 (\ref{eq:ampf}) of the diagrams in Figure \ref{fig:diag}, one would need to take into account the factors $(n+1)^{1/2}$ or $n^{1/2}$ accompanying the creation or annihilation of an internal phonon in a mode already thermally populated (these factors in the modes of incoming or outgoing phonons are indeed included in equation (\ref{eq:dnq})); furthermore, the new diagrams allowed by this thermal occupation of the intermediate modes should be added to Figure \ref{fig:diag}. This issue has been addressed for collisions $2\phi\to 2\phi$ in reference \cite{Annalen}, which finds that the transition amplitude has the same value in the thermal bath and in vacuum as long as one remains on the energy shell $E_\ii=E_\ff$. In Section \ref{ssec:soupe}, we extend this conclusion to collisions $2\phi\rightarrow3\phi$ (still at the dominant order in $\hat{V}$).

\subsection{A subtlety of the subdominant order}\label{ssec:crise}

The simplest way to conduct the perturbative study of the transition probability amplitude from $|\mathrm{i}\rangle=|\qq_1,\qq_2\rangle$ to $|\textrm{f}\rangle=|\kk_1,\kk_2,\kk_3\rangle$ is to start from its exact expression \cite{Bordeaux}, 
\begin{equation}
\langle\ff|\hat{U}(t)|\mathrm{i}\rangle = \int_{+\infty+\mathrm{i} 0^+}^{-\infty+\mathrm{i} 0^+} \frac{\mathrm{d} z}{2\mathrm{i}\pi} \eee^{-\mathrm{i} z t/\hbar} \langle\ff|\hat{G}(z)|\mathrm{i}\rangle
\end{equation}
 with $\hat{U}(t)$ as the evolution operator during $t>0$, then expand the resolvent of the complete Hamiltonian $\hat{G}(z)=(z\mathrm{Id}-\hat{H})^{-1}$ into powers of $\hat{V}$, $\hat{G}=\hat{G}_0+\hat{G}_0\hat{V}\hat{G}_0+\hat{G}_0\hat{V}\hat{G}_0\hat{V}\hat{G}_0+\ldots$ (the notation is that of Section \ref{sec2}). At the third order in $\hat{V}$, which is of interest to us, we obtain 
\begin{multline}
\langle\ff|\hat{U}(t)|\mathrm{i}\rangle\Big|^{(3)}= \int_{+\infty+\mathrm{i} 0^+}^{-\infty+\mathrm{i} 0^+} \frac{\mathrm{d} z}{2\mathrm{i}\pi} \frac{\exp(-\mathrm{i} z t/\hbar)}{(z-E_\ii)(z-E_\ff)} 
\langle\ff|\hat{V}\hat{G}_0(z)\hat{V}\hat{G}_0(z)\hat{V}|\mathrm{i}\rangle \\
= \int_{+\infty+\mathrm{i} 0^+}^{-\infty+\mathrm{i} 0^+} \frac{\mathrm{d} z}{2\mathrm{i}\pi} \frac{\exp(-\mathrm{i} z t/\hbar)}{(z-E_\ii)(z-E_\ff)} \sum_{\lambda,\mu} \frac{\langle\ff|\hat{V}|\mu\rangle \langle\mu|\hat{V}|\lambda\rangle \langle\lambda|\hat{V}|\mathrm{i}\rangle}{(z-E_\lambda)(z-E_\mu)}
\end{multline}
where the sum, as in Equation (\ref{eq:fmix}), is over the finite number of possible intermediate states. We now need to close the integration contour with a semicircle at infinity in the lower half of the complex plane, and apply the residue theorem in the general case of pairwise distinct energies: 
\begin{multline}
\label{eq:avcro}
\langle\ff|\hat{U}(t)|\mathrm{i}\rangle\Big|^{(3)} = \sum_{\lambda,\mu} \langle\ff|\hat{V}|\mu\rangle \langle\mu|\hat{V}|\lambda\rangle \langle\lambda|\hat{V}|\mathrm{i}\rangle
\left[
\frac{\exp(-\mathrm{i} E_\mathrm{i} t/\hbar)}{(E_\ii-E_\ff) (E_\ii-E_\lambda)(E_\ii-E_\mu)}
+\frac{\exp(-\mathrm{i} E_\textrm{f} t/\hbar)}{(E_\ff-E_\ii) (E_\ff-E_\lambda)(E_\ff-E_\mu)} \right. \\
\left. +\frac{\exp(-\mathrm{i} E_\lambda t/\hbar)}{(E_\lambda-E_\ii) (E_\lambda-E_\ff)(E_\lambda-E_\mu)}
+\frac{\exp(-\mathrm{i} E_\mu t/\hbar)}{(E_\mu-E_\ii) (E_\mu-E_\ff)(E_\mu-E_\lambda)}
\right]
\end{multline}
 The rest resembles the derivation of the usual Fermi golden rule: the square of the amplitude begins to diverge linearly with time, which allows us to define an elementary transition rate $\delta\Gamma_{\mathrm{i}\to\textrm{f}}$, when two of the energies are very close. However, recall that the perturbation $\hat{V}$ is a three-phonon interaction and does not conserve energy on a concave acoustic branch; it follows that $E_\lambda$ is well separated from $E_\ii$, $E_\mu$ is well separated from $E_\ff$, and $E_\lambda$ and $E_\mu$ are well separated from each other; on the other hand, we may have $E_\ii=E_\ff$, $E_\ii=E_\mu$, or $E_\lambda=E_\ff$ (but not two of these possibilities at the same time).

Let’s start with the expected case of a crossing between $E_\ii$ and $E_\ff$. We can then approximate $E_\ii$ and $E_\ff$ by their half-sum $\bar{E}=(E_\ii+E_\ff)/2$ in the expression in brackets in (\ref{eq:avcro}), except in the factor $E_\ii-E_\ff$ and under the exponentials (the time $t$ diverges): 
\begin{equation}
\label{eq:croisif}
\Big[\ldots\Big]_{\rm Eq.\, (\ref{eq:avcro})} \simeq \frac{-2\mathrm{i}\eee^{-\mathrm{i} \bar{E}t/\hbar}\sin[(E_\ii-E_\ff)t/2\hbar]}{(E_\ii-E_\ff)(\bar{E}-E_\lambda)(\bar{E}-E_\mu)} 
+\frac{1}{E_\lambda-E_\mu}\left(\frac{\eee^{-\mathrm{i} E_\lambda t/\hbar}}{(E_\lambda-\bar{E})^2}-\frac{\eee^{-\mathrm{i} E_\mu t/\hbar}}{(E_\mu-\bar{E})^2} \right)
\end{equation}
 In the sense of distributions, $\sin^2(\Omega t)/(\Omega^2 t)$ tends toward $\pi\delta(\Omega)$ when $t\to +\infty$, so the square of the modulus of the first contribution in (\ref{eq:croisif}) diverges linearly in time; conversely, there is no divergence in time in the square of the modulus of the second contribution, or in the interference term between the two contributions (for example, $\sin(\Omega t)/\Omega$ tends toward $\pi\delta(\Omega)$). We arrive at the expected generalized golden rule: 
\begin{equation}
\label{eq:attendu}
\left|\langle\ff|\hat{U}(t)|\mathrm{i}\rangle\Big|^{(3)}\right|^2 \stackrel{E_\mathrm{i}\simeq E_\textrm{f}}{\underset{t\to +\infty}{\sim}} \frac{2\pi t}{\hbar} \delta(E_\ii-E_\ff) \left|\sum_{\lambda,\mu} \frac{\langle\ff|\hat{V}|\mu\rangle \langle\mu|\hat{V}|\lambda\rangle \langle\lambda|\hat{V}|\mathrm{i}\rangle}{(\bar{E}-E_\lambda)(\bar{E}-E_\mu)}\right|^2
\end{equation}

 Let us continue with the unusual case of a crossing between $E_\ii$ and $E_\mu$ (the case of a crossing between $E_\ff$ and $E_\lambda$ is handled symmetrically). First, noting Figure \ref{fig:diag}, we observe that resonance $E_\ii=E_\mu$ can indeed occur, but only in diagrams (a), (c) and (e), where the intermediate state $|\mu\rangle$ has two phonons.\footnote{Indeed, on a concave rising acoustic branch, the merging of two phonons always decreases the energy and the splitting into two phonons always increases it, meaning that the energy varies in the same direction as the number of phonons under the effect of $\hat{H}_3^{(\pm)}$. } Next, let us perform the same manipulations in equation (\ref{eq:avcro}), this time setting $\bar{E}=(E_\ii+E_\mu)/2$, to obtain:\footnote{One could bring the sum over $\mu$ and the Dirac distribution inside the square modulus, at the cost of the somewhat unorthodox introduction of the square root $\delta^{1/2}$, to account for degeneracy, that is, the existence of several intermediate states $|\mu\rangle$ of the same energy. } 
\begin{equation}
\label{eq:drame}
\left|\langle\ff|\hat{U}(t)|\mathrm{i}\rangle\Big|^{(3)}\right|^2 \stackrel{E_\mathrm{i}\simeq E_\mu}{\underset{t\to +\infty}{\sim}} \frac{2\pi t}{\hbar}\sum_\mu \delta(E_\ii-E_\mu) \left|\sum_{\lambda} \frac{\langle\ff|\hat{V}|\mu\rangle \langle\mu|\hat{V}|\lambda\rangle \langle\lambda|\hat{V}|\mathrm{i}\rangle}{(\bar{E}-E_\lambda)(\bar{E}-E_\ff)}\right|^2
\end{equation}
 This expression seems to predict the existence of a transition $|\mathrm{i}\rangle\to|\textrm{f}\rangle$ with a non-zero rate occurring without energy conservation (if $E_\ii=E_\mu$, we necessarily have $E_\textrm{f}\neq E_\ii$ because $E_\textrm{f}\neq E_\mu$, see above). We would therefore be facing a serious “energy crisis.”

Fortunately, there is nothing real to this; it is merely the result of a misinterpretation. Equation (\ref{eq:drame}) is not a contribution to the transition rate $2\phi\to 3\phi$ between $|\mathrm{i}\rangle$ and $|\textrm{f}\rangle$, but results from the transition $2\phi\to 2\phi$ between $|\mathrm{i}\rangle$ and $|\mu\rangle$. This is paradoxical: since state $|\textrm{f}\rangle$ has three phonons and state $|\mu\rangle$ having two phonons, these states are orthogonal, and it is unclear how a transition from $|\mathrm{i}\rangle$ to $|\mu\rangle$ could increase the probability amplitude of presence in $|\textrm{f}\rangle$.
 But that would be to forget that the phonons introduced so far, far from being exact elementary excitations of the system, are merely a zero-order approximation in $\hat{V}$: under the incessant effect of the cubic Hamiltonian $\hat{V}$, each bare phonon interacts with itself and becomes dressed with so-called captive or virtual phonons \cite{erratique}. State $|\mu\rangle$ is therefore only the zero-order approximation in $\hat{V}$ of a two-phonon dressed state $|\tilde\mu\rangle$, which is written at first order as follows, according to standard perturbation theory: 
\begin{equation}
|\tilde\mu\rangle = |\mu\rangle + \sum_\nu |\nu\rangle \frac{\langle\nu|\hat{V}|\mu\rangle}{E_\mu-E_\nu}+O(\hat{V}^2)
\end{equation}
 The three-bare-phonon state $|\textrm{f}\rangle$ is not orthogonal to the two-phonon dressed states $|\tilde\mu\rangle$, and the transitions $2\phi\to 2\phi$ therefore indirectly induce a transition from $|\mathrm{i}\rangle$ to $|\textrm{f}\rangle$ with amplitude 
\begin{equation}
\label{eq:indi}
\langle f|\hat{U}(t)|\mathrm{i}\rangle\Big|_{2\phi\to 2\phi} = \sum_\mu \langle\ff|\tilde{\mu}\rangle \langle\tilde{\mu}|\hat{U}(t)|\mathrm{i}\rangle \simeq \sum_\mu \frac{\langle\ff|\hat{V}|\mu\rangle}{E_\mu-E_\textrm{f}} \langle\mu|\hat{U}(t)|\mathrm{i}\rangle\Big|^{(2)}
\end{equation}
It remains to recall the generalized Fermi golden rule at second order in $\hat{V}$, 
\begin{equation}
\label{eq:dgim}
\left|\langle\mu|\hat{U}(t)|\mathrm{i}\rangle\Big|^{(2)}\right|^2 \underset{t\to +\infty}{\sim}\frac{2\pi t}{\hbar} \delta(E_\ii-E_\mu) \left|\sum_\lambda \frac{\langle\mu|\hat{V}|\lambda\rangle \langle\lambda|\hat{V}|\mathrm{i}\rangle}{\bar{E}-E_\lambda}\right|^2
\end{equation}
 with $\bar{E}=(E_\ii+E_\mu)/2$ as before, and to approximate the remaining denominator $E_\mu-E_\ff$ by $\bar{E}-E_\ff$, to recognize the disturbing result (\ref{eq:drame}) in the square of the modulus of (\ref{eq:indi}).

In conclusion, at the dominant order in $\hat{V}$, the transition rate $2\phi\to 3\phi$ for a concave acoustic branch is correctly given by the expected golden rule (\ref{eq:attendu}), which is used in sections \ref{sec1} and \ref{sec2}. In particular, on the energy shell $E_\ii=E_\ff$, there is no particular divergence to be concerned about because there is no possible cancellation of an energy denominator $\bar{E}-E_\lambda$ or $\bar{E}-E_\mu$ in the effective transition amplitude, even if the transitions $2\phi\to 2\phi$ may conserve the energy-momentum. The situation is therefore considerably simpler than that of even a partially convex acoustic branch, where the calculation of the collision rate $2\phi\to 2\phi$, which is subdominant relative to $1\phi\leftrightarrow 2\phi$, requires non-perturbative regularization (the finite lifetime of intermediate phonons due to three-phonon processes is accounted for by assigning them a complex energy) \cite{Ukr}.\footnote{Let us be more precise in the concave case. The intermediate phonons, being part of a thermal bath, ultimately acquire a finite lifetime there due to collisions $2\phi\to 2\phi$. This gives them a complex energy, with an imaginary part of the order of $T^{7}$ as stated in section \ref{sec0}. This is negligible at low temperatures compared to the scale of the energy denominators $E_\ii-E_\lambda$, $E_\ff-E_\mu$, of the order of $T^3$ at small angles as in (\ref{eq:denomt3}). In particular, we can find in equation (\ref{eq:attendu}) times $t$ that are sufficiently long for the width $\hbar/t$ of the Dirac pseudo-delta to be negligible, but sufficiently short not to trigger the instability $2\phi\to 2\phi$.}

\subsection{In a thermal bath rather than in vacuum}\label{ssec:soupe}

The thermal occupation of the internal phonon modes in Figure \ref{fig:diag} has two consequences: (i) it introduces a new mechanism, the absorption of intermediate phonons (ultimately compensated by their re-emission),
 which is of course impossible at zero temperature, and (ii) it modifies the transition amplitudes via the bosonic amplitudes $n^{1/2}$ or $(1+n)^{1/2}$, where $n$ is the number of internal phonons. Fortunately, upon reflection we find that, for each diagram in Figure \ref{fig:diag}, the elementary three-phonon processes (the vertices of the diagram) remain exactly the same, but must now be combined in any order; in other words, each diagram at zero temperature gives rise to a class of $3!=6$ diagrams at non-zero temperature, derived by permutation of the three vertices (the case of diagrams (e) and (f) is a special case; see below). The calculation of the new amplitudes therefore remains fairly simple, since the matrix elements of the cubic interaction $\hat{V}$ can be factored out.

Let us examine the case of diagram (a) in detail. Let $\qq_{12}=\qq_1+\qq_2$ and $\kk_{jl}=\kk_j+\kk_l$ be the wave vectors of the internal phonons, and let $A_1=\langle\qq_{12}|\hat{V}|\qq_1,\qq_2\rangle$, $A_2=\langle\kk_i,\kk_{jl}|\hat{V}|\qq_{12}\rangle$, and $A_3=\langle\kk_j,\kk_l|\hat{V}|\kk_{jl}\rangle$ denote the zero-temperature matrix elements of the elementary three-phonon processes, in the order in which they appear in Figure \ref{fig:diag}, and let $\delta_1=\hbar\omega_{\qq_{12}}-\hbar\omega_{\qq_1}-\hbar\omega_{\qq_2}$, $\delta_2=\hbar\omega_{\kk_i}+\hbar\omega_{\kk_{jl}}-\hbar\omega_{\qq_{12}}$, and $\delta_3=\hbar\omega_{\kk_j}+\hbar\omega_{\kk_l}-\hbar\omega_{\kk_{jl}}$ denote the corresponding energy variations. The amplitude of the diagrams of class (a) in the presence of the thermal bath is written as a sum over the permutations $\sigma$ of the elementary processes, 
\begin{equation}
\label{eq:moda}
\mathcal{A}_{\mathrm{i}\to\textrm{f}}^{(a)} = \sum_i\sum_{\sigma\in S_3} \mathcal{A}^{(a)}_{\sigma(1)\sigma(2)\sigma(3)}
\end{equation}
 with the index of each term being the order of the actions chosen, from left to right in the diagram; as one will have understood, the case in vacuum--without a thermal bath--corresponds to the identity permutation and to the index $123$, as shown in Figure \ref{fig:diag}. We immediately focus on the energy shell, hence the relation 
\begin{equation}
\label{eq:mag}
\delta_1+\delta_2+\delta_3=0
\end{equation}
which we will make good use of in what follows. The calculation yields\footnote{In these expressions, the denominators are simply $\delta_{\sigma(1)}(\delta_{\sigma(1)}+\delta_{\sigma(2)})=-\delta_{\sigma(1)}\delta_{\sigma(3)}$.}
\begin{align}
\label{eq:deb1}
\mathcal{A}^{(a)}_{123}&=\frac{A_1A_2A_3}{-\delta_1\delta_3}(1+n_{\qq_{12}})(1+n_{\kk_{jl}}) \quad \quad &\mathcal{A}^{(a)}_{132}&=\frac{A_1A_2A_3}{-\delta_1\delta_2} n_{\kk_{jl}} (1+n_{\qq_{12}}) \\
\label{eq:deb2}
\mathcal{A}^{(a)}_{213}&=\frac{A_1A_2A_3}{-\delta_2\delta_3}n_{\qq_{12}} (1+n_{\kk_{jl}}) \quad \quad &\mathcal{A}^{(a)}_{231}&= \frac{A_1A_2A_3}{-\delta_2\delta_1} n_{\qq_{12}} (1+n_{\kk_{jl}}) \\
\label{eq:deb3}
\mathcal{A}^{(a)}_{312}&=\frac{A_1A_2A_3}{-\delta_3\delta_2}n_{\kk_{jl}}(1+n_{\qq_{12}}) \quad \quad &\mathcal{A}^{(a)}_{321}&= \frac{A_1A_2A_3}{-\delta_3\delta_1} n_{\qq_{12}} n_{\kk_{jl}}
\end{align}
which, after grouping the terms and then using (\ref{eq:mag}) to replace the factor $(\delta_1+\delta_3)$ with $-\delta_2$, 
\begin{multline}
\mathcal{A}_{\mathrm{i}\to\textrm{f}}^{(a)} = -\sum _i \frac{A_1A_2A_3}{\delta_1\delta_2\delta_3}\left\{(\delta_1+\delta_3)\left[n_{\kk_{jl}}(1+n_{\qq_{12}})+n_{\qq_{12}}(1+n_{\kk_{jl}})\right]\right. \\
\left.+\delta_2\left[(1+n_{\qq_{12}})(1+n_{\kk_{jl}})+n_{\qq_{12}}n_{\kk_{jl}}\right]\right\}=-\sum_i \frac{A_1A_2A_3}{\delta_1\delta_3}
\end{multline}
 Remarkably, we find the result in vacuum (\ref{eq:ampa}), regardless of the occupation numbers $n_{\qq_{12}}$ and $n_{\kk_{jl}}$ of the internal phonon modes (they do not need to be thermal) .

The same reasoning can be applied to diagrams (b) through (d), leading to the same conclusion. Finally, the remaining diagrams (e) and (f) should be grouped into a single class (ef), since the two intermediate phonons in these diagrams are equal,
 $\kk_{i\beta}= \kk_i-\qq_\beta$ and $\kk_{jl}=\kk_j+\kk_l$, and the three elementary processes involved are exactly the same, with matrix elements $A_1=\langle\kk_{jl},\kk_{i\beta}|\hat{V}|\qq_\alpha\rangle$, $A_2=\langle\kk_i|\hat{V}|\qq_\beta,\kk_{i\beta}\rangle$, $A_3=\langle\kk_j,\kk_l|\hat{V}|\kk_{jl}\rangle$ and the corresponding energy changes $\delta_1=\hbar\omega_{\kk_{jl}}+\hbar\omega_{\kk_{i\beta}}-\hbar\omega_{\qq_\alpha}$, $\delta_2=\hbar\omega_{\kk_i} -\hbar\omega_{\qq_\beta}-\hbar\omega_{\kk_{i\beta}}$, $\delta_3=\hbar\omega_{\kk_j} +\hbar\omega_{\kk_l}-\hbar\omega_{\kk_{jl}}$, a phenomenon that does not occur with the other diagrams. We then have, following the model of (\ref{eq:moda}), 
\begin{equation}
\mathcal{A}_{\mathrm{i}\to\textrm{f}}^{(ef)} = \sum_\alpha\sum_i\sum_{\sigma\in S_3} \mathcal{A}^{(ef)}_{\sigma(1)\sigma(2)\sigma(3)}
\end{equation}
 In vacuum, the identity permutation yields diagram (e) of Figure \ref{fig:diag}, the permutation $\tau_{23}$--transposition of indices $2$ and $3$--yields diagram (f), and the other permutations yield zero. In the thermal bath, following the lexicographic order as in (\ref{eq:deb1}), (\ref{eq:deb2}), (\ref{eq:deb3}) for $\sigma(1),\sigma(2),\sigma(3)$, we obtain 
\begin{multline}
\mathcal{A}_{\mathrm{i}\to\textrm{f}}^{(ef)} = \sum_\alpha\sum_i (-A_1A_2A_3) \left[\frac{1}{\delta_1\delta_3} (1+n_{\kk_{jl}})(1+n_{\kk_{i\beta}})
+\frac{1}{\delta_1\delta_2} (1+n_{\kk_{jl}})(1+n_{\kk_{i\beta}})+\frac{1}{\delta_2\delta_3} n_{\kk_{i\beta}} (1+n_{\kk_{jl}})\right.  \\
\left.+\frac{1}{\delta_1\delta_2} n_{\kk_{i\beta}} n_{\kk_{jl}} + \frac{1}{\delta_2\delta_3} n_{\kk_{jl}} (1+n_{\kk_{i\beta}}) 
+\frac{1}{\delta_1\delta_3} n_{\kk_{jl}} n_{\kk_{i\beta}}\right]
\end{multline}
After grouping the terms and then using (\ref{eq:mag}) to replace the factor $(\delta_2+\delta_3)$ with $-\delta_1$, we get 
\begin{multline}
\mathcal{A}_{\mathrm{i}\to\textrm{f}}^{(ef)} = -\sum_\alpha\sum_i \frac{A_1A_2A_3}{\delta_1\delta_2\delta_3} \left\{(\delta_2+\delta_3)\left[(1+n_{\kk_{jl}})(1+n_{\kk_{i\beta}})+n_{\kk_{i\beta}} n_{\kk_{jl}}\right]\right. \\
\left. +\delta_1\left[n_{\kk_{i\beta}}(1+n_{\kk_{jl}})+n_{\kk_{jl}}(1+n_{\kk_{i\beta}})\right]\right\}=
\sum_\alpha\sum_i \frac{A_1A_2A_3}{\delta_2\delta_3}
\end{multline}
We thus find exactly the result in vacuum (\ref{eq:ampef}).

In conclusion, on the energy shell, the five-phonon collision amplitude $2\phi\to 3\phi$ is, at the dominant order in the cubic interaction $\hat{V}$, unaffected by the presence of a thermal bath, more precisely by the fact that the internal phonon modes have non-zero occupation numbers, whether thermal or not. The approximation of “collision in vacuum” is exact at this order.

\section{Collision amplitude $2\phi\to 3\phi$ at low temperature}\label{ann:lim}

Here we establish the dominant behavior (\ref{eq:resann}) of the scattering amplitude (\ref{eq:fmix}) on the energy shell in the limit of low wave numbers and small angles $O(T)$ for the incoming $\qq_1,\qq_2$ and outgoing $\kk_1,\kk_2,\kk_3$ vectors. We perform some manipulations on the expressions (\ref{eq:ampa})--(\ref{eq:ampf}) for the various contributions in Section \ref{ssec:qta}, before explicitly taking the limit in Section \ref{ssec:plpa}.

\subsection{Clever transformations}\label{ssec:qta}

\paragraph{Combining contributions (e) and (f), then (c)}
 It is quite natural to combine expressions (\ref{eq:ampe}) and (\ref{eq:ampf}) because (i) they have the same summation domain, (ii) they have the same numerator, (iii) they have the common factor $\Delta E_1^{(e)}(\lambda)=\Delta E_1^{(f)}(\lambda)=\veps_{\qq_\alpha}-\veps_{\kk_j+\kk_l}-\veps_{\kk_i-\qq_\beta}$ in the denominator (the quantities $\Delta E_1(\lambda)$ and $\Delta E_2(\mu)$ are those from equation (\ref{eq:fmix}))
 . All that remains is to add 
\begin{equation}
\frac{1}{\Delta E_2^{(e)}(\mu)}+\frac{1}{\Delta E_2^{(f)}(\mu)}=\frac{1}{\veps_{\kk_j}+\veps_{\kk_l}-\veps_{\kk_j+\kk_l}}+\frac{1}{\veps_{\kk_i}-\veps_{\qq_\beta}-\veps_{\kk_i-\qq_\beta}}\stackrel{E_\ii=E_\textrm{f}}{=}\frac{\Delta E_1^{(e,f)}(\lambda)}{\Delta E_2^{(e)}(\mu) \Delta E_2^{(f)}(\mu)}
\end{equation}
The factors $\Delta E_1^{(e,f)}(\lambda)$ simplify, and we arrive at the nice result 
\begin{equation}
\label{eq:ampef}
\mathcal{A}_{\mathrm{i}\to\textrm{f}}^{(e)}+\mathcal{A}_{\mathrm{i}\to\textrm{f}}^{(f)}=\sum_\alpha\sum_i \frac{\langle\kk_i|\hat{V}|\qq_\beta,\kk_i-\qq_\beta\rangle \langle\kk_j,\kk_l|\hat{V}|\kk_j+\kk_l\rangle \langle\kk_i-\qq_\beta,\kk_j+\kk_l|\hat{V}|\qq_\alpha\rangle}{(\veps_{\kk_j}+\veps_{\kk_l}-\veps_{\kk_j+\kk_l})(\veps_{\kk_i}-\veps_{\qq_\beta}-\veps_{\kk_i-\qq_\beta})}
\end{equation}
We then notice that if, in the range $\mathcal{A}_{\mathrm{i}\to\textrm{f}}^{(c)}$ written as in (\ref{eq:ampc}), we perform the transposition $\alpha\leftrightarrow\beta$ and change the order of the factors in the numerator, we obtain an expression that closely resembles (\ref{eq:ampef}), 
\begin{equation}
\label{eq:ampcbis}
\mathcal{A}_{\mathrm{i}\to\textrm{f}}^{(c)}=\sum_\alpha\sum_i \frac{\langle \kk_i,\qq_\beta-\kk_i|\hat{V}|\qq_\beta\rangle\langle \kk_j,\kk_l|\hat{V}|\kk_j+\kk_l\rangle\langle\kk_j+\kk_l|\hat{V}|\qq_\beta-\kk_i,\qq_\alpha\rangle}{(\veps_{\kk_j}+\veps_{\kk_l}-\veps_{\kk_j+\kk_l})(\veps_{\qq_\beta}-\veps_{\kk_i}-\veps_{\qq_\beta-\kk_i})}
\end{equation}
which will allow us later to combine contributions (c), (e), and (f) in the small-angle limit.

\paragraph{Transformation of contribution (d), then (b)}
 In expression (\ref{eq:ampd}) for $\mathcal{A}_{\mathrm{i}\to\textrm{f}}^{(d)}$, we perform the transposition $(\alpha,i) \leftrightarrow (\beta,j)$ on the dummy variables, which changes neither the index $l$ nor the summation domain; we obtain 
\begin{equation}
\label{eq:ampdbis}
\mathcal{A}_{\mathrm{i}\to\textrm{f}}^{(d)}=\sum_\alpha\sum_{i\neq j}\frac{\langle\kk_l|\hat{V}|\qq_\beta-\kk_j,\qq_\alpha-\kk_i\rangle \langle\kk_i,\qq_\alpha-\kk_i|\hat{V}|\qq_\alpha\rangle\langle\kk_j,\qq_\beta-\kk_j|\hat{V}|\qq_\beta\rangle}{(\veps_{\kk_l}-\veps_{\qq_\beta-\kk_j}-\veps_{\qq_\alpha-\kk_i})(\veps_{\qq_\beta}-\veps_{\kk_j}-\veps_{\qq_\beta-\kk_j})}
\end{equation}
 The numerator and the factor $\Delta E_2^{(d)}(\mu)$ in the denominator are also unchanged. Next, let us replace $\mathcal{A}_{\mathrm{i}\to\textrm{f}}^{(d)}$ with the half-sum of its old form (\ref{eq:ampd}) and its new form (\ref{eq:ampdbis}); we are left with the task of adding 
\begin{multline}
\frac{1}{2}\left(\frac{1}{\Delta E_1^{(d)}(\lambda)}+\frac{1}{\Delta E_1^{(d)}(\lambda)|_{(\alpha,i) \leftrightarrow (\beta,j)}}\right)
=\frac{1}{2}\left(\frac{1}{\veps_{\qq_\alpha}-\veps_{\kk_i}-\veps_{\qq_\alpha-\kk_i}}+\frac{1}{\veps_{\qq_\beta}-\veps_{\kk_j}-\veps_{\qq_\beta-\kk_j}}\right)\\
\stackrel{E_\ii=E_\textrm{f}}{=} \frac{\frac{1}{2}\Delta E_2^{(d)}(\mu)}{(\veps_{\qq_\alpha}-\veps_{\kk_i}-\veps_{\qq_\alpha-\kk_i}) (\veps_{\qq_\beta}-\veps_{\kk_j}-\veps_{\qq_\beta-\kk_j})}
\end{multline}
 The factors $\Delta E_2^{(d)}(\mu)$ simplify, and we obtain 
\begin{equation}
\label{eq:ampdter}
\mathcal{A}_{\mathrm{i}\to\textrm{f}}^{(d)}=\frac{1}{2}\sum_\alpha\sum_{i\neq j}\frac{\langle\kk_l|\hat{V}|\qq_\alpha-\kk_i,\qq_\beta-\kk_j\rangle \langle\kk_i,\qq_\alpha-\kk_i|\hat{V}|\qq_\alpha\rangle\langle\kk_j,\qq_\beta-\kk_j|\hat{V}|\qq_\beta\rangle}{(\veps_{\qq_\alpha}-\veps_{\kk_i}-\veps_{\qq_\alpha-\kk_i}) (\veps_{\qq_\beta}-\veps_{\kk_j}-\veps_{\qq_\beta-\kk_j})}
\end{equation}
 Let us perform a similar renumbering in the amplitude (\ref{eq:ampb}) of diagram (b): first, we swap the silent variables $l$ and $j$ (which does not change $i$ or the summation domain), then we perform the transposition $(\alpha,i)\leftrightarrow(\beta,j)$ (which does not change $l$).\footnote{The whole process amounts to performing a direct circular permutation on the indices $(\alpha,\beta)$ and $(i,j,l)$. } We arrive at 
\begin{equation}
\label{eq:ampbbis}
\mathcal{A}_{\mathrm{i}\to\textrm{f}}^{(b)}=\sum_\alpha\sum_{i\neq j} \frac{\langle\kk_l,\kk_i-\qq_\alpha|\hat{V}|\qq_\beta-\kk_j\rangle\langle\kk_i|\hat{V}|\qq_\alpha,\kk_i-\qq_\alpha\rangle\langle\kk_j,\qq_\beta-\kk_j|\hat{V}|\qq_\beta\rangle}{(\veps_{\kk_i}-\veps_{\qq_\alpha}-\veps_{\kk_i-\qq_\alpha})(\veps_{\qq_\beta}-\veps_{\kk_j}-\veps_{\qq_\beta-\kk_j})
}
\end{equation}
 after changing the order of the factors in the numerator. This resembles the new form (\ref{eq:ampdter}) of $\mathcal{A}_{\mathrm{i}\to\textrm{f}}^{(d)}$ up to a factor $1/2$, and will allow us to combine contributions (b) and (d) in the small-angle limit.

\subsection{Taking the small-angle limit at low temperature}\label{ssec:plpa}

From now on, the wave numbers of the incoming and outgoing phonons, as well as all angles between them, tend linearly toward zero with temperature, as in equation (\ref{eq:ech}).

In this limit, we must perform a calculation to obtain an equivalent of the various contributions to the transition amplitude $\mathcal{A}_{\mathrm{i}\to\textrm{f}}$ in the case where the angles of the internal phonons with the incoming or outgoing phonons also tend toward zero rather than toward $\pi$--it is indeed under this condition that all the energy denominators $\Delta E_1(\lambda)$ and $\Delta E_2(\mu)$ are in $O(T^3)$ rather than in $O(T)$, and that we fully benefit from the small-denominator effect. We keep track of the constraints that may result on the wave numbers $q_\alpha$ and $k_i$ using the Heaviside function $Y$. For example, in diagram (a) of Figure \ref{fig:diag}, there are no constraints; the internal vectors $\qq_1+\qq_2$ and $\kk_j+\kk_l$ automatically become collinear with and in the same direction as the incoming and outgoing vectors when all angles (\ref{eq:angles}) tend toward zero. In contrast, in diagram (b), this is only the case if $q_\alpha>k_i$ and $k_l>q_\beta$; therefore, a factor $Y(q_\alpha-k_i) Y(k_l-q_\beta)$ must be included in the numerator in the form (\ref{eq:ampb}), which becomes $Y(q_\beta-k_j) Y(k_i-q_\alpha)$ in the form (\ref{eq:ampbbis}) that we will use in practice.

In the numerator of the amplitudes, we use the limit behavior at zero angle (\ref{eq:h3m}) of the matrix elements rewritten as follows, without forgetting the factor $2$ arising from bosonic exchange symmetry, \footnote{The terms of $\hat{H}_3^{(-)}$ contributing to (\ref{eq:elem}) are indeed of indices $(\qq,\qq')=(\kk,\kk')$ and $(\qq,\qq')=(\kk',\kk)$.} 
\begin{equation}
\label{eq:elem}
\langle \kk+\kk'|\hat{V}|\kk,\kk'\rangle \underset{T\to 0^+}{\sim} \frac{3 mc^2(1+\Lambda)}{2^{3/2}(\rho\mathcal{V})^{1/2}}\epsilon^{3/2}[\bar{k}\kb'(\kb+\kb')]^{1/2}
\end{equation}
 In the denominator of the amplitudes, we use relation (14) from reference \cite{insuffisance}, written in the present concave case in the form 
\begin{equation}
\label{eq:denomt3}
\veps_\qq+\veps_{\qq'}-\veps_{\qq+\qq'} \underset{T\to 0^+}{\sim} \frac{1}{2} k_{\rm B}T |\gamma|\epsilon^2 \bar{q} \qb' (\qb+\qb') \left[\frac{\bar{\theta}^2}{(\qb+\qb')^2}+\frac{3}{4}\right]
\end{equation}
 with the angle $\theta=\widehat{(\qq,\qq')}=O(T)$ and $\thb$ in its reduced form. It immediately follows that 
\begin{eqnarray}
\veps_{\qq_1}+\veps_{\qq_2}-\veps_{\qq_1+\qq_2} &\underset{T\to 0^+}{\sim}& \frac{1}{2} k_{\rm B}T |\gamma|\epsilon^2 \qb_1 \qb_2 (\qb_1+\qb_2) \left[\frac{\bar{\theta}^2_{12}}{(\qb_1+\qb_2)^2}+\frac{3}{4}\right] \\
\veps_{\kk_j}+\veps_{\kk_l}-\veps_{\kk_j+\kk_l} &\underset{T\to 0^+}{\sim}& \frac{1}{2} k_{\rm B}T |\gamma|\epsilon^2 \kb_j \kb_l (\kb_j+\kb_l) \left[\frac{\bar{\chi}^2_{jl}}{(\kb_j+\kb_l)^2}+\frac{3}{4}\right]
\end{eqnarray}
 holds without any conditions on the wave numbers. Under the condition $\kb_l>\qb_\beta$ ensuring that the vectors $\qq_\beta$ and $\kk_l-\qq_\beta$ are nearly collinear and in the same direction, it also follows that 
\begin{equation}
\veps_{\qq_\beta}+\veps_{\kk_l-\qq_\beta}-\veps_{\kk_l}\stackrel{\kb_l>\qb_\beta}{\underset{T\to 0^+}{\sim}} \frac{1}{2} k_{\rm B}T |\gamma|\epsilon^2 (\kb_l-\qb_\beta) \qb_\beta \kb_l \left(\frac{\bar{\theta}^2}{\kb_l^2}+\frac{3}{4}\right)
\end{equation}
 holds with the angle $\theta=\widehat{(\qq_\beta,\kk_l-\qq_\beta)}=O(T)$. Using the norm of a vector product, \\$\sin\theta ={|(\kk_l-\qq_\beta)\wedge \qq_\beta|}/{|\kk_l-\qq_\beta|\, q_\beta}={k_l \sin\psi_{\beta l}}/{|\kk_l-\qq_\beta|}$, we find that $\theta\sim k_l\psi_{\beta l}/(k_l-q_\beta)$ and therefore that 
\begin{equation}
\veps_{\qq_\beta}+\veps_{\kk_l-\qq_\beta}-\veps_{\kk_l} \stackrel{\kb_l>\qb_\beta}{\underset{T\to 0^+}{\sim}} \frac{1}{2} k_{\rm B}T |\gamma|\epsilon^2 (\kb_l-\qb_\beta) \qb_\beta \kb_l \left[\frac{\bar{\psi}^2_{\beta l}}{(\kb_l-\qb_\beta)^2}+\frac{3}{4}\right]
\end{equation}
Similarly, by formally swapping $\qq$ and $\kk$, we have 
\begin{equation}
\veps_{\kk_i}+\veps_{\qq_\alpha-\kk_i}-\veps_{\qq_\alpha} \stackrel{\qb_\alpha>\kb_i}{\underset{T\to 0^+}{\sim}} \frac{1}{2} k_{\rm B}T |\gamma|\epsilon^2 (\qb_\alpha-\kb_i) \qb_\alpha \kb_i \left[\frac{\bar{\psi}^2_{\alpha i}}{(\qb_\alpha-\kb_i)^2}+\frac{3}{4}\right]
\end{equation}

We apply the previous expressions to the potentially optimized forms (\ref{eq:ampa}), (\ref{eq:ampef}), (\ref{eq:ampcbis}), (\ref{eq:ampdter}), and (\ref{eq:ampbbis}) of the transition amplitudes. We observe that we can factor out a quantity 
\begin{equation}
\mathcal{A}_0=\left[\frac{3 mc^2(1+\Lambda)\epsilon^{3/2}}{2^{3/2}(\rho\mathcal{V})^{1/2}}\right]^3\frac{1}{(\frac{1}{2}k_{\rm B}T|\gamma|\epsilon^2)^2(\bar{q}_1\bar{q}_2\bar{k}_1\bar{k}_2\bar{k}_3)^{1/2}}
\end{equation}
which can be trivially rewritten as in the stated expression (\ref{eq:resann}). We then obtain \footnote{In particular, we note that the product of all the internal wave numbers appears systematically in the numerator and denominator, and simplifies in the ratio.} 
\begin{eqnarray}
\label{eq:equivampa}
\mathcal{A}_{\mathrm{i}\to\textrm{f}}^{(a)} &\underset{T\to 0^+}{\sim}& \mathcal{A}_0\sum_i \frac{\kb_i}{\left[\frac{\bar{\theta}^2_{12}}{(\qb_1+\qb_2)^2}+\frac{3}{4}\right]\left[\frac{\bar{\chi}^2_{jl}}{(\kb_j+\kb_l)^2}+\frac{3}{4}\right]} \\
\label{eq:equivampef}
\mathcal{A}_{\mathrm{i}\to\textrm{f}}^{(e)}+\mathcal{A}_{\mathrm{i}\to\textrm{f}}^{(f)} &\underset{T\to 0^+}{\sim}& -\mathcal{A}_0 \sum_\alpha\sum_i \frac{\qb_\alpha\,Y(\kb_i-\qb_\beta)}{\left[\frac{\bar{\chi}^2_{jl}}{(\kb_j+\kb_l)^2}+\frac{3}{4}\right]\left[\frac{\bar{\psi}^2_{\beta i}}{(\kb_i-\qb_\beta)^2}+\frac{3}{4}\right]} \\
\label{eq:equivampc}
\mathcal{A}_{\mathrm{i}\to\textrm{f}}^{(c)} &\underset{T\to 0^+}{\sim}& -\mathcal{A}_0 \sum_\alpha\sum_i \frac{\qb_\alpha\,Y(\qb_\beta-\kb_i)}{\left[\frac{\bar{\chi}^2_{jl}}{(\kb_j+\kb_l)^2}+\frac{3}{4}\right]\left[\frac{\bar{\psi}^2_{\beta i}}{(\qb_\beta-\kb_i)^2}+\frac{3}{4}\right]}\\
\label{eq:equivampd}
\mathcal{A}_{\mathrm{i}\to\textrm{f}}^{(d)} &\underset{T\to 0^+}{\sim}& \frac{1}{2} \mathcal{A}_0\sum_\alpha \sum_{i\neq j} \frac{\kb_l\,Y(\qb_\alpha-\kb_i)\,Y(\qb_\beta-\kb_j)}{\left[\frac{\bar{\psi}^2_{\alpha i}}{(\qb_\alpha-\kb_i)^2}+\frac{3}{4}\right] \left[\frac{\bar{\psi}^2_{\beta j}}{(\qb_\beta-\kb_j)^2}+\frac{3}{4}\right]}\\
\label{eq:equivampb}
\mathcal{A}_{\mathrm{i}\to\textrm{f}}^{(b)} &\underset{T\to 0^+}{\sim}& \mathcal{A}_0\sum_\alpha \sum_{i\neq j}\frac{\kb_l\,Y(\kb_i-\qb_\alpha)\,Y(\qb_\beta-\kb_j)}{\left[\frac{\bar{\psi}^2_{\alpha i}}{(\qb_\alpha-\kb_i)^2}+\frac{3}{4}\right] \left[\frac{\bar{\psi}^2_{\beta j}}{(\qb_\beta-\kb_j)^2}+\frac{3}{4}\right]} 
\end{eqnarray}
The amplitude $\mathcal{A}_{\mathrm{i}\to\textrm{f}}^{(a)}$ in (\ref{eq:equivampa}) directly yields the first sum of equation (\ref{eq:Abar}). In the sum $(\mathcal{A}_{\mathrm{i}\to\textrm{f}}^{(e)}+\mathcal{A}_{\mathrm{i}\to\textrm{f}}^{(f)})+\mathcal{A}_{\mathrm{i}\to\textrm{f}}^{(c)}$ of (\ref{eq:equivampef}) and (\ref{eq:equivampc}), the Heaviside functions add up to one, $Y(\kb_i-\qb_\beta)+Y(\qb_\beta-\kb_i)=1$, and we find the second sum of equation (\ref{eq:Abar}) after transposition $\alpha\leftrightarrow\beta$. Finally, before combining it with $\mathcal{A}_{\mathrm{i}\to\textrm{f}}^{(d)}$, we need to work a bit on the amplitude $\mathcal{A}_{\mathrm{i}\to\textrm{f}}^{(b)}$ in (\ref{eq:equivampb}): by performing the exchange $(\alpha,i)\leftrightarrow(\beta,j)$ in the right-hand side--which changes neither the index $l$ nor the denominator--and by taking the half-sum of the right-hand side and its new expression thus obtained, we arrive at 
\begin{equation}
\label{eq:equivampbbis}
\mathcal{A}_{\mathrm{i}\to\textrm{f}}^{(b)} \underset{T\to 0^+}{\sim} \frac{1}{2}\mathcal{A}_0\sum_\alpha \sum_{i\neq j}\frac{\kb_l\left[Y(\kb_i-\qb_\alpha)\,Y(\qb_\beta-\kb_j) + Y(\qb_\alpha-\kb_i) Y(\kb_j-\qb_\beta)\right]}{\left[\frac{\bar{\psi}^2_{\alpha i}}{(\qb_\alpha-\kb_i)^2}+\frac{3}{4}\right]
\left[\frac{\bar{\psi}^2_{\beta j}}{(\qb_\beta-\kb_j)^2}+\frac{3}{4}\right]}
\end{equation}
Now, at the zero order in $\epsilon$, we have $\qb_1+\qb_2=\kb_1+\kb_2+\kb_3$ by conservation of energy or momentum, so that the inequality $\kb_i>\qb_\alpha=\kb_i+\kb_j+\kb_l-\qb_\beta$ implies that $\qb_\beta>\kb_j+\kb_l\geq \kb_j$ and we can write $Y(\kb_i-\qb_\alpha)\,Y(\qb_\beta-\kb_j)=Y(\kb_i-\qb_\alpha)$ in (\ref{eq:equivampbbis}). It follows that, in the grouping $\mathcal{A}_{\mathrm{i}\to\textrm{f}}^{(b)}+\mathcal{A}_{\mathrm{i}\to\textrm{f}}^{(d)}$ of (\ref{eq:equivampbbis}) and (\ref{eq:equivampd}), Heaviside functions appear that complement each other in cascade to one, $Y(\kb_i-\qb_\alpha)+Y(\qb_\alpha-\kb_i) Y(\kb_j-\qb_\beta)+Y(\qb_\alpha-\kb_i)\,Y(\qb_\beta-\kb_j)= Y(\kb_i-\qb_\alpha)+Y(\qb_\alpha-\kb_i)=1$. In the new summand obtained, which is invariant under exchange $(\alpha,i)\leftrightarrow(\beta,j)$, the parts $\sum_{i<j}$ and $\sum_{j<i}$ yield equal contributions after summation over $\alpha$: they are doubled, causing the overall factor $1/2$ to disappear, and we recover the third and final sum of equation (\ref{eq:Abar}). This establishes the announced result (\ref{eq:resann}), which holds unconditionally for the internal wave vectors.

\section{Simplifications of the integral expressions of $C_\phi$ and $I_{\rm ang}$}\label{ann:simp}

\subsection{Taking advantage of scale invariance}

We note that the angular integral $I_{\rm ang}$ defined by equation (\ref{eq:Iang}) is a positively homogeneous function of degree eight of its five variables, in the sense that 
\begin{equation}
\label{eq:sca}
I_{\rm ang}(\lambda\qb_1,\lambda\qb_2,\lambda\kb_1,\lambda\kb_2,\lambda\kb_3)=\lambda^8 I_{\rm ang}(\qb_1,\qb_2,\kb_1,\kb_2,\kb_3) \quad\forall\lambda>0
\end{equation}
 To see this, let us multiply, as in the left-hand side of (\ref{eq:sca}), the five numbers of $\qb_\alpha$ and $\kb_i$ by the scale factor $\lambda>0$ and perform the variable changes $\bar{\theta}_{12}\to \lambda \bar{\theta}_{12}$, $Z_1\to \lambda Z_1$, and $Z_2\to \lambda Z_2$ in (\ref{eq:Iang}),
 which already yields a factor of $\lambda^4/\lambda=\lambda^3$ in the prefactor of (\ref{eq:Iang}) and a factor of $(\lambda^2)^3$ through the integration elements $\mathrm{d}\bar{\theta}_{12}\,\bar{\theta}_{12} \dd^2Z_1\dd^2Z_2$. Furthermore, under the effect of these transformations, all reduced angles (\ref{eq:tangred}) are multiplied by $\lambda$, the reduced amplitude $\bar{\mathcal{A}}_{\mathrm{i}\to\textrm{f}}$ is multiplied by $\lambda$ in (\ref{eq:Abar}), so its square by $\lambda^2$--the denominators remain unchanged, the numerators are multiplied by $\lambda$--, the quantities $u$ and $v$ are multiplied by $\lambda^3$ in (\ref{eq:defu}, \ref{eq:defv}), so that $\delta(u-v)\to \lambda^{-3}\delta(u-v)$. As with $\lambda^3(\lambda^2)^3\lambda^2/\lambda^3=\lambda^8$, we arrive at the result (\ref{eq:sca}) announced.

Let’s exploit the scale invariance (\ref{eq:sca}) of $I_{\rm ang}$ in the definition (\ref{eq:cphi}) of the coefficient $C_\phi(z_\phi)$, by adding a factor $\delta(\qb_1+\qb_2-\lambda)$ to the integrand and an outer integral (on the far left) $\int_0^{+\infty}\mathrm{d}\lambda$, according to a well-known trick, 
\begin{multline}
C_\phi(z_\phi)=\int_0^{+\infty}\mathrm{d}\lambda \int_0^{+\infty} \prod_{\alpha=1}^{2} \mathrm{d}\qb_\alpha \prod_{i=1}^{3}\mathrm{d}\kb_i\, \delta(\qb_1+\qb_2-\lambda)\delta(\qb_1+\qb_2-\kb_1-\kb_2-\kb_3) I_{\rm ang}(\qb_1,\qb_2,\kb_1,\kb_2,\kb_3)  \\
\times\eee^{-(\qb_1+\qb_2)} \prod_{\alpha=1}^{2} (1+\bar{n}_{q_\alpha}^{\rm lin}) \prod_{i=1}^{3}
(1+\bar{n}_{k_i}^{\rm lin})
\end{multline}
then performing the change of variables $\qb_\alpha\to\lambda\qb_\alpha$ and $\kb_i\to\lambda\kb_i$ on the five reduced wave numbers. The integration elements yield a factor $\lambda^5$, each Dirac delta yields a factor $1/\lambda$, and $I_{\rm ang}$ yields a factor $\lambda^8$. This leaves \footnote{We have taken the liberty of replacing $\qb_1+\qb_2$ with $1$ in the argument of the second Dirac delta and under the exponential as a prefactor of $S$, as permitted by the first Dirac delta, which also allows us to restrict the integration domain over the wave numbers to $[0,1]^5$.} 
\begin{multline}
\label{eq:cphisca}
C_\phi(z_\phi)=\int_0^1 \mathrm{d}\qb_1\mathrm{d}\qb_2\mathrm{d}\kb_1\mathrm{d}\kb_2\mathrm{d}\kb_3\, \delta(\qb_1+\qb_2-1) \delta(\kb_1+\kb_2+\kb_3-1) I_{\rm ang}(\qb_1,\qb_2,\kb_1,\kb_2,\kb_3)\\
\times \int_0^{+\infty} \mathrm{d}\lambda\,\lambda^{11} \eee^{-\lambda} S(\lambda\qb_1,\lambda\qb_2,\lambda\kb_1,\lambda\kb_2,\lambda\kb_3)
\end{multline}
 with the compact notation for the bosonic statistical part, 
\begin{multline}
\label{eq:defS}
S(\qb_1,\qb_2,\kb_1,\kb_2,\kb_3)\equiv\prod_{\alpha=1}^{2} (1+\bar{n}_{q_\alpha}^{\rm lin}) \prod_{i=1}^{3} (1+\bar{n}_{k_i}^{\rm lin}) \\
=\left[\left(1-z_\phi\eee^{-\qb_1}\right)\left(1-z_\phi\eee^{-\qb_2}\right)\left(1-z_\phi\eee^{-\kb_1}\right)\left(1-z_\phi\eee^{-\kb_2}\right)\left(1-z_\phi\eee^{-\kb_3}\right)\right]^{-1}
\end{multline}
This new form of $C_\phi(z_\phi)$ has two advantages: (i) it allows for the decoupling of the integration of thermal factors, which are the only ones to depend on the fugacity, and (ii) it allows the evaluation of $I_{\rm ang}$ to be restricted to the manifold 
\begin{equation}
\label{eq:defW}
\mathcal{W}=\{(\qb_1,\qb_2,\kb_1,\kb_2,\kb_3)\in[0,1]^5 | \qb_1+\qb_2=\kb_1+\kb_2+\kb_3=1\}
\end{equation}
which leads to simplifications in the expressions of the quantities $\bar{\mathcal{A}}_{\mathrm{i}\to\textrm{f}}$, $u$, and $v$, for example 
\begin{align}
v & \stackrel{\mathcal{W}}{=} \frac{3}{4}\left[(1-\kb_1)(1-\kb_2)(1-\kb_3)-(1-\qb_1)(1-\qb_2)\right] \\
\label{eq:uW}
u &\stackrel{\mathcal{W}}{=} \qb_1\qb_2\thb^2_{12}-\frac{1}{\kb_3} (Z_1^*,Z_2^*) M \binom{Z_1}{Z_2} 
\end{align}
with
\begin{equation}
\label{eq:defM}
M\stackrel{\mathcal{W}}{=}\begin{pmatrix} \kb_1(1-\kb_2) & \kb_1\kb_2\\ \kb_1\kb_2 & \kb_2(1-\kb_1)\end{pmatrix}\quad\mbox{and}\quad \mathrm{det}\, M\stackrel{\mathcal{W}}{=} \kb_1 \kb_2 \kb_3
\end{equation}

\subsection{Taking advantage of exchange symmetry}\label{ssec:symbos}

In the quintuple integral (\ref{eq:cphisca}) over the reduced wave numbers yielding $C_\phi(z_\phi)$, the integrand must be a symmetric function of the first two $\qb_1,\qb_2$ and a symmetric function of the last three $\kb_1,\kb_2,\kb_3$. This naturally reflects the exchange symmetry of the initial state $|\mathrm{i}\rangle$ and that of the final state $|\textrm{f}\rangle$ in the collision $2\phi\to 3\phi$. Only the symmetry of $I_{\rm ang}$ is not obvious: it is not immediately apparent from the definition (\ref{eq:Iang}), due to the special role played by the wave vectors $\kk_1$ and $\kk_2$ in the parameterization (\ref{eq:k3}) from which it derives, but we will explicitly verify that it is satisfied in section \ref{ssec:chvar}. This symmetry allows us to restrict ourselves to the ordered domain $\qb_1<\qb_2$ at the cost of adding a factor $2$ and to $\kb_1<\kb_2<\kb_3$ at the cost of adding a factor $3!=6$, as we will now detail.

 Let us begin with the variables $\qb_\alpha$: by implying the dependence on $\kb_i$, we are faced with the integral 
\begin{equation}
I=\int_0^1\mathrm{d}\qb_1 \int_0^1\mathrm{d}\qb_2\,\delta(\qb_1+\qb_2-1) f(\qb_1,\qb_2)=2 \int_0^1\mathrm{d}\qb_2 \int_0^{\qb_2}\mathrm{d}\qb_1 \delta(\qb_1+\qb_2-1) f(\qb_1,\qb_2)
\end{equation}
 where the function $f$ is symmetric, and we have reduced ourselves, as promised, to the ordered domain in the second equality. But it is possible to simplify the notation further. Let us integrate the Dirac delta over $\qb_1$, which amounts to imposing that $\qb_1=1-\qb_2$, which is clearly positive; the constraint $\qb_1<\qb_2$ then leads to $1-\qb_2<\qb_2$ and thus to $\qb_2>1/2$. Conversely, if $\qb_2>1/2$, we automatically have $\qb_1=1-\qb_2<1/2<\qb_2$. We are thus left with 
\begin{equation}
I=2\int_{1/2}^{1} \mathrm{d}\qb_2\, f(1-\qb_2,\qb_2)
\end{equation}

Let us proceed in the same way for the variables $\kb_i$: by implying the dependence on $\qb_\alpha$, we are faced with the integral 
\begin{multline}
J=\int_0^1\mathrm{d}\kb_1 \int_0^1\mathrm{d}\kb_2 \int_0^1\mathrm{d}\kb_3\,\delta(\kb_1+\kb_2+\kb_3-1) g(\kb_1,\kb_2,\kb_3) \\
=6 \int_0^1\mathrm{d}\kb_3 \int_0^{\kb_3}\mathrm{d}\kb_2 \int_0^{\kb_2}\mathrm{d}\kb_1\,\delta(\kb_1+\kb_2+\kb_3-1) g(\kb_1,\kb_2,\kb_3)
\end{multline}
 with a symmetric function $g$ and a restriction to the ordered domain. To simplify further, let us integrate the Dirac delta over $\kb_1$ as before, which amounts to imposing $\kb_1=1-\kb_2-\kb_3$. Since $\kb_1<\kb_3$ and $\kb_2<\kb_3$, we deduce that $1<3\kb_3$ and thus $\kb_3>1/3$. Since $\kb_1<\kb_2$, we conclude that $\kb_2>(1-\kb_3)/2$; this is compatible with condition $\kb_2<\kb_3$ since we have $\kb_3>1/3$. Conversely, the inequality $2\kb_2>1-\kb_3$ can also be written as $2(1-\kb_1-\kb_3)>1-\kb_3$ or even $\kb_1<(1-\kb_3)/2$, and the inequality $\kb_1<\kb_2$ is automatically satisfied. Finally, we must remember to verify that $\kb_1=1-\kb_2-\kb_3$ remains positive, so that $\kb_2<1-\kb_3$ holds, which is a condition independent of the inequality $\kb_2<\kb_3$ and more restrictive than it as soon as $\kb_3>1/2$ holds. Finally, we are left with 
\begin{equation}
J=6\int_{1/3}^{1}\mathrm{d}\kb_3 \int_{(1-\kb_3)/2}^{\min(1-\kb_3,\kb_3)}\mathrm{d}\kb_2\, g(1-\kb_2-\kb_3,\kb_2,\kb_3) 
\end{equation}

By carrying out these simplifications in parallel in (\ref{eq:cphisca}), we arrive at the maximally reduced expression we sought 
\begin{multline}
\label{eq:cphifin}
C_\phi(z_\phi)=12
\int_{1/3}^{1}\mathrm{d}\kb_3 \int_{(1-\kb_3)/2}^{\min(1-\kb_3,\kb_3)}\mathrm{d}\kb_2\,\int_{1/2}^{1} \mathrm{d}\qb_2 I_{\rm ang}(1-\qb_2,\qb_2,1-\kb_2-\kb_3,\kb_2,\kb_3)\\
\times\int_0^{+\infty} \mathrm{d}\lambda\, \lambda^{11} \eee^{-\lambda} S(\lambda(1-\qb_2),\lambda\qb_2,\lambda(1-\kb_2-\kb_3),\lambda\kb_2,\lambda\kb_3)
\end{multline}

\subsection{Simplifying changes of angular variables}\label{ssec:chvar}

In this section \ref{ssec:chvar}, we focus immediately on the manifold $\mathcal{W}$ (\ref{eq:defW}). The guiding idea is to further simplify the expression of the quantity $u$ (\ref{eq:uW}) as an argument of the Dirac delta in $I_{\rm ang}$ (\ref{eq:Iang}). To this end, we perform a change of variables on the complex quantities $Z_1$ and $Z_2$: we reduce the Hermitian form defining $u$ as much as possible by placing ourselves in the orthonormal eigenbasis 
\begin{equation}
\mathbf{e}_+=\binom{a}{b} \quad\quad  \mathbf{e}_-=\binom{-b}{\phantom{+}a}
\end{equation}
 of the matrix $M$ and by introducing into the amplitudes in the denominator the square root of the corresponding eigenvalues $m_+>m_->0$ and $M$ in the denominator of the amplitudes, and the square root of $\kb_3$, 
\begin{equation}
\binom{Z_1}{Z_2}= \left(\frac{\kb_3}{m_+}\right)^{1/2} \zeta_+ \mathbf{e}_+ + \left(\frac{\kb_3}{m_-}\right)^{1/2} \zeta_- \mathbf{e}_- \quad\mbox{that is} \quad
\left\{
\begin{array}{lcl}
\displaystyle Z_1 &=& \displaystyle\frac{a\kb_3^{1/2}}{m_+^{1/2}} \zeta_+ -\frac{b\kb_3^{1/2}}{m_-^{1/2}}\zeta_-\\
&&\\
\displaystyle Z_2  &=& \displaystyle\frac{b\kb_3^{1/2}}{m_+^{1/2}} \zeta_++\frac{a\kb_3^{1/2}}{m_-^{1/2}}\zeta_-
\end{array}
\right.
\label{eq:cgz12}
\end{equation}
 in the numerator, where $\zeta_+$ and $\zeta_-$ describe $\mathbb{C}$. Starting from the expression (\ref{eq:defM}) of $M$ on the manifold, we find that 
\begin{equation}
\label{eq:mab}
m_\pm=s-p\pm |\delta+\mathrm{i} p| \quad ; \quad a=\frac{1}{\sqrt{2}}\left(1+\frac{\delta}{|\delta+\mathrm{i} p|}\right)^{1/2} \quad ;\quad b=\frac{1}{\sqrt{2}}\left(1-\frac{\delta}{|\delta+\mathrm{i} p|}\right)^{1/2}
\end{equation}
 with the notation shortcuts 
\begin{equation}
s=\frac{1}{2}(\kb_1+\kb_2) \quad ; \quad \delta=\frac{1}{2}(\kb_1-\kb_2) \quad ; \quad p=\kb_1\kb_2
\end{equation}
 Furthermore, the Dirac delta allows integration over the reduced positive angle $\thb_{12}$, by virtue of the identity 
\begin{equation}
\delta(u-v)=\delta\left(\qb_1\qb_2\thb_{12}^2-|\zeta_+|^2-|\zeta_-|^2-v\right)=\frac{\delta\left(\thb_{12}-\thb_{12}^{(0)}\right)}{2\qb_1\qb_2\thb_{12}^{(0)}} \quad \mbox{where}\quad \thb_{12}^{(0)}=\frac{\left(|\zeta_+|^2+|\zeta_-|^2+v\right)^{1/2}}{(\qb_1\qb_2)^{1/2}}
\end{equation}
 assuming that this value of $\thb_{12}$, which cancels the argument of the Dirac delta, is positive real. Given the Jacobian of the change of variables (\ref{eq:cgz12}), 
\begin{equation}
\frac{D(\re Z_1,\im Z_1,\re Z_2,\im Z_2)}{D(\re\zeta_+,\im\zeta_+,\re\zeta_-,\im\zeta_-)}=\left[\frac{(a^2+b^2)\kb_3}{(m_+ m_-)^{1/2}}\right]^2=\frac{\kb_3}{\kb_1\kb_2}
\end{equation}
 it remains in (\ref{eq:Iang}) (with $Y$ the Heaviside function) 
\begin{equation}
\label{eq:Iangjol}
\boxed{
I_{\rm ang}(\qb_1,\qb_2,\kb_1,\kb_2,\kb_3)\stackrel{\mathcal{W}}{=}\frac{1}{2}\int_{\mathbb{C}}\dd^2\zeta_+\int_{\mathbb{C}}\dd^2\zeta_- |\bar{\mathcal{A}}_{\mathrm{i}\to\textrm{f}}|^2_{\thb_{12}=\thb_{12}^{(0)}} Y(|\zeta_+|^2+|\zeta_-|^2+v)}
\end{equation}
 In the numerics, we will switch to phase-modulus representation $\zeta_\pm=\rho_\pm\exp(\mathrm{i}\phi_\pm)$, then to polar coordinates to integrate over $(\rho_+,\rho_-)$, setting $\rho_+=R\cos\alpha$ and $\rho_-=R\sin\alpha$ with $R=(|\zeta_+|^2+|\zeta_-|^2)^{1/2}>0$ and $\alpha\in[0,\pi/2]$, in which case \footnote{\label{note:parite} The integrand is invariant under the change from $(\zeta_+,\zeta_-)$ to $(\zeta_+^*,\zeta_-^*)$, and thus under the change from $(\phi_+,\phi_-)$ to $(2\pi-\phi_+,2\pi-\phi_-)$. We can then simply numerically integrate over $\phi_+\in[0,\pi]$ after removing the factor $1/2$ in front of the integral. This reduces the time for calculation by a factor of two.} 
\begin{equation}
\label{eq:fnum}
I_{\rm ang}(\qb_1,\qb_2,\kb_1,\kb_2,\kb_3)\stackrel{\mathcal{W}}{=} \frac{1}{2} \int_0^{2\pi}\mathrm{d}\phi_+ \int_0^{2\pi}\mathrm{d}\phi_- \int_0^{\pi/2} \mathrm{d}\alpha\,\cos\alpha\sin\alpha \int_0^{+\infty} \mathrm{d} R\, R^3 |\bar{\mathcal{A}}_{\mathrm{i}\to\textrm{f}}|^2_{\thb_{12}=\thb_{12}^{(0)}} Y(R^2+v) 
\end{equation}
 with the positive real value of $\thb_{12}$ (if it exists) 
\begin{equation}
\label{eq:thb120num}
\thb_{12}^{(0)}\stackrel{\mathcal{W}}{=}\frac{(R^2+v)^{1/2}}{(\qb_1\qb_2)^{1/2}}
\end{equation}
 Finally, let us verify the exchange symmetry of $I_{\rm ang}$ used in section \ref{ssec:symbos}: 
\begin{itemize}\item[(i)] It is easy to see that $I_{\rm ang}$ is invariant under the transformation $(\qb_1,\qb_2,\zeta_+,\zeta_-) \to (\qb_2,\qb_1,-\zeta_+,-\zeta_-)$ applied to (\ref{eq:Iangjol}) . Indeed, the form of the integral in (\ref{eq:Iangjol}) is clearly preserved, and the quantities $v$ and $\thb_{12}^{(0)}$ are clearly unchanged. Furthermore, since the coefficients $m_\pm$,
 $a$, and $b$ do not change either, the original complex variables are transformed as follows in (\ref{eq:cgz12}), $(Z_1,Z_2)\to (-Z_1,-Z_2)$, and so $Z_3\to -Z_3$ in (\ref{eq:chvar}), which swaps the reduced angles $\bar{\psi}_{1i}$ and $\bar{\psi}_{2i}$ without affecting the reduced angles $\bar{\chi}_{ij}$ in (\ref{eq:tangred} ). We deduce that the reduced amplitude $\bar{\mathcal{A}}_{\mathrm{i}\to\textrm{f}}$ (\ref{eq:Abar}) does not change (after summation over the index $\alpha$, of course), since $\qb_\alpha$ and $\qb_\beta$ are interchanged, as are $\psib_{\alpha i}$ and $\psib_{\beta i}$. The symmetry under exchange of $\qb_1$ and $\qb_2$ is thus confirmed. \item[(ii)] It is straightforward to establish the symmetry under exchange of $\kb_1$ and $\kb_2$. It suffices to verify, in a manner similar to point (i), that $I_{\rm ang}$ is invariant under the transformation $(\kb_1,\kb_2,\zeta_+,\zeta_-) \to (\kb_2,\kb_1,\zeta_+,-\zeta_-)$: the quantities $v,\thb_{12}^{(0)},m_+,m_-$ are invariant, but the coordinates of the eigenvector $\mathbf{e}_+$ are swapped in (\ref{eq:mab}), $a\leftrightarrow b$; it follows that the complex variables are interchanged, $(Z_1,Z_2)\to (Z_2,Z_1)$, as do the angles $\bar{\psi}_{\alpha 1}$ and $\bar{\psi}_{\alpha 2},\, \alpha\in\{1,2\}$ or the angles $\bar{\chi}_{13}$ and $\bar{\chi}_{23}$, while the complex $Z_3$ and the other reduced angles $\bar{\psi}_{13},\bar{\psi}_{23},\bar{\chi}_{12}$ remain unchanged; the amplitude $\bar{\mathcal{A}}_{\mathrm{i}\to\textrm{f}}$ as a whole is preserved.
\item[(iii)] The case of symmetry by swapping $\kb_2$ and $\kb_3$ is more intricate. Upon reflection, we introduce the transformation $(\kb_2,\kb_3,\zeta_+,\zeta_-) \to (\kb_3,\kb_2,\tilde{\zeta}_+,\tilde{\zeta}_-)$ with 
\begin{equation}
\label{eq:defP}
\binom{\tilde{\zeta}_+}{\tilde{\zeta}_-}=P \binom{\zeta_+}{\zeta_-} \quad\mbox{and}\quad P=
\begin{pmatrix}
\frac{\tilde{m}_+^{1/2}}{\kb_2^{1/2}} & 0 \\ 
0 & \frac{\tilde{m}_-^{1/2}}{\kb_2^{1/2}}
\end{pmatrix}
\,
\begin{pmatrix} 
\phantom{+}\tilde{a} & \tilde{b}\\ 
-\tilde{b} & \tilde{a}
\end{pmatrix}
\,
\begin{pmatrix}
1 & 0 \\ 
-\frac{\kb_1}{\kb_3} & -\frac{\kb_2}{\kb_3}
\end{pmatrix} 
\,
\begin{pmatrix}
a & -b \\ b & \phantom{+}a
\end{pmatrix} 
\,
\begin{pmatrix}
\frac{\kb_3^{1/2}}{m_+^{1/2}} & 0 \\
0 & \frac{\kb_3^{1/2}}{m_-^{1/2}}
\end{pmatrix}
\end{equation}
The quantities $\tilde{a},\tilde{b},\tilde{m}_{\pm}$ follow from $a,b,m_\pm$ by swapping $\kb_2$ and $\kb_3$. We verify that $P$ is an orthogonal matrix with determinant $-1$, by laboriously calculating $P ({}^{\rm t}P)$ after showing that 
\begin{equation}
Q ({}^{\rm t}Q)= \begin{pmatrix}\kb_1^{-1} & 0\\0 & \kb_3^{-1}\end{pmatrix} - \begin{pmatrix} 1 & 1 \\ 1 & 1 \end{pmatrix}
\end{equation}
 where the matrix $Q$ is the product of the last three matrices entering equation (\ref{eq:defP}).\footnote{\label{note:interm} Let us give three intermediate results: $m_+a^2+m_-b^2\stackrel{\mathcal{W}}{=}\kb_1(1-\kb_2)$, $m_-a^2+m_+b^2\stackrel{\mathcal{W}}{=}\kb_2(1-\kb_1)$ and $ab(m_+-m_-)\stackrel{\mathcal{W}}{=}\kb_1\kb_2$, obtained by writing the diagonal matrix elements $M_{11}$ and $M_{22}$ and the off-diagonal element $M_{12}$ in two different ways: (i) as in the definition (\ref{eq:defM}) and (ii) via the spectral decomposition $M=m_+|\mathbf{e}_+\rangle\langle\mathbf{e}_+|+m_-|\mathbf{e}_-\rangle\langle\mathbf{e}_-|$ in Dirac notation. Furthermore, the determinant of a product is the product of the determinants, hence $\mathrm{det}\, P=-(a^2+b^2)(\tilde{a}^2+\tilde{b}^2)(\tilde{m}_+\tilde{m}_-/m_+m_-)^{1/2}$. Now, $a^2+b^2=\tilde{a}^2+\tilde{b}^2=1$ by construction and $m_+m_-=\tilde{m}_+\tilde{m}_-\stackrel{\mathcal{W}}{=}\kb_1\kb_2\kb_3$ from (\ref{eq:defM}).} The transformation (\ref{eq:defP}) therefore has a unit Jacobian, and we have $|\tilde{\zeta}_+|^2+|\tilde{\zeta}_-|^2 = |\zeta_+|^2+|\zeta_-|^2$; $\thb_{12}^{(0)}$ remains unchanged and the structure of the integral (\ref{eq:Iangjol}) over $\zeta_+,\zeta_-$ is preserved. This transformation is actually very simple in the original complex variables $Z_i$ (\ref{eq:chvar}), $(Z_1,Z_2)\to (Z_1,Z_3)$, and $Z_3\to Z_2$; in fact, this is exactly how it was constructed;\footnote{The last two matrices in (\ref{eq:defP}) take us from $(\zeta_+,\zeta_-)$ to $(Z_1,Z_2)$ as in (\ref{eq:cgz12}); the middle matrix changes $(Z_1,Z_2)$ to $(Z_1,Z_3)$; the first two matrices return to the zeta variables via the inverse transformation of (\ref{eq:cgz12}) after swapping $\kb_2$ and $\kb_3$.} as a result, in (\ref{eq:tangred}) $\psib_{11}$, $\bar{\psi}_{21}$, and $\bar{\chi}_{23}$ remain unchanged, but $\psib_{\alpha 2}$ and $\psib_{\alpha 3},\,\alpha\in\{1,2\}$ are swapped, just like $\chib_{12}$ and $\chib_{13}$; the reduced amplitude $\bar{\mathcal{A}}_{\mathrm{i}\to\textrm{f}}$ is preserved in (\ref{eq:Abar}), and the exchange symmetry of $I_{\rm ang}$ is confirmed. 
\end{itemize}
 
\section{Some details on the numerical calculation of $C_\phi$}\label{ann:num}

\subsection{Numerical integration over wave numbers and angles $\phi_\pm,\alpha$; final result}

We use the expression (\ref{eq:cphifin}) for $C_\phi(z_\phi)$, and we perform the calculation in parallel for the $n_\phi\simeq 100$ required values of the fugacity $z_\phi$ distributed across $[0,1]$, so as not to have to evaluate $n_\phi$ times the angular integral $I_{\rm ang}$, which does not depend on it and is time-consuming. We introduce a small parameter $\eta_{qk}$ controlling the discretization of the integrals over the reduced wave numbers in the midpoint method. In the integral over $\kb_3$, we take the step size $\mathrm{d}\kb_3\simeq (2/3) \eta_{qk}$ (we wrote $\simeq$ because we need to adjust $\mathrm{d}\kb_3$ slightly to have an integer number of steps within the length $2/3$ of the integration interval, when $\eta_{qk}$ is not the inverse of an integer). In the next integral, over $\kb_2$, we use $\mathrm{d}\kb_2\simeq \eta_{qk}$, a value rounded here as well to obtain a non-zero integer number of steps. In the third integral, over $\qb_2$, we identify $\kb_i$ and $1-\kb_i$ as notable points, $i\in\{1,2,3\}$--at least those within the interval $[1/2,1]$--and the cancellation point of $v$, $\qb_2(v=0)=[1+(1-4\sigma)^{1/2}]/2$, if $\sigma\equiv (1-\kb_1)(1-\kb_2)(1-\kb_3)<1/4$; we then integrate over each subinterval $[\qb_2^{\rm inf},\qb_2^{\rm sup}]$ bounded by these points and by the integration limits $1/2$ and $1$.\footnote{If $\qb_2=\kb_i$ or $\qb_1=1-\qb_2=\kb_i$, we encounter a division by zero in (\ref{eq:Abar}). The case $v=0$ is also special, as suggested by the function $Y(R^2+v)$ in (\ref{eq:fnum}) and as will be confirmed below. If the subinterval has width $\qb_2^{\rm sup}-\qb_2^{\rm inf}<\eta_{qk}^2$, we simply ignore it; otherwise, we take an integration step $\mathrm{d}\qb_2\simeq \eta_{qk}$ rounded to ensure an integer number of steps at least equal to one.} Finally, the integral over the scaling factor is calculated, again using the midpoint method, with a step size $\mathrm{d}\lambda=1/n_\phi\simeq 0.01$ (so that $1-z_\phi$ is always zero or greater than $\mathrm{d}\lambda$) and a cutoff $\lambda_{\rm max}=40$.

We numerically evaluate the angular integral $I_{\rm ang}$ using the form (\ref{eq:fnum}). The integration over the angles $\phi_\pm$ and $\alpha$ is controlled by the integer $n_{\alpha\phi}\gg1$, in the sense that we use the steps $\mathrm{d}\alpha=\pi/2n_{\alpha\phi}$ and $\mathrm{d}\phi_+=\mathrm{d}\phi_-=\mathrm{d}\phi=2\pi/n_{\alpha\phi}$; so we have $\mathrm{d}\alpha\ll\mathrm{d}\phi$. We shift the regular grids in $\phi_+$ and $\phi_-$ by half a step to avoid having $\phi_+=\phi_-[\pi]$, for a reason that will become clear below.\footnote{\label{note:decale} We also recall Note \ref{note:parite}. In practice, we therefore take $n_{\alpha\phi}$ even and $\phi_+\in\{\mathrm{d}\phi/2,3\mathrm{d}\phi/2,\ldots,(n_{\alpha\phi}-1)\mathrm{d}\phi/2\}, \phi_-\in\{0,\mathrm{d}\phi,\ldots,(n_{\alpha\phi}-1)\mathrm{d}\phi\}$.} The integral over $R$ is calculated analytically as explained in section \ref{ssec:intR}.

In our numerical calculation of $C_\phi(z_\phi)$, starting from the reference point $(\eta_{qk}=1/40,n_{\alpha\phi}=60)$, (i) we double $n_{\alpha\phi}$ at fixed $\eta_{qk}=1/40$ until the result changes by less than $10^{-3}$ in relative value (and this uniformly for $z_\phi\in[0,1]$), and (ii) we halve $\eta_{qk}$ at fixed $n_{\alpha\phi}=60$ until the result again changes by less than $10^{-3}$ in relative value. Convergence is reached at point (i) for $n_{\alpha\phi}=240$, and at point (ii) for $\eta_{qk}=1/160$. Performing the calculation directly for $(\eta_{qk}=1/160,n_{\alpha\phi}=240)$ would be time-consuming; we prefer to extrapolate from the reference point as follows: 
\begin{multline}
\label{eq:cphiextrap}
C_\phi(z_\phi) \simeq C_\phi(z_\phi)\Big|^{\eta_{qk}=1/40}_{n_{\alpha\phi}=60}+ 
\left(C_\phi(z_\phi)\Big|^{\eta_{qk}=1/40}_{n_{\alpha\phi}=240}-C_\phi(z_\phi)\Big|^{\eta_{qk}=1/40}_{n_{\alpha\phi}=60}\right)  \\
+
\left(C_\phi(z_\phi)\Big|^{\eta_{qk}=1/160}_{n_{\alpha\phi}=60}-C_\phi(z_\phi)\Big|^{\eta_{qk}=1/40}_{n_{\alpha\phi}=60}\right) 
\end{multline}
which leads to figure \ref{fig2}. For comparison, we also perform the calculation for $(\eta_{qk}=1/80,n_{\alpha\phi}=120)$; the half-sum and half-difference with (\ref{eq:cphiextrap}) lead to the values and error bars given for $z_\phi=0$ and $z_\phi=1$ in (\ref{eq:val01}).

\subsection{Analytical integration over the hyperradius $R$}\label{ssec:intR}

First, we write the reduced collision amplitude (\ref{eq:Abar}) in the synthetic form 
\begin{equation}
\label{eq:syn}
\bar{\mathcal{A}}_{\mathrm{i}\to\textrm{f}}= \sum_{n=1}^{15} \frac{C_n}{\Big(\left|A_nR+\lambda_n\bar{\theta}_{12}\right|^2 +\frac{3}{4}\Big)\Big(\left|B_nR+\mu_n\bar{\theta}_{12}\right|^2+\frac{3}{4}\Big)}
\end{equation}
where the various complex $A_n,B_n$ or real $\lambda_n,\mu_n$ quantities appear in Table \ref{tab:qtes}; then we expand the square $\big|\bar{\mathcal{A}}_{\mathrm{i}\to \textrm{f}}\big|^2$ into 15 square terms $n=n'$ and 105 rectangular terms $n<n'$ (without forgetting the factor of two in these double products) to reduce ourselves in (\ref{eq:fnum}) to 
\begin{equation}
\label{eq:K}
K=\int_0^{+\infty} \frac{\mathrm{d} R\, R^3 Y(R^2+v)}{\prod_{m=1}^{4} \left(|D_m R + \nu_m\bar{\theta}_{12}^{(0)}|^2+\frac{3}{4}\right)}
\end{equation}
where $D_m\in\mathbb{C}, \nu_m\in\mathbb{R}$. In the case $v\neq 0$ where we are (see above), we perform the strictly increasing change of variables 
\begin{equation}
\label{eq:chgt}
R=\frac{|v|^{1/2}}{2}\left(x-\frac{s_v}{x}\right) \quad \mbox{where}\quad s_v\equiv v/|v|\in\{-1,1\}\quad\mbox{and}\quad x\in[1,+\infty[
\end{equation}
 We then verify that the minimum value taken by $R$ in (\ref{eq:chgt}) is 0 if $v>0$ and $|v|^{1/2}$ if $v<0$,
 exactly as required by the Heaviside function in (\ref{eq:K}), and that the value (\ref{eq:thb120num}) of the reduced polar angle ensuring energy conservation, 
\begin{equation}
\bar{\theta}_{12}^{(0)} \stackrel{\mathcal{W}}{=} \frac{|v|^{1/2}}{2(\qb_1\qb_2)^{1/2}}\left(x+\frac{s_v}{x}\right)
\end{equation}
is free of square roots. We are thus reduced to the integral of a rational fraction, 
\begin{equation}
\label{eq:Kbis}
K=\frac{v^2}{16}\int_1^{+\infty} \frac{\mathrm{d} x\, x^3(x^4-1)(x^2-s_v)^2}{\prod_{m=1}^{4} \left(a_m x^4+b_m x^2+c_m\right)} \stackrel{X=x^2}{=}
\frac{v^2}{32} \int_1^{+\infty} \mathrm{d} X\, \frac{X(X-1)^{2+s_v}(X+1)^{2-s_v}}{\prod_{m=1}^{4} \left(a_m X^2+b_m X +c_m\right)}
\end{equation}
with the following notations 
\begin{equation}
\label{eq:defabc}
a_m=\left|\check{D}_m+\check{\nu}_m\right|^2 \quad ; \quad b_m=\frac{3}{4}+2s_v\left(\check{\nu}_m^2-\left|\check{D}_m\right|^2\right)\quad ; \quad c_m=\left|\check{D}_m-\check{\nu}_m\right|^2
\end{equation}
 and 
\begin{equation}
\label{eq:renor}
\check{D}_m=\frac{|v|^{1/2}}{2}D_m \quad ; \quad \check{\nu}_m=\frac{|v|^{1/2}}{2(\qb_1\qb_2)^{1/2}}\nu_m
\end{equation}
The poles $X_i$ of the rational fraction $R(X)$ in the integrand of (\ref{eq:Kbis}) are obtained by solving a quadratic equation; we verify that no pole can be a positive real number (otherwise the underlying $x_i$ would be real and we would have $\big|DR+\nu\thb_{12}^{(0)}\big|^2=-3/4$ with $R$ and $\thb_{12}^{(0)}$ being real, which is impossible); on the other hand, $X_i$ can be a negative real number (the underlying $x_i$ is purely imaginary) or have a non-zero imaginary part, belonging to a pair of complex conjugate poles $X_i$ and $X_i^*$. For the calculation of the integral, we must decompose the rational fraction into simple elements over $\mathbb{C}$ and use the natural antiderivative\footnote{We agree that the last sum in (\ref{eq:primi}) is empty if $n_i=1$. Here the roots $X_i$ are pairwise distinct, and the $n_i$ are their multiplicities.} 
\begin{equation}
\label{eq:primi}
F(X)=\int\mathrm{d} X \, R(X) = \int\mathrm{d} X \, \sum_{i} \sum_{k=1}^{n_i} \frac{c_k(i)}{(X-X_i)^k} = \sum_i c_1(i)\ln(X-X_i) + \sum_{k=2}^{n_i} \frac{c_k(i)}{1-k} \frac{1}{(X-X_i)^{k-1}}
\end{equation}
 to obtain in the general case where $R(X)=o(1/X)$ as $X \to \infty$ and where we therefore have $\sum_i c_1(i)=0$:
\begin{equation}
\label{eq:analy}
K=\frac{v^2}{32}\left[-F(1)\right]
\end{equation}
(special cases to the contrary are discussed in connection with equations (\ref{eq:deuxnuls}) and (\ref{eq:patho})).

 \begin{table}[t]
\begin{center}
\begin{tabular}{||c||ccc||}
\hline
      & $n=1$ & $n=2$ & $n=3$ \\
\hline
$A_nR$ & $0$ & $0$ & $0$  \\
$\lambda_n$ & $ {1}$ & ${1}$ & ${1}$ \\
\hline
$B_nR$ & $\displaystyle\frac{Z_2^{\phantom 1}-Z_3}{1-\kb_1}$ & $\displaystyle\frac{Z_3-Z_1}{1-\kb_2}$ & $\displaystyle\frac{Z_1-Z_2}{1-\kb_3}$  \\
$\mu_n$ & $0$ &  $0$ & $0$  \\
\hline
$C_n$ & ${\kb_1}^{\phantom 1}$ & $\kb_2$ & $\kb_3$  \\
\hline
\end{tabular}

\begin{tabular}{||c||ccc|ccc||}
\hline
      & $n=4$ & $n=5$ & $n=6$ & $n=7$ & $n=8$ & $n=9$ \\
\hline
$A_nR$ &  $\displaystyle\frac{Z_1^{\phantom 1}}{|\qb_1-\kb_1|}$ & $\displaystyle\frac{Z_2}{|\qb_1-\kb_2|}$ & $\displaystyle\frac{Z_3}{|\qb_1-\kb_3|}$ & $\displaystyle\frac{Z_1}{|\qb_2-\kb_1|}$ & $\displaystyle\frac{Z_2}{|\qb_2-\kb_2|}$ & $\displaystyle\frac{Z_3}{|\qb_2-\kb_3|}$ \\
$\lambda_n$ & $\displaystyle\frac{\qb_2}{|\qb_1-\kb_1|}$ & $\displaystyle\frac{\qb_2}{|\qb_1-\kb_2|}$ & $\displaystyle\frac{\qb_2}{|\qb_1-\kb_3|}$ & $\displaystyle\frac{-\qb_1}{|\qb_2-\kb_1|}$ & $\displaystyle\frac{-\qb_1}{|\qb_2-\kb_2|}$ & $\displaystyle\frac{-\qb_1}{|\qb_2-\kb_3|_{\phantom 0}}$ \\
\hline
$B_nR$ & $\displaystyle\frac{Z_2^{\phantom 1}-Z_3}{1-\kb_1}$ & $\displaystyle\frac{Z_3-Z_1}{1-\kb_2}$ & $\displaystyle\frac{Z_1-Z_2}{1-\kb_3}$ & $\displaystyle\frac{Z_2-Z_3}{1-\kb_1}$ & $\displaystyle\frac{Z_3-Z_1}{1-\kb_2}$ & $\displaystyle\frac{Z_1-Z_2}{1-\kb_3}$ \\
$\mu_n$ & $0$ & $0$ & $0$ & $0$ & $0$ & $0$ \\
\hline
$C_n$ & $-\qb_2$ & $-\qb_2$ & $-\qb_2$ & $-\qb_1$ & $-\qb_1$ &$-\qb_1$ \\
\hline
\end{tabular}

\begin{tabular}{||c||ccc|ccc||}
\hline
    & $n=10$ & $n=11$ & $n=12$ & $n=13$ & $n=14$ & $n=15$ \\
\hline 
$A_nR$ & $\displaystyle\frac{Z_1^{\phantom 1}}{|\qb_1-\kb_1|}$ & $\displaystyle\frac{Z_1}{|\qb_1-\kb_1|}$ & $\displaystyle\frac{Z_2}{|\qb_1-\kb_2|}$ & $\displaystyle\frac{Z_1}{|\qb_2-\kb_1|}$ & $\displaystyle\frac{Z_1}{|\qb_2-\kb_1|}$ & $\displaystyle\frac{Z_2}{|\qb_2-\kb_2|}$ \\
$\lambda_n$ & $\displaystyle\frac{\qb_2}{|\qb_1-\kb_1|}$ & $\displaystyle\frac{\qb_2}{|\qb_1-\kb_1|}$ & $\displaystyle\frac{\qb_2}{|\qb_1-\kb_2|}$ & $\displaystyle\frac{-\qb_1}{|\qb_2-\kb_1|}$ & $\displaystyle\frac{-\qb_1}{|\qb_2-\kb_1|}$ & $\displaystyle\frac{-\qb_1}{|\qb_2-\kb_2|_{\phantom 0}}$ \\
\hline 
$B_nR$ & $\displaystyle\frac{Z_2^{\phantom 1}}{|\qb_2-\kb_2|}$ & $\displaystyle\frac{Z_3}{|\qb_2-\kb_3|}$ & $\displaystyle\frac{Z_3}{|\qb_2-\kb_3|}$ & $\displaystyle\frac{Z_2}{|\qb_1-\kb_2|}$ & $\displaystyle\frac{Z_3}{|\qb_1-\kb_3|}$ & $\displaystyle\frac{Z_3}{|\qb_1-\kb_3|_{\phantom 0}}$ \\ 
$\mu_n$ & $\displaystyle\frac{-\qb_1}{|\qb_2-\kb_2|}$ & $\displaystyle\frac{-\qb_1}{|\qb_2-\kb_3|}$ & $\displaystyle\frac{-\qb_1}{|\qb_2-\kb_3|}$ & $\displaystyle\frac{\qb_2}{|\qb_1-\kb_2|}$ & $\displaystyle\frac{\qb_2}{|\qb_1-\kb_3|}$ & $\displaystyle\frac{\qb_2}{|\qb_1-\kb_3|}$ \\
\hline
$C_n$ & ${\kb_3}^{\phantom 1}$ & $\kb_2$ & $\kb_1$ & $\kb_3$ & $\kb_2$ & $\kb_1$ \\
\hline
\end{tabular}
\end{center}
\caption{Coefficients in the synthetic form (\ref{eq:syn}) of the reduced collision amplitude $2\phi\to 3\phi$. The numbering from one to fifteen follows the order of the sums and the lexicographic order of their terms in its expression (\ref{eq:Abar}). For simplicity, we have focused on the variety $\mathcal{W}$ (\ref{eq:defW}), as we do in calculation.}
\label{tab:qtes}
\end{table}

 In practice, the analytical expression (\ref{eq:analy}) taken from (\ref{eq:primi}) is quite tricky to use (as is the one giving the roots of the trinomial $aX^2+bX+c$ in the form $(-b\pm\Delta^{1/2})/2a, \Delta=b^2-4ac$, when $ac/b^2$ is less than the numerical precision--it is better then to express the roots by their Taylor expansion in powers of $ac/b^2$). Suppose, in fact--this is the worst-case scenario--that the eight poles $X_i$ are pairwise distinct but very close to a fixed real point $x_0<1$, which amounts to taking $X_i=x_0+\eta Y_i, Y_i\,\mbox{fixed}, \eta\to 0^+$, and consider the simplified integral 
\begin{equation}
\label{eq:defKsimp}
K_{\rm simp}(\eta)=\int_1^{+\infty}\mathrm{d} X\, R_{\rm simp}(X) \quad \mbox{with}\quad R_{\rm simp}(X) = \frac{1}{\prod_{i=1}^{8} (X-X_i)} = \sum_{i=1}^{8} \frac{c_1(i)}{X-X_i}
\end{equation}
The coefficients of the simple elements are written as $c_1(i)=C(i)/\eta^7$, with $C(i)=\prod_{j\neq i} (Y_i-Y_j)$ independent of $\eta$. According to equations (\ref{eq:primi}, \ref{eq:analy}), we then have 
\begin{equation}
\label{eq:dur}
K_{\rm simp}(\eta)=-\frac{1}{\eta^7}\sum_{i=1}^{8} C(i)\ln(1-x_0-\eta Y_i)
\end{equation}
 The problem is that the integral $K_{\rm simp}(\eta)$ is, according to (\ref{eq:dur}), a sum of terms individually of order $\eta^{-7}$, whereas, according to its definition (\ref{eq:defKsimp}), it has a finite limit $\int_1^{+\infty}\mathrm{d} X/(X-x_0)^8=(1/7)(1-x_0)^{-7}$ when $\eta\to 0^+$: in the Taylor expansion of the sum (\ref{eq:dur}) in powers of $\eta$, exact compensations must occur from order $\eta^0$ to $\eta^6$ inclusive. On a computer with non-infinitesimal numerical precision $\pi_{\rm num}$, the analytical form is therefore usable only if 
\begin{equation}
\eta^7 \gg  \pi_{\rm num}, \quad\mbox{so}\quad \left\{\begin{array}{lcl} 
\eta\gtrsim 10^{-2}  & \mbox{in double precision} & \pi_{\rm num}\approx 10^{-15} \\
&&\\
\eta\gtrsim 10^{-4}  & \mbox{in quadruple precision}& \pi_{\rm num}\approx 10^{-31}
\end{array}
\right.
\end{equation}
In practice, it is therefore better to merge poles $X_i$ that are less than $\eta$ apart before using (\ref{eq:analy}).\footnote{We equate two poles $X_i$ and $X_j$ to their average weighted by their multiplicities if $|X_i-X_j|<\eta\,\mathrm{max}\,(1,(|X_i|+|X_j|)/2)$. } Since quadruple precision is time-consuming and $\eta=10^{-2}$ is not sufficiently small (this choice would lead to merging complex conjugate poles with real parts strictly greater than one, which would in turn lead to an incorrect result)\footnote{For example, for a fixed $x_0>1$ and $\eta\to 0^+$, we have $\int_1^{+\infty} \mathrm{d} X \, \phi(X)/[(X-x_0-\mathrm{i}\eta)(X-x_0+\mathrm{i}\eta)]=\pi\phi(x_0)/\eta+\phi(x_0)/(1-x_0)-\phi'(x_0)\ln(x_0-1)+\ldots$, where $\phi(X)$ is a regular function at $x_0$ and we have isolated the contribution of the poles $x_0\pm \mathrm{i}\eta$ to the integral; the coalescence and indiscriminate use of (\ref{eq:primi})--with real-part extraction--would cause us to miss the divergent part $\propto 1/\eta$.}, we have adopted a compromise strategy: (i) calculations are generally performed in double precision, with a coalescence distance of $\eta=10^{-3}$; (ii) in the case of multiplicity greater than or equal to four or an absurd result (a negative value of $K$), we switch locally and temporarily to quadruple precision; (iii) this strategy is validated for numerical grids that are not too large, $n_{\alpha\phi}=60$ and $\eta_{qk}=1/40$, by a calculation performed entirely in quadruple precision with $\eta=10^{-4}$ (the computation time is half a day on 128 cores, and the result differs from double precision by less than $2\cdot 10^{-6}$ in relative value).

Another difficulty arises in the analytical evaluation of (\ref{eq:Kbis}) using (\ref{eq:analy}), namely the cancellation of one or more coefficients $a_m$, more precisely their quasi-cancellation in the numerical calculation (in the sense that, for example, $|a_m|<10^{-12}(|b_m|+|c_m|)$): 
\begin{itemize}\item[(i)] If a single coefficient $a_m$ is quasi-zero, it is directly replaced by zero in (\ref{eq:Kbis}) without triggering a divergence at infinity. We are simply reduced to a rational fraction $R_7(X)$ with a denominator of degree 7, whose primitive is calculated as in (\ref{eq:primi}), and we can take $K\simeq (v^2/32)[-F_7(1)]$. \item[(ii)] In the case where two coefficients $a_m$ are nearly zero, for example $a_1$ and $a_2$, setting them to zero leads to a rational fraction $R_6(X)$ with a denominator of degree 6; since the numerator is of degree 5,
 this triggers a logarithmic divergence at infinity. Nevertheless, we show that we can use the expression \footnote{Let us choose for this a cutoff $\Lambda$ much larger than unity and than the bounded poles $X_i, 1\leq i\leq 6$ of $R_8(X)$, which can be approximated by the poles of $R_6(X)$, but much smaller than the divergent poles $X_0$ and $X_0'$ of $R_8(X)$. In the integral on the right-hand side of (\ref{eq:Kbis}), we can then approximate the integrand by $R_6(X)$ over the interval $X\in [0,\Lambda]$ and by $1/[a_1a_2a_3a_4 X (X-X_0)(X-X_0')]$ over the interval $X\in[\Lambda,+\infty[$. It remains to apply (\ref{eq:primi}) to each interval, noting that $F_6(\Lambda)\simeq (\ln\Lambda)/(b_1 b_2 a_3 a_4)$.} 
\begin{equation}
\label{eq:deuxnuls}
K\simeq \frac{v^2}{32}\left[-F_6(1) + \frac{X_0\ln(-X_0')-X_0'\ln(-X_0)}{X_0-X_0'} \frac{1}{b_1 b_2 a_3 a_4}\right]
\end{equation}
 by introducing an error that tends to zero with $(a_1^2+a_2^2)^{1/2}$. Here, $X_0=-b_1/a_1$ and $X_0'=-b_2/a_2$ are an approximation of the divergent poles in the original rational fraction $R_8(X)$, and the function $F_6(X)$ is the antiderivative of $R_6(X)$ given by equation (\ref{eq:primi}) after performing a partial fraction decomposition on $R_6(X)$; the first contribution in (\ref{eq:deuxnuls}) is therefore merely the result of a blind--and incorrect--application of the analytical expression (\ref{eq:analy}) to $R_6(X)$. The special case $X_0=X_0'$ is handled using L'Hospital's rule (we replace the first fraction of the second term in (\ref{eq:deuxnuls}) with its limit when $X_0'\to X_0$, that is, with $-1+\ln(-X_0)$). Of course, formula (\ref{eq:deuxnuls}) fails if the two coefficients $a_1$ and $a_2$ are exactly zero; as seen in the definition (\ref{eq:defabc}) of $a_m$ and in their explicit values taken from Table \ref{tab:qtes}, this event is highly improbable (and does not occur in our numerical calculation) when the phases $\phi_+$ and $\phi_-$ differ modulo $\pi$, hence the shifted grids in Note \ref{note:decale}.
\item[(iii)] If three or four coefficients $a_m$ are nearly zero, the divergence becomes linear or quadratic in the inverse distance of these coefficients from zero. But this requires both $\check{A}_n+\check{\lambda}_n\simeq 0$ and $\check{B}_n+\check{\mu}_n\simeq 0$ for one of the fifteen terms of (\ref{eq:syn}), the renormalizations marked by a Czech accent being those of equation (\ref{eq:renor}), or if we prefer $A_n+\lambda_n/(\qb_1\qb_2)^{1/2}\simeq 0$ and $B_n+\mu_n/(\qb_1\qb_2)^{1/2}\simeq 0$. We then show that strict double cancellation does not occur, and that there is no problem. Let us verify this for the term $n=10$: the double cancellation imposes $Z_1|_{R=1}=-(\qb_2/\qb_1)^{1/2}$ and $Z_2|_{R=1}=(\qb_1/\qb_2)^{1/2}$. Given equation (\ref{eq:cgz12}) and the phase-modulus representation above (\ref{eq:fnum}), this constitutes a Cramer system for the two unknowns $\cos\alpha \exp(\mathrm{i}\phi_+)$ and $\sin\alpha \exp(\mathrm{i}\phi_-)$, which is easily solved. Since the sum of the squares of the two unknowns must equal one, we obtain, recalling Note \ref{note:interm}, condition 
\begin{equation}
\label{eq:patho}
\frac{\qb_1(1-\kb_1)}{(1-\qb_1)\kb_1} +\frac{(1-\qb_1)(1-\kb_2)}{\qb_1\kb_2} -2 \stackrel{\mathcal{W}}{=} \frac{1-\kb_1-\kb_2}{\kb_1\kb_2}
\end{equation}
This is ultimately a quadratic equation in $\qb_1$; we verify that the two roots are simply $\qb_1=\kb_1$ and $\qb_1=1-\kb_2$ (i.e., $\qb_2=\kb_2$ on the manifold $\mathcal{W}$). This result extends to the other terms of (\ref{eq:syn})--since the reduced wave numbers are nonzero, we always encounter a condition $\kb_i=\qb_\alpha, i\in\{1,2,3\},\alpha\in\{1,2\}$ for $n$ greater than or equal to 3. Then (i) this condition is violated on the numerical grid of $\qb_2$, which excludes the notable points $\kb_i$ and $1-\kb_i$, and (ii) in the limit $\mathrm{d}\qb_2\to 0^+$, this condition would tend to be satisfied, but the corresponding term in the amplitude $\bar{\mathcal{A}}_{\mathrm{i}\to\textrm{f}}$ is canceled out by the divergence of $1/(\kb_i-\qb_\alpha)^2$ in the denominator. 
\end{itemize}
\bibliographystyle{crunsrt}

\nocite{*}

\bibliography{english}
\end{document}